%% file: jfm.tex
\documentclass[lineno]{jfm}

\usepackage{graphicx}
\usepackage{floatrow}
\usepackage{subfig}
\usepackage[justification=justified]{caption}
\usepackage{newtxtext}
\usepackage{newtxmath}
\usepackage{natbib}
\usepackage{orcidlink}
\usepackage{hyperref}
\hypersetup{
    colorlinks = true,
    urlcolor   = blue,
    citecolor  = black,
}
\usepackage{bm}

\newcommand{\RomanNumeralCaps}[1]
\linenumbers
\usepackage{float}
\usepackage{longtable,tabularx}
\usepackage{booktabs}
\usepackage{multirow}
\usepackage{upgreek}
\usepackage{tikz,pgfplots}
\usepackage{pgfplotstable}
\usetikzlibrary{plotmarks}
\usetikzlibrary{shapes}
\usetikzlibrary{calc,shadows}\usepackage{comment}
\usetikzlibrary{positioning,spy, intersections}
\usepackage{upgreek}
\usetikzlibrary{arrows.meta}

\usepackage{color}

\pgfplotsset{ 
  compat=1.8, 
  legend style =
  {font=\small\sffamily},
  label style = {font=\small\sffamily}
}

\definecolor{redd}{rgb}{0.9373, 0.4196, 0.3490} 
\definecolor{greenn}{rgb}{ 0.4078, 0.5922, 0.1608}
\definecolor{greennn}{rgb}{0.0039, 0.5216, 0.4431}
\definecolor{bluee}{rgb}{0.0549, 0.3451, 0.7686}
\definecolor{pat_grey_1}{rgb}{0, 0, 0}
\definecolor{pat_grey_2}{rgb}{0.4902,0.4902,0.4902}
\definecolor{pat_grey_3}{rgb}{0.7059,0.7059,0.7059}
\definecolor{pat_purple}{rgb}{0.51, 0.048, 0.67}
\definecolor{pat_cyan}{rgb}{0.302, 0.85,1}
\definecolor{pat_orange}{rgb}{1,0.2,0.051}
\definecolor{pat_blue}{rgb}{0.35,0.54,0.93}
\definecolor{pat_green}{rgb}{0.38, 0.9, 0.29}
\definecolor{pat_green2}{rgb}{0, 0.6, 0.3}
\definecolor{blue_particle}{rgb}{0.37,0.57,0.78}

\newcommand{\greenlinesolidthicker}{\raisebox{2pt}{\tikz{\draw[-,pat_green,solid,line width = 1.2pt](0,0) -- (5mm,0);}}}

\newcommand{\bluelinesolidthicker}{\raisebox{2pt}{\tikz{\draw[-,pat_blue,solid,line width = 1.2pt](0,0) -- (5mm,0);}}}

\newcommand{\purplelinesolidthicker}{\raisebox{2pt}{\tikz{\draw[-,pat_purple,solid,line width = 1.2pt](0,0) -- (5mm,0);}}}

\newcommand{\starfillbluethree}{\raisebox{-2pt}{%
  \tikz{
  \node[star, star points=6, minimum size=8pt,
          inner sep=0pt, star point ratio=1.75, fill=pat_blue, draw, pat_blue](key) {};  
}}}

\newcommand{\starfillgreenthree}{\raisebox{-2pt}{%
  \tikz{
  \node[star, star points=6, minimum size=8pt,
          inner sep=0pt, star point ratio=1.75, fill=pat_green, draw, pat_green](key) {};  
}}}

\newcommand{\starfillpurplethree}{\raisebox{-2pt}{%
  \tikz{
  \node[star, star points=6, minimum size=8pt,
          inner sep=0pt, star point ratio=1.75, fill=pat_purple, draw, pat_purple](key) {};  
}}}

\newcommand{\starhollowblack}{\raisebox{-2pt}{%
  \tikz{
  \node[star, star points=6, minimum size=8pt,
          inner sep=0pt, star point ratio=1.75, draw](key) {};  
}}}

\newcommand{\redlinedashedthickk}{\raisebox{2pt}{\tikz{\draw[-,redd,dashed,line width = 1.2pt](0,0) -- (5mm,0);}}}

\newcommand{\blacklinedashedthickk}{\raisebox{2pt}{\tikz{\draw[-,black,dashed,line width = 1.2pt](0,0) -- (5mm,0);}}}

\newcommand{\blacklinesolidthickk}{\raisebox{2pt}{\tikz{\draw[-,black,solid,line width = 1.2pt](0,0) -- (5mm,0);}}}

\newcommand{\trianglehollow}{\raisebox{0pt}{\tikz{\draw[black,solid,line width = 0.5pt](0.mm,0) -- (2.5mm,0mm)--(1.3mm,2.0mm)--(0mm,0mm) ;}}}

\newcommand{\trianglefillPurple}{\raisebox{0pt}{\tikz{\draw[pat_purple,solid,line width = 0pt,fill=pat_purple](0.mm,0) -- (2.5mm,0mm)--(1.3mm,2.0mm)--(0mm,0mm) ;}}}

\newcommand{\righttrianglefillPurple}{\raisebox{0pt}{\tikz{\draw[rotate=270, pat_purple,solid,line width = 0pt,fill=pat_purple](0mm,0) -- (2.5mm,0mm)--(1.3mm,2.0mm)--(0mm,0mm) ;}}}

\newcommand{\rectanglefillPurplelarge}{\raisebox{0pt}{\tikz{\draw[pat_purple,solid,line width = 0pt,fill={rgb:pat_purple,1;white,0}](2.mm,0) rectangle (4mm,2mm)}}}

\newcommand{\rectanglehollow}{\raisebox{0pt}{\tikz{\draw[black,solid,line width = 0.5pt](2.mm,0) rectangle (3.5mm,1.5mm)}}}

\newcommand{\Cross}{$\mathbin{\tikz [x=1.4ex,y=1.4ex,line width=.2ex, black] \draw (0,0) -- (0.8,0.8) (0,0.8) -- (0.8,0);}$}

\newcommand{\triangleCrossPurple}[1][]{\begin{tikzpicture}[#1]
\draw[black,solid,line width = 0pt,fill={rgb:red,0.51;green, 0.048; blue, 0.67;white,0}](0,0) -- (1.2,0)--(0.6,1)--(0,0);
\draw [white,solid,line width = 1pt](0.2,0)--(1,0.75);
\draw [white,solid,line width = 1pt](0,0.85) -- (1,0);
\end{tikzpicture}}

\newcommand{\SquareCrossPurple}[1][]{\begin{tikzpicture}[#1]
\draw[black,solid,line width = 0pt,fill={rgb:red,0.51;green, 0.048; blue, 0.67;white,0}](-0.09,-0.09) -- (1,-0.09)--(1,1)--(-0.09,1)--(-0.09,-0.09);
\draw [white,solid,line width = 1pt](0,0)--(0.9,0.9);
\draw [white,solid,line width = 1pt](0,0.9) -- (0.9,0);
\end{tikzpicture}}

\newcommand{\plussign}{$\mathbin{\tikz [x=1.4ex,y=1.4ex,line width=.2ex, black] \draw (0,0.5) -- (1,0.5) (0.5,0) -- (0.5,1);}$}

\newcommand{\asterisksign}{$\mathbin{\tikz [x=1ex,y=1ex,line width=.2ex, black] \draw (-0.205,0.5) -- (1.205,0.5) (0.5,-0.205) -- (0.5,1.205) (0,0) -- (1,1) (0,1) -- (1,0);}$}

\newcommand{\circlefillPurple}{\raisebox{0pt}{\tikz{\draw[pat_purple,solid,line width = 0.5pt,fill={rgb:pat_purple,1;white,0}](3.mm,0.8mm) circle (1mm);}}}

\newcommand{\circlefillBluetwo}{\raisebox{0pt}{\tikz{\draw[pat_blue,solid,line width = 0.5pt,fill={rgb:pat_blue,1;white,0}](3.mm,0.8mm) circle (1mm);}}}

\newcommand{\circlefillGreentwo}{\raisebox{0pt}{\tikz{\draw[pat_green,solid,line width = 0.5pt,fill={rgb:pat_green,1;white,0}](3.mm,0.8mm) circle (1mm);}}}

\newcommand{\rectanglefillbluevel}{\raisebox{0pt}{\tikz{\draw[blue,solid,line width = 0.75pt,fill=blue](2.mm,0) rectangle (3.5mm,1.5mm)}}}

\newcommand{\rectanglefillbluetvel}{\raisebox{0pt}{\tikz{\draw[black,solid,line width = 0.75pt,fill=blue, fill opacity=0.4](2.mm,0) rectangle (3.5mm,1.5mm)}}}

\newcommand{\rectanglefillredvel}{\raisebox{0pt}{\tikz{\draw[red,solid,line width = 0.75pt,fill=red](2.mm,0) rectangle (3.5mm,1.5mm)}}}

\newcommand{\rectanglefillredtvel}{\raisebox{0pt}{\tikz{\draw[black,solid,line width = 0.75pt,fill=red, fill opacity=0.4](2.mm,0) rectangle (3.5mm,1.5mm)}}}

\title{Experimental and numerical investigation of inertial particles in underexpanded jets}

\author{Meet Patel\aff{1}
  \corresp{\email{meetm@umich.edu}},
  Juan Sebastian Rubio\aff{2},
  David Shekhtman\aff{3},
  Nick Parziale\aff{3},
  Jason Rabinovitch\aff{3},
  Rui Ni\aff{2},
  Jesse Capecelatro\aff{1,4}
  }

\affiliation{\aff{1}Department of Aerospace Engineering, University of Michigan, Ann Arbor, MI 48109, USA
\aff{2}Department of Mechanical Engineering, Johns Hopkins University, Baltimore, MD 21218, USA
\aff{3}Department of Mechanical Engineering, Stevens Institute of Technology, Hoboken, NJ 07030, USA
\aff{4}Department of Mechanical Engineering, University of Michigan, Ann Arbor, MI 48109, USA}

\begin{document}
\newlength\figureheight
\newlength\figurewidth
\newlength\mylinewidth
\newlength\mymarksize

\setlength\mylinewidth{0.4pt}

\maketitle
\nolinenumbers
\begin{abstract}
Experiments and numerical simulations of inertial particles in underexpanded jets are performed. The structure of the jet is controlled by varying the nozzle pressure ratio, while the influence of particles on emerging shocks and rarefaction patterns is controlled by varying the particle size and mass loading. Ultra-high-speed schlieren and Lagrangian particle tracking are used to experimentally determine the two-phase flow quantities. Three-dimensional simulations are performed using a high-order, low dissipative discretization of the gas phase while particles are tracked individually in a Lagrangian manner. A simple two-way coupling strategy is proposed to handle interphase exchange in the vicinity of shocks. Velocity statistics of each phase are reported for a wide range of pressure ratios, particle sizes, and volume fractions. The extent to which particles affect the location of the Mach disk are quantified and compared to previous work from the literature. Furthermore, a semi-analytic model is presented based on a one-dimensional Fanno flow that takes into account volume displacement by particles and interphase exchange due to drag and heat transfer. The percent shift in Mach disk is found to scale with the mass loading, nozzle pressure ratio, interphase slip velocity, and inversely with the particle diameter.
\end{abstract}

\section{Introduction} \label{sec:intro}
Compressible flows containing inertial (i.e. heavy) particles can be found in many engineering applications and natural phenomena. Coal dust explosions~\citep{Coal_Dust_explosion}, volcanic eruptions~\citep{volcanic,lube2020multiphase}, solid propellant combustion in rocket engines~\citep{davenas2012solid} and plume-surface interactions during the powered descent of spacecraft~\citep{Mehta_PSI, BALAKRISHNAN_PSI,capecelatro2022modeling} are few such examples. In all these cases, the flows exhibit strong coupling between gas-phase compressibility, turbulence, and solid particles. This work deals with particle-laden underexpanded jets as a canonical flow configuration for studying the transport of particles through shocks and their back coupling on the gas phase.

Dedicated studies on particle-laden compressible jets date back to the 1960s, primarily motivated by solid propellant-based rocket combustion \citep{bailey1961gas,hoglund1962recent,marble1963nozzle,lewis1964normal,bauer1965normal,jarvinen1967underexpanded}. The location of the normal shock wave (or Mach disk), $L_{\rm MD}$, is a key quantity since it affects the radiation of the plume and downstream structure of the jet~\citep{Franquet2015}. Experiments by \cite{lewis1964normal} revealed an upstream movement of the Mach disk by as much as 30\% with the addition of particles. The movement was found to increase with increasing particle-to-gas mass fraction, $\mathit{\Phi_m}$, and be independent of nozzle pressure ratio, defined as $\eta_0\equiv p_0/p_\infty$, where $p_0$ and $p_\infty$ are the total and ambient pressures, respectively. They proposed an empirical correlation for the percent change in $L_{\rm MD}$ that depends only on $\mathit{\Phi_m}$ and the nozzle exit Mach number, $M_e$. Semi-analytic models were proposed in the ensuing years that treat the two-phase mixture as an equivalent perfect gas so the usual one-dimensional gas dynamics can be applied \citep[e.g.][]{bauer1965normal,jarvinen1967underexpanded,Marble1970}. The models assume that the particles are sufficiently small so that the slip velocity between the phases is negligible. The results showed reasonable agreement to the experiments of \citet{lewis1964normal}. 

Since the 1960s, advancements in experimental diagnostics and numerical methods have provided new insights into the effect particles have on the structure of underexpanded jets. Numerical simulations of rocket exhaust plumes by \cite{dash1985analysis} showed that decreasing the particle size resulted in the \textit{downstream} movement of the Mach disk, i.e. an increase in $L_{\rm MD}$ counter to the observations by \citet{lewis1964normal}. In contrast, two-dimensional axisymmetric simulations of \cite{sommerfeld1994structure} showed an upstream movement of the Mach disk that was more pronounced for smaller particles. Because the Stokes number scales with the square of the particle diameter, it was hypothesized that smaller particles exhibit a greater radial spread, and thus affect a larger portion of the shock structure. Reynolds-averaged Navier-Stokes (RANS) simulations of \citet{carcano2013semi} predicted a similar upstream movement of the Mach disk. Recently, \citet{Ejtehadi2018} simulated conditions similar to \citet{sommerfeld1994structure} using an Eulerian-based two-fluid model. The Mach disk was found to move downstream for small particles and low concentrations, consistent with \citet{dash1985analysis}. In a recent experimental study by \citet{jain2024experimental}, the upstream movement of the Mach disk was found to have a strong dependence on the nozzle pressure ratio, counter to what was found in previous works. Thus, there appears to be a lack of consensus regarding both the extent and direction of the Mach disk shift caused by particles. These studies are summarized in table~\ref{table:one}.

\begin{table}[ht]
  \begin{center}
  \def~{\hphantom{0}}
  \begin{tabular}{lcccccccc}
       &$\eta_0$  & $d_p$   & $\mathit{\mathit{\Phi_m}}$ & $\mathit{\mathit{\Phi_v}}$ $(\times 10^{-3})$ &  ${ {L}}_{\rm MD}/D_e$ &$\rho_p$& $M_e$ \\ 
       &--  & $\upmu {\rm m}$   & -- &-- &  --  &${\rm kg}/{\rm m^3}$& -- \\ [3pt] \hline
       \multirow{2}{*}{\cite{sommerfeld1994structure}} &33 & 26 &$0-1.08$ &  \multirow{2}{*}{1} &  $3.79-2.30$ & \multirow{2}{*}{2500} &\multirow{2}{*}{1} \\
        &28 & 45 &$0-1.08$&  & $3.45-2.51$  && \\   \hline
       \cite{Ejtehadi2018} &29.8 & 26 &$0-1.08$ & 1 & $3.71-3.695$ &2500& 1\\ \hline
       \multirow{2}{*}{\cite{lewis1964normal}} & $*$ & 28 &$0-0.6$ &  \multirow{2}{*}{$*$} &    \multirow{2}{*}{$*$} &\multirow{2}{*}{2700} & 1.75\\
        & $*$ & 28 &$0-1.08$&  &      && 2.905\\   \hline
      {\cite{carcano2013semi}} &{58} &{10} &$0.21-1.73$ & $0.5-4$ &  $4.94-4.16$  &{2500} & {1} \\  \hline
         \multirow{5}{*}{\cite{jain2024experimental}} &3.46 & \multirow{5}{*}{116} & $2.25-2.75$
        & $2.3-2.9$ & $0.6-0.7$  &\multirow{5}{*}{2500} & \multirow{5}{*}{1} \\ 
        & 4.16 &  & $1.85-2.25$ & $2.3-2.8$ &  $0.7-0.9$ &  & \\
        & 4.85 &  & $1.6-2$ &$2.3-2.9$ &   $0.8-0.95$  & & \\
        & 5.54 &  & $1-1.4$&$1.7-2.3$ &  $1-1.2$   &&\\
        & 6.23 &  & $0.9-1.35$& $1.7-2.5$  & $0.95-1.3$ &&\\\hline
  \end{tabular}
  \caption{Summary of previous studies on particle-laden underexpanded jets. `$*$' indicates the data was not reported and could not be inferred.}
  \label{table:one}
  \end{center}
\end{table}

To date, a theoretical understanding of the gas-particle dynamics in compressible jets remains limited. The extent to which particles influence the carrier-phase turbulence is typically characterized by $\mathit{\Phi_m}$ (or volume fraction, $\mathit{\Phi_v}$) and particle diameter, $d_p$. For incompressible flows, two-way coupling becomes apparent when the particle diameter is significantly larger than the Kolmogorov length scale, or when the mass loading is non-negligible \citep[see][]{Balachandar2010}. 

In low-speed, incompressible jets, particle dispersion is entirely controlled by the Stokes number \citep{chung1988simulation,longmire1992structure,lau2014influence,li2011direct,monroe2021role}, defined as $St=\tau_p/\tau_f$, where $\tau_p$ and $\tau_f$ are characteristic time scales of the particles and fluid, respectively. When $St\ll 1$, particles match the fluid dispersion rate and conversely, when $St\gg 1$, particle dispersion lags that of the fluid. At intermediate values, particles are capable of dispersing faster than the fluid and being ejected outside the jet. The spatial distribution of particles in an incompressible jet is strongly influenced by the underlying vortex ring structures \citep{longmire1992structure,monroe2021role}. Particles, irrespective of their size, tend to preferentially accumulate in regions in the jet where the streamwise velocity is greater than the mean \citep{li2011direct}.  At a fixed mass loading, small Stokes number particles are more successful at modulating three-dimensional vortex structures than at intermediate or large Stokes numbers \citep{li2011direct}. This seems to be consistent with the findings of \citet{sommerfeld1994structure} for the modulation of shock structures in underexpanded jets.

Compared to incompressible flows, the analysis of particle-laden compressible flows is complicated by the manifestation of additional length- and time-scales, such as those arising from acoustic and shock waves \citep{capecelatro2024gas}. Direct numerical simulations (DNS) of homogeneous, compressible turbulence conducted by \citet{xia2016modulation} demonstrated that dilute suspensions of heavy particles tend to suppress gas-phase dilatation. This suppression leads to weaker shocklets and lower turbulent Mach numbers. DNS of inertial particles in a spatially developing compressible turbulent boundary layer by \citet{xiao2020eulerian} revealed a unique preferential concentration mechanism specific to compressible flows. Namely, larger particles have a tendency to accumulate in regions of low gas-phase density in the inner zones and high-density regions in the outer zones, while small particles remain in regions of low density.

Experiments and numerical simulations involving water injection in high-speed jets have shown that particles are capable of modulating acoustic radiation, resulting in changes to the near-field and far-field sound pressure levels (SPL) \citep{henderson2010fifty,krothapalli2003turbulence,buchta2019sound}. This has been attributed to a combination of interphase momentum exchange, work due to volume displacement caused by the disperse phase and latent heat due to evaporation. In recent years, ultra-high-speed holographic velocimetry has revealed that individual particles alter the shock structure of underexpanded jets as they pass through them \citep{ingvorsen2012ultra,Buchmann2012}. Meanwhile, the mechanisms contributing to alterations in the structure of compressible jets are not well known.

In this study, a series of high-resolution experiments and simulations of particle-laden underexpanded jets are performed to quantify two-phase flow statistics and better understand the effects of two-way coupling for a range of nozzle pressure ratios, particle sizes, and volume fractions. The following section describes the particle-laden jet configuration and provides details on the experimental facility. The governing equations and discretization of the numerical simulations are then given in \S\ref{sec:comp}. Next, comparisons between the experiments and simulations and analysis of the Mach disk characteristics are given in \S\ref{sec:results}. Finally, a semi-analytic model describing the effect of particles on the Mach disk location is given in \S\ref{sec:model}.

\section{Particle-laden jet configuration}\label{sec:config}

\begin{figure}
\centering
\includegraphics[width = 1\textwidth]{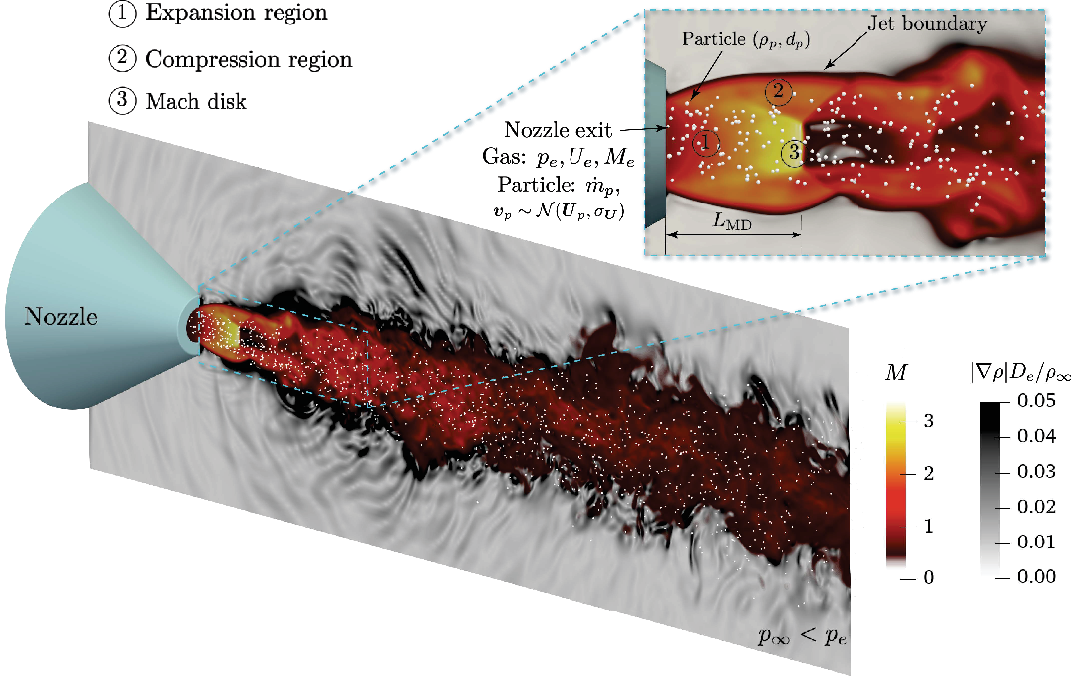}
\caption{Numerical simulation of the underexpanded particle-laden jet showing the nozzle (blue), particles (white), local gas-phase Mach number (red/yellow) and density gradient (grey scale).}
\label{fig:schematic}
\end{figure}

The particle-laden underexpanded jet configuration is shown in figure~\ref{fig:schematic}. A high-pressure gas is discharged from a sonic nozzle with an exit diameter $D_e=2$ mm. As the gas exits the nozzle, it undergoes rapid expansion and acceleration, giving rise to supersonic flow downstream of the nozzle exit (refer to region 1 in the figure). Expansion waves initiated at the nozzle exit approach the jet boundary and subsequently reflect back towards the jet axis as weak compression waves (region 2 in the figure). When these waves coalesce, they form oblique shocks that meet at the jet axis. For nozzle pressure ratios $\eta_0\gtrapprox 4$ \cite[see][]{Franquet2015}, the oblique shocks no longer meet at the jet axis, and a Mach disk emerges (region 3 in the figure). A shear layer forms at the triple point, where the Mach disk and reflected shock merge.

Particles of diameter $d_p$ and density $\rho_p$ are injected upstream of the nozzle with a prescribed mass flow rate, $\dot{m}_p$. The mass loading is defined as the ratio of specific masses between the particles and the fluid, i.e. $\mathit{\Phi_m} = \rho_p\mathit{\Phi_v}/(\rho_e(1-\mathit{\Phi_v}))$, where $\mathit{\Phi_v}$ is the particle volume fraction and $\rho_e$ is the gas-phase density at the nozzle exit that can be determined from isentropic relations (see Appendix~\ref{app:nozzle}). In this study, the mass loading is defined according to the ratio of mass flow rates, $\mathit{\Phi_m}=\dot{m}_p/\dot{m}_f$, since it is easier to measure experimentally. With this, the average volume fraction at the exit of the nozzle can be determined according to $\mathit{\Phi_v}=\mathit{\Phi_m} \rho_e/[\rho_p(1+\mathit{\Phi_m}\rho_e/\rho_p)]$.

The velocity mismatch between the phases gives rise to a slip velocity that determines the particle Reynolds number ${{Re}_p}$ and Mach number ${{M}_p}$ at the nozzle exit. The jet Reynolds number is defined as $Re=\rho_e U_e D_e/\mu$, where $\mu$ is the dynamic viscosity of the gas at the nozzle exit. The Stokes number is defined as $St=\tau_p/\tau_f$, where $\tau_p{=}\rho_p d_p^2/(18 \mu F_d)$ is the particle response time due to drag, $\tau_f=D_e/U_e$ is the characteristic fluid time scale based on the exit parameters and $F_d$ is the non-dimensional drag correlation of \citet{osnes2023comprehensive} (see \S\ref{subsec:particle_subsec}) based on the exit conditions.

A summary of the relevant parameters considered in this study are given in table~\ref{table: params}. 

\begin{table}
  \begin{center}
  \def~{\hphantom{0}}
  \begin{tabular}{llcccccc}
    Case && A1 & A2& B1 & B2 & B3 &B4 \\
    \midrule
    \multicolumn{8}{c}{Physical parameters} \\
    $p_0$ & Total (tank) pressure ($\mathrm{kPa}$)   & 345 & 345 & 655 & 655 & 655 & 655\\
    $\overline{d}_p$& Mean particle diameter ($\upmu\mathrm{m}$) & 29 & 29 & 42 & 42 & 96 & 96\\
    $\dot{m}_p$& Particle mass flow rate (g/s) & 0.4 & 0.96 & 1.02 & 2.2 & 3.7 & 4.2\\
    $U_p$& Mean particle injection &\multirow{ 2}{*}{ 159} & \multirow{ 2}{*}{154.8} & \multirow{ 2}{*}{151.3} &\multirow{ 2}{*} {145.9} & \multirow{ 2}{*}{114} & \multirow{ 2}{*}{113}   \\
    & velocity ($\mathrm{m}/\mathrm{s}$) & \multicolumn{6}{c}{} \\
    \multicolumn{8}{c}{} \\
   \multicolumn{8}{c}{Non-dimensional parameters} \\
    $\eta_0$ &Nozzle pressure ratio& 3.4 & 3.4 & 6.46 & 6.46 & 6.46 & 6.46 \\
    $\mathit{\Phi_v}$& Particle volume fraction & \multirow{ 2}{*}{0.29} & \multirow{ 2}{*}{0.77} & \multirow{ 2}{*}{0.84} & \multirow{ 2}{*}{1.7} & \multirow{ 2}{*}{3.1} &\multirow{ 2}{*}{3.5} \\
    & ($\times 10^{-3}$) & \multicolumn{6}{c}{} \\
    $\mathit{\Phi_m}$ & Mass loading & 0.15  & 0.37  & 0.21  & 0.44 &  0.78 & 0.88 \\
    $St$ &Stokes number & 34.16 & 31.85 & 30.23 & 28.6 & 49.02 & 48.41  \\
    $Re$ &Jet Reynolds number & 9.03 & 9.03 & 17.2 & 17.2 & 17.2 & 17.2 \\
     & ($\times 10^4$) & \multicolumn{6}{c}{} \\
    \bottomrule
  \end{tabular}
  \caption{Summary of parameters considered in the current work. In each case, gas is discharged through a nozzle  with diameter $D_e=2$ mm into ambient conditions with an exit velocity $U_e=312$ m/s and Mach number $M_e=1$.}
  \label{table: params}
  \end{center}
\end{table}

\section{Experimental setup} \label{sec: Experimental setup}

\subsection{High-speed jet facility}
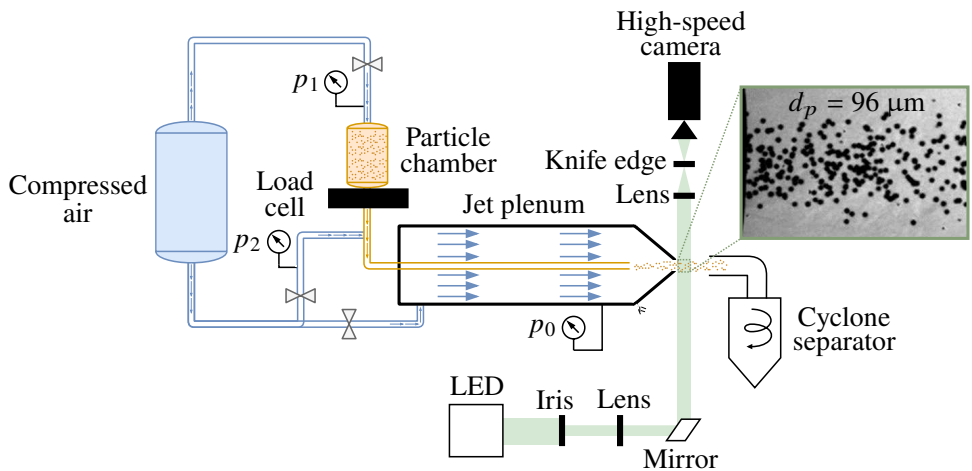
\begin{figure}
\centering
\subfloat{%
  \begin{tikzpicture}
        \node[anchor=north west,inner sep=0pt] at (0,0){\includegraphics[width= .8\textwidth]{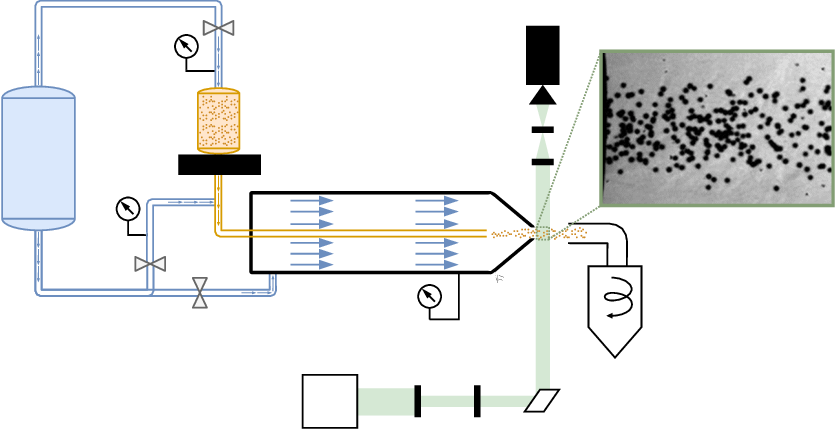}\label{fig: experimental schematic2}};
        \node[] at (-6ex,-12ex) {Compressed};
        \node[] at (-6ex,-14ex) {air};
        \node[] at (23.5ex,-8ex) {Particle};
        \node[] at (23.5ex,-10ex) {chamber};
        \node[] at (29.5ex,-13.5ex) {Jet plenum};
        \node[] at (31ex,-23.2ex) {$p_0$};
        \node[] at (25.8ex,-27.7ex) {LED};
        \node[] at (32ex,-28.8ex) {Iris};
        \node[] at (39ex,-12.5ex) {Lens};
        \node[] at (36ex,-10ex) {Knife edge};
        \node[] at (37.5ex,-28.8ex) {Lens};
        \node[] at (42ex,-33.5ex) {Mirror};
        \node[] at (56ex,-5.5ex) {$d_p = 96 \ \upmu$m};
        \node[] at (42ex,1ex) {High-speed};
        \node[] at (42ex,-1ex) {camera};
        \node[] at (55ex,-22.5ex) {Cyclone};
        \node[] at (55ex,-24.5ex) {separator};
        \node[] at (12.2ex,-3.8ex) {$p_1$};
        \node[] at (10.5ex,-11.5ex) {Load};
        \node[] at (10.5ex,-13.5ex) {cell};
        \node[] at (7.5ex,-16.5ex) {$p_2$};
    \end{tikzpicture}
}
\caption{Schematic of the particle-laden underexpanded jet facility at Johns Hopkins University. Blue lines correspond the air flow and orange lines correspond to the particle flow.}
\label{fig: experimental schematic}
\end{figure}

Figure~\ref{fig: experimental schematic} shows a schematic of the high-speed particle-laden jet facility at Johns Hopkins University. The facility consists of the air supply system and jet plenum, particle injection system, and high-speed imaging system. Compressed air flows from a high-pressure tank into a jet manifold with three pressure regulators that allow for control of the stagnation pressure $p_0$ and both $p_1$ and $p_2$, with the latter used for the particle injection system. To measure $p_0$, a pressure probe is inserted upstream of the nozzle exit in the constant area section of the jet plenum, where the flow speed is subsonic. Due to the large diameter of the jet plenum (25.4 mm) and the low-speed flow, the difference between the static and stagnation pressure is negligible. For simplicity, we treat the measured static pressure as equivalent to the stagnation pressure. To control the nozzle pressure ratio $\eta_0$, the pressure regulator controlling the flow entering the jet plenum is set to the desired $p_0$. The facility is capable of achieving pressure ratios within the range of $1.89 \leq \eta_0 \leq 6.86$, sufficient to study underexpanded and highly underexpanded jets. The flow is accelerated to sonic conditions using a commercial stainless-steel converging nozzle (CCP-1, Ikeuchi, Inc.) with an exit diameter of $D_e = 2$ mm.

\subsection{Particle characteristics}
Particles are injected into the flow stream by applying a pressure difference between $p_1$ and $p_2$, which forces the stationary particles within the particle chamber to enter a feed tube that extends from the particle chamber to the end of the subsonic section of the jet plenum. Particles leave the chamber through an orifice diameter that is $\sim 10d_p$ and enter a feed tube with an inner diameter of  $\sim 30d_p$. Once the particles leave the feed tube, they mix with the subsonic gas and enter the converging section of the nozzle, where the flow is accelerated to sonic speeds at the nozzle exit. Particles are collected downstream of the nozzle using a cyclone separator (Oneida, Inc.). The particle mass flow rate is measured using a single-point load cell (PW4CM-2KG, HBM), positioned beneath the feeding chamber. The electrical signal from the load cell is captured via a 16-bit data acquisition system (NI-9215, National Instruments), which is converted to a mass flow rate after obtaining a calibration relationship a-priori. Further details on the mass flow rate calculation can be found in \citet{kim2020dynamics}. 

The particles used in this study are clear poly(methyl methacrylate) acrylic microspheres (PMMA, Cospheric LLC) with a density of $\rho_p = 1211 \ \mathrm{kg/m^3}$. Three particle sizes were considered with mean diameters of $\overline{d}_p=29$, $42$ and $96$ $\upmu$m, corresponding to Stokes numbers in the range $29<St<57$ (see Table~\ref{table: params}). The particle number density was determined from the laser diffraction data acquired at NASA's Jet Propulsion Laboratory, California Institute of Technology, with deionized water as the dispersant. Measurements were carried out at brief, one-second intervals over a minute, ensuring precise analysis. 

\subsection{High-speed imaging technique}
Ultra-high-speed schlieren is used to visualise the gas phase shock structures and particles. The optical setup shown in figure~\ref{fig: experimental schematic} consists of a 100 W LED that passes through an iris to collimate the light, followed by a series of lenses that eventually collimate the light through the measurement plane. A horizontal knife-edge was placed in front of the camera at the focal point to allow for the visualisation of the changes in the density gradient in the vertical direction. In figure \ref{fig: experimental schematic} a sample zoomed-in image is shown for Case B3, where slight shock structures are visible near the jet boundary, clearly showing the effect of the knife edge.

In order to resolve the micron-sized particles in time and space, it was necessary to maximize the signal-to-noise ratio at the imaging sensor and minimize motion blur \citep{versluis2013high,buchmann2014ultra}. The optimal frame rate for time-resolved imaging is defined as $f=nU/S$, where $f$ is the frame rate, $n\geq2$ is the number of samples to fulfill the sampling criterion, $U$ is the velocity scale and $S$ is a length scale. In this study, Case A1 is expected to have the fastest particles due to the small size of $\overline{d}_p = 29 \ \upmu$m. Assuming the particles reach velocities of $U = 200$ m/s and knowing a-priori that these particles move approximately 6 pixels frame-to-frame, results in a length scale $S = 210 \ \upmu$m. Using these parameters, the optimal frame rate required for study is 1.87 MHz. Cases B1--B4 have larger particles that move slower, so this frame rate should suffice. Additionally, to avoid motion blur, the minimum exposure time is expressed as $\tau\leq d_p/U$, resulting in a required exposure time of $\tau \leq 145$ ns. To satisfy both $f$ and $\tau$ requirements, the Shimadzu HPV-X2 camera (FTCMOS2 sensor) was used, which enabled particle images to be acquired at 2 MHz at exposure times of 200 ns (camera limit). Although the required exposure time is lower than 200 ns, the particle images acquired for Case A1 and A2 were found to be sufficient for tracking. The resulting spatial resolutions are $17 \ \upmu$m/px for Case A1 and A2, $23 \ \upmu$m/px for Case B1 and B2, and $21 \ \upmu$m/px for Case B3 and B4, with an estimated depth of field of 1 mm. The resulting field of view is $5.6D_e$ in length and $3.5D_e$ in height, sufficient to capture the particle dynamics as they pass through a pair of shock cells downstream of the nozzle. The camera outputs a total of 256 frames per run, necessitating approximately three to four experiments per case to ensure velocity statistics are adequately converged.

To obtain particle positions and velocities, a Lagrangian particle tracking algorithm was applied to the acquired images. From the determined particle positions, particle trajectories were determined based on the nearest neighbor distance in the subsequent frames, followed by a convolution with a Gaussian smoothing kernel to estimate the particle velocities in each frame. Further details on this particle tracking algorithm can be found in \citet{ouellette2006quantitative,kelley2011onset}. 



\section{Simulation details}{\label{sec:comp}}

\subsection{Flow configuration}{\label{sec:flowconfig}}
A schematic view of the flow configuration is shown in figure~\ref{fig:schematic}. The computational domain is $Lx \times Ly \times Lz = 30D_e \times 15D_e \times 15D_e$ discretized on a Cartesian mesh with $nx \times ny \times nz = 1201 \times 201 \times 201$ grid points. The mesh is uniform in the streamwise ($x$) direction with grid spacing $\Delta x=D_e/40$. Grid stretching is applied in the spanwise ($y$ and $z$) directions using the mapping proposed by \citet{vishnampet2015practical} such that the grid spacing varies smoothly from $\Delta y_{\text{min}}=\Delta z_{\text{min}}=D_e/40$ at the centreline to $\Delta y_{\text{max}}=\Delta z_{\text{max}}=D_e/6$ at the lateral boundaries. The maximum point-to-point relative change in the grid spacing is $< 4\%$. It should be noted that the minimum grid spacing is approximately three time smaller than the diameter of the largest particles considered.

The nozzle extends $4.35D_e$ into the domain. The inner conical contour (converging section) follows a hyperbolic tangent function (see Appendix~\ref{app:nozzle} for details). Particles are injected into the flow at the nozzle exit with a prescribed mass flow rate $\dot{m}_p$. Their velocity and diameter are randomly sampled from distributions informed by the experiments. More details on the velocity and size distribution can be found in \S\ref{subsec:vel_inj}.

The numerical simulations are solved in a volume-filtered Eulerian-Lagrangian framework using a class of high-order, energy stable finite difference operators \citep{SHALLCROSS2020103138,jCODE}. Details on the governing equations and discretization are provided below.


\subsection{Gas-phase description}\label{sec:gas_sec}
The volume-filtered compressible Navier-Stokes equations describing the gas phase can be expressed compactly as
\begin{equation}\label{eq:main_gas}
    \frac{\partial \bm{Q}}{\partial t}+\frac{\partial}{\partial x_i}\left[\alpha\left(\bm{F}_i^I-\bm{F}_i^V\right)\right]=\bm{S},
\end{equation}
where $\bm{Q} = [ \alpha \rho, \alpha\rho u_i,\alpha\rho E ]^{T}$ is the vector of conserved variables, $\bm{F}_i^V$ and $\bm{F}_i^I$ are the viscous and inviscid fluxes, and $\bm{S}$ contains source terms that account for two-way coupling with the particles, given by
\begin{equation*}
    \bm{F}^I_i=\begin{bmatrix}
\rho u_i \\
\rho u_1 u_i + p \delta_{i1} \\
\rho u_2 u_i + p \delta_{i2} \\
\rho u_3 u_i + p \delta_{i3} \\
u_i(\rho E + p ) 
\end{bmatrix},
\bm{F}^V_i=\begin{bmatrix}
0 \\
\tau_{1i} \\
\tau_{2i} \\
\tau_{3i} \\
u_j \tau_{ij} - q_i 
\end{bmatrix}, 
 \bm{S}=\begin{bmatrix}
0 \\
\left(p \delta_{i1}-\tau_{i1}\right)\frac{\partial\alpha}{\partial x_i}+\mathcal{F}_1 \\
\left(p \delta_{i2}-\tau_{i2}\right)\frac{\partial\alpha}{\partial x_i}+\mathcal{F}_2\\
\left(p \delta_{i3}-\tau_{i3}\right)\frac{\partial\alpha}{\partial x_i}+\mathcal{F}_3\\
\left(\tau_{ij}-p \delta_{ij}\right)\frac{\partial}{\partial x_i}(\alpha_p u_{p,j})+q_i\frac{\partial\alpha}{\partial x_i}+ \\ u_{p,i}\mathcal{F}_i + \mathcal{Q}
\end{bmatrix}.
\end{equation*}

The conserved variables include the local gas-phase volume fraction $\alpha$, density $\rho$, velocity $u_i$ (in direction $i$), and total energy $E$. In the source term, $\alpha_p=1-\alpha$ and $u_{p,i}$ are the volume fraction and velocity of the particle phase in an Eulerian frame, respectively, defined explicitly in \S\ref{sec:2way}. The thermodynamic pressure is defined as $p = (\gamma-1)(\rho E - \rho u_i u_i/2)$, where $\gamma=1.4$ is the ratio of specific heats. The viscous stress tensor is given by
\begin{equation}{\label{eq: visc_stress}}
    \tau_{ij}  = \mu \left( \dfrac{\partial u_i}{\partial x_j} + \dfrac{\partial u_j}{\partial x_i}\right) + \left(\mu_b-2/3\mu\right) \dfrac{\partial u_k}{\partial x_k} \delta_{ij},
\end{equation}
where $\delta_{ij}$ is the Dirac delta function and the bulk viscosity $\mu_b = 0.6\mu$ is chosen as a model for air \citep{ref:Sharma2023}. The shear viscosity varies with temperature based on a power law according to $\mu  = \mu_{\text{$\infty$}}\left(T/T_{\text{$\infty$}}\right)^{2/3}$, where $\mu_{\text{$\infty$}} = 1.8\times10^{-5}~\mathrm{Pa}\cdot{\rm s}$ and $T_{\text{$\infty$}} = 298~\mathrm{K}$ are the reference shear viscosity and temperature, respectively. Temperature is obtained from the ideal gas law, $T = p/(\rho R)$, where $R$ is the gas constant. The heat flux is defined as
\begin{equation}{\label{eq: thermal_cond}}
q_i=-\kappa\dfrac{\partial T}{\partial x_i},
\end{equation}
where $\kappa$ is the thermal conductivity of the gas. Finally, $\mathcal{F}_i$ and $\mathcal{Q}$ are interphase momentum and heat exchange terms that will be defined in \S\ref{sec:2way}. 

Spatial derivatives are approximated using fourth-order, narrow-stencil finite difference operators that satisfy the summation-by-parts (SBP) property \citep{strand1994summation}. Kinetic energy preservation is achieved using a skew-symmetric-type splitting of the inviscid flux~\citep{pirozzoli2011stabilized}, extended to account for the effect of particles. Specifically, the convective fluxes appearing in \eqref{eq:main_gas} are expressed in split form as
\begin{equation}
    \frac{\partial \alpha \rho u_i\varphi}{\partial x_i}=\frac{1}{2}\frac{\partial \alpha  \rho u_i\varphi}{\partial x_i}+\frac{1}{2}\varphi\frac{\partial \alpha  \rho u_i}{\partial x_i}+ \alpha  \rho u_i\frac{\partial \varphi}{\partial x_i},
\end{equation}
where $\varphi$ is a generic transported scalar that is unity for the continuity equation, $u_j$ for the momentum equation, and $E+p/\rho$ for the total energy equation. This provides nonlinear stability at low Mach numbers.

To evaluate second and mixed derivatives, first derivative operators are applied consecutively, necessitating the use of artificial dissipation to damp the highest wavenumber components supported by the grid. High-order accurate SBP dissipation operators are used that provide artificial viscosity based on a sixth-order derivative~\citep{mattsson2004stable}. In addition, localized artificial diffusivity is used as a means of shock capturing following the formulation in \citet{kawai2010assessment}. In this approach, $\mu_b$ and $\kappa$ appearing in \eqref{eq: visc_stress} and \eqref{eq: thermal_cond} are augmented based on a modified Ducros-type sensor. Full details on the shock capturing implementation are provided in Appendix~\ref{app:shock}.

The SBP scheme is combined with the simultaneous approximation treatment (SAT) at the domain boundaries to facilitate an energy estimate~\citep{carpenter1994time,nordstrom2005well}. Non-reflecting characteristic boundary conditions \citep{svard2007stable} are enforced at each of the domain boundaries. No-slip, adiabatic boundary conditions are enforced at the surface of the nozzle via a ghost-point immersed boundary method \citep{chaudhuri2011use,khalloufi2023drag}. Further details on the nozzle profile and integration of the immersed boundary method in the SBP-SAT framework can be found in Appendix~\ref{app:numerics}. 

The equations are advanced in time using a standard fourth-order Runge-Kutta scheme, resulting in the usual Courant-Friedrichs-Lewy (CFL) restrictions on the simulation time step $\Delta t$. The CFL is taken as the maximum between the acoustic CFL,  ${\rm CFL}_a=\max\left(|\bm{u}|+c\right)\Delta t/\Delta$ and the viscous CFL, ${\rm CFL}_v = \max\left(2\mu,\mu_b,\kappa\right)\Delta t/\Delta^2$, where $\Delta$ is the local grid spacing and $c=\sqrt{\gamma p/\rho}$ is the local sound speed. An addition time step restriction is applied to ensure the source terms due to drag are resolved~\citep{patel2022}.

\subsection{Particle-phase description}{\label{subsec:particle_subsec}}

The particle equations of motion are given by
\begin{equation}
\frac{{\mathrm d}\boldsymbol{x}_p^{(i)}}{{\mathrm d}t} = \boldsymbol{v}_p^{(i)},
\end{equation}
and
\begin{equation}\label{eq:dvdt}
m_p\frac{{\mathrm d}\boldsymbol{v}_p^{(i)}}{{\mathrm d}t}={V_p}\nabla \cdot \left( \boldsymbol{\tau} - p \boldsymbol{\mathrm I}\right) + {{\boldsymbol{F}}_{\text{drag}}^{(i)}} + {{\boldsymbol{F}}_{\text{lift}}^{(i)}} + {{\boldsymbol{F}}_{\text{am}}^{(i)}},
\end{equation}
where $\boldsymbol{\mathrm I}$ is the identity tensor, $\boldsymbol{x}_p^{(i)}=(x_p^{(i)},y_p^{(i)},z_p^{(i)})$ and $\boldsymbol{v}_p^{(i)}=(v_{p,x}^{(i)},v_{p,y}^{(i)},v_{p,z}^{(i)})$ are the position and velocity of particle $i$, respectively, $m_p$ is the mass of the particle and $V_p$ is its volume. ${\boldsymbol{F}}_{\text{drag}}^{(i)}$,  ${\boldsymbol{F}}_{\text{lift}}^{(i)}$, and ${\boldsymbol{F}}_{\text{am}}^{(i)}$ are the force contributions due to drag, lift, and added-mass, respectively.

The unsteady aerodynamic forces acting on a particle (i.e. the pressure gradient and added mass) are typically neglected in low-speed, gas-solid flows due to the high density ratio. However, under sufficient gas-phase acceleration, such as in flows with shocks, the unsteady contributions can dominate \citep{parmar2009modeling,ling2011importance} and are thus considered here. Due to the low volume fractions considered, inter-particle collisions are neglected.

The quasi-steady drag force is given by
\begin{equation}\label{eq: drag}
\dfrac{{\boldsymbol{F}}_{\text{drag}}^{(i)}}{m_p}=\frac{1}{\tau_p}\left(\boldsymbol{u}-{\boldsymbol{v}}_p^{(i)}\right),
\end{equation}
where $\tau_p{=}\rho_p d_p^2$/$\left(18\mu F_d\right)$ is the particle response time due to drag and $F_d{=}F_d(\alpha_p,Re_p, M_p)$ is the non-dimensional drag correlation of \citet{osnes2023comprehensive}. Here, $Re_p=\rho|\boldsymbol{u}-\boldsymbol{{v}}_p^{(i)}|d_p$/$\mu$ is the particle Reynolds number and  $M_p=|\boldsymbol{u}-\boldsymbol{{v}}_p^{(i)}|$/$c$ is the particle Mach number.

The Saffman lift force is modeled according to~\citep{mclaughlin1991inertial}
\begin{equation}\label{eq: lift}
    {{\boldsymbol{F}}_{\text{lift}}^{(i)}} = \dfrac{9.69 \sqrt{\rho \mu}}{\pi \rho_p d_p} \dfrac{(\boldsymbol{u}-\boldsymbol{v}^{(i)}_p) \times \boldsymbol{\omega}}{\sqrt{|\boldsymbol{\omega}|}},
\end{equation}
where $\boldsymbol{\omega}$ is the gas-phase vorticity. 

Following the formulation proposed by \citet{parmar2010}, added mass is expressed as
\begin{equation}\label{eq: addedmass}
    \boldsymbol{F}_{\text{am}}^{(i)} = V_pC_M(M_p, \alpha_p)\left[\dfrac{{\rm D}(\rho \boldsymbol{u})}{{\rm D}t}-\dfrac{{\rm d}(\rho_p \boldsymbol{v}_p^{(i)})}{{\rm d}t} \right],
\end{equation}
where $C_M = C_{M,0}\eta_1(M_p)\eta_2(\alpha_p)$ is the added-mass coefficient \citep{ling2011importance} and $C_{M,0}=0.5$ is the value in the zero Mach number and zero volume fraction limit. The Mach number correction of \citet{parmar2008unsteady} and volume fraction correction of \citet{sangani1991added} are employed, expressed as
\begin{equation}{\label{eq:parmar_C_M}}
    \eta_1(M_p)= 
\begin{cases}
    1 + 1.8 M_p + 7.6M^4_p & \text{if } M_p < 0.6,\\
    2.633,              & \text{otherwise}
\end{cases}
\quad\text{and}\quad \eta_2(\alpha_p) = \dfrac{1 + 2\alpha_p}{1-\alpha_p}.
\end{equation}

The evolution of particle temperature is given by
\begin{equation}\label{eq: parttemp}
    m_p C_{p,p} \dfrac{{\mathrm d}T^{(i)}_p}{{\mathrm d}t} = {q}^{(i)}_{\text{inter}},
\end{equation}
where $C_{p,p}$ is heat capacity of the particle, $T^{(i)}_p$ is its temperature, and  $q^{(i)}_{\text{inter}}$ is the interphase heat exchange given by
\begin{equation}\label{eq: hxparticle}
    q^{(i)}_{\text{inter}} = \dfrac{6V_p\kappa \mathit{Nu}}{d_p^2}\left(T-T^{(i)}_p\right),
\end{equation}
where $\mathit{Nu}$ is the Nusselt number that is modeled using the correlation of \citet{Gunn}. 

Particles are advanced in time simultaneously with the fluid via a fourth-order Runge-Kutta scheme. Special care needs to be taken when evaluating fluid quantities appearing in \eqref{eq: drag}--\eqref{eq: hxparticle} at the location of the particle, namely $\alpha$, $\rho$, $\boldsymbol{u}$, $\mu$, $\nabla \cdot \left( \boldsymbol{\tau} - p \boldsymbol{\mathrm I}\right)$, $\boldsymbol{\omega}$ and $T$, especially near shocks where the flow is nearly discontinuous (see figure~\ref{fig: filter_challenges}). These details will be given in \S\ref{sec:2way}.

\subsection{Two-way coupling}\label{sec:2way}
Particle information (drag, heat exchange, volume fraction, etc.) is projected to the Eulerian grid using the two-step filtering approach proposed by \citet{capecelatro2013euler}. The gas-phase volume fraction is computed according to
\begin{equation}{\label{eq: vf2way}}
    \alpha( \bm{x},\textit{t}) =1- \sum_{1}^{N_p}\mathcal{G}\left(|\bm{x}-\bm{x}^{(i)}_p|\right) V_p,
\end{equation}
where $\mathcal{G}$ is a Gaussian filter kernel with a characteristic length $\delta_f=4\overline{d}_p$ taken to be the full width at half maximum, $N_p$ is the total number of particles and $\bm{x}$ is the position on the Eulerian grid. Interphase momentum exchange appearing in the source term of \eqref{eq:main_gas} is given by 
\begin{equation}{\label{eq: momentum2way}}
    \bm{\mathcal{F}} =- \sum_{i=1}^{N_p} \mathcal{G}\left(|\bm{x}-\bm{x}_p^{(i)}|\right)\left(\bm{F}_{\text{drag}}^{(i)} + \bm{F}_{\text{lift}}^{(i)} + \bm{F}_{\text{am}}^{(i)}\right).
\end{equation}
Similarly, the work done by drag appearing in the energy equation is expressed as
\begin{equation}{\label{eq:energy2way}}
    \bm{u}_p \bm{\cdot} \bm{\mathcal{F}} =-  \sum_{i=1}^{N_p} \mathcal{G}\left( {|\bm{x}-\bm{x}_p^{(i)}|} \right)\left(\bm{F}_{\text{drag}}^{(i)} + \bm{F}_{\text{lift}}^{(i)} + \bm{F}_{\text{am}}^{(i)}\right) \bm{\cdot} \bm{v}_p^{(i)}
    \end{equation}
and the interphase heat exchange term is given by
    \begin{equation}{\label{eq: heat2way}}
    \mathcal{Q} =- \sum_{i=1}^{N_p} \mathcal{G}\left(|\bm{x}-\bm{x}_p^{(i)}|\right){{q}}_{\text{inter}}^{(i)}.
\end{equation}

The models employed in the particle equation of motion \eqref{eq:dvdt} were formulated using correlations reliant on `far-field' fluid quantities, such as velocity, temperature, and volume fraction that remain unaffected by the presence of the particle. In order for these models to be applicable in a two-way coupled simulation, the self-induced disturbance must be removed. This has been an active area of research in recent years \citep[e.g.][]{horwitz2016accurate,horwitz2018correction,balachandar2019self,liu2019self,pakseresht2020correction,evrard2020euler,pakseresht2021disturbance}. However, they have not yet been extended to compressible flows. Various interpolation techniques have been proposed for particles in the vicinity of shocks \citep{jacobs2009high,kozak2020weno}, though their application to finite size particles necessitating corrections for self-induced disturbances have not yet been explored. Recently,~\citep{superspnic_EL_filter} proposed a weighted average-based interpolation scheme with weights biased away from the particle center to interpolate far-field conditions. This was reported to restore the undisturbed quantities.

\tikzset{%
    FARROW1/.style={line width=1pt, arrows={-latex[angle=40:2.5mm]}},
    FARROW2/.style={line width=1pt, arrows={latex[angle=45:2.5mm]-latex[angle=45:2.5mm]}},
    FARROWsmall2/.style={line width=0.75pt, arrows={latex[angle=20:1mm]-latex[angle=20:1mm]}},
    FARROWsmall3/.style={line width=0.75pt, arrows={-latex[angle=20:1mm]}},
    FARROWsmall4/.style={line width=0.75pt, arrows={latex[angle=10:1pt]-}},
}
\begin{center}
    \begin{tikzpicture}[node distance=0.1cm, spy using outlines={thick, rectangle, pat_green2, magnification=6,
 connect spies}]
    \node[anchor=south west,inner sep=0] (image) at (0,0) {%
    \centering
        \includegraphics[width=0.35\textwidth]{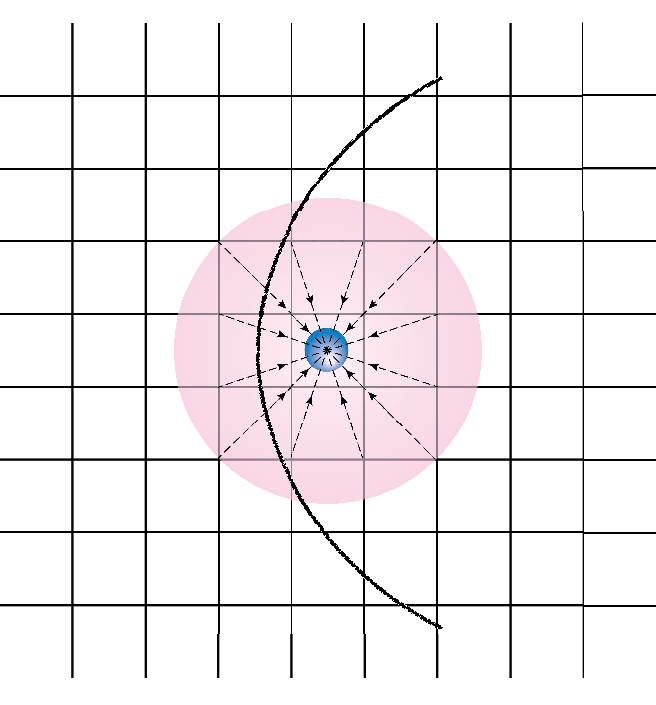}
    };

    \begin{scope}[x={(image.south east)},y={(image.north west)}]
        \draw[FARROWsmall2, color= black] (0.28, 0.0) -- (0.725, 0.0);
        \draw[FARROWsmall2, color= black] (0.1, 0.0) -- (0.23, 0.0);
        \draw[FARROWsmall2, color= black] (-0.025, 0.14) -- (-0.025, 0.25);
        \draw[FARROWsmall3, color= black] (0.39, 0.81) -- (0.5, 0.75);
        \draw[FARROWsmall4,dashed,line width = 1pt](1.1,0.75) -- (1.2, 0.75);
        \node[fill = none,anchor=west] at (1.18, 0.75) {~\small{Interpolation}\strut};
        \filldraw[blue_particle] (1.12, 0.65) circle (4pt) node[anchor=west]{};
        \node[fill = none,anchor=west] at (1.18, 0.65) {~\small{Particle}\strut};
        \node[fill = none, opacity=0.8,align=left,anchor=base]at (0.5, -0.05) { \color{black} \small{${\delta}_{f}$}};
        \node[fill = none, opacity=0.8,align=left,anchor=base]at (0.17, -0.05) { \color{black} \small{${\Delta}\mathrm{x}$}};
        \node[fill = none, opacity=0.8,align=left,anchor=base]at (-0.075, 0.175) { \color{black} \small{${\Delta}\mathrm{y}$}};
        \node[fill = white, opacity=0.8,align=left,anchor=base]at (0.29, 0.8) { \color{black} \small{shock}};
    \end{scope}
    \end{tikzpicture}
    \captionof{figure}{Schematic of a classical interpolation scheme for a particle in the vicinity of a shock.} 
    \label{fig: filter_challenges}
\end{center}

A simple approach to remove the effect of self-induced disturbances is to filter the fluid field prior to interpolation \citep{evrard2020euler}. A similar strategy is followed in this work. Any gas-phase quantity, ${\zeta}(\boldsymbol{x},t)$, appearing in \eqref{eq: drag}--\eqref{eq: hxparticle} (i.e. $\alpha$, $\rho$, $\boldsymbol{u}$, $\mu$, $\boldsymbol{\omega}$ and $T$) is evaluated at the particle position via convolution with a Gaussian filter according to
\begin{equation}\label{eq:ufilt}
    \tilde{\zeta}[\boldsymbol{x}_p^{(i)}(t)] = \int_V{\mathcal{G}(|\boldsymbol{x}-\boldsymbol{x}_p^{(i)}|){\zeta}(\boldsymbol{x} ,t)\,\mathrm{d}V}.
\end{equation}
However, a direct application of \eqref{eq:ufilt} can be computationally expensive as it requires each particle to loop through (potentially) many surrounding grid points~\citep{capecelatro2013euler}. Instead, this is performed in two steps according to
\begin{equation}\label{eq:interp}
    \tilde{\zeta}[\boldsymbol{x}_p^{(i)}(t)] \approx \sum_{k=1}^{n}\mathcal{W}(|\boldsymbol{x}-\boldsymbol{x}_p^{(i)}|) \overline{\zeta}(\boldsymbol{x} ,t)\,\mathrm{d}V\quad\text{with}\quad\overline{\zeta}(\boldsymbol{x}, t) = \zeta(\boldsymbol{x}, t) * \mathcal{G},
\end{equation}
where $\mathcal{W}$ corresponds to weights of a tri-linear interpolation scheme that is localized to the $n$ nearest grid cells of particle $i$. The $\overline{(\cdot)}$ notation denotes a filtered quantity on the grid that can be solved efficiently via an alternating-direct-implicit scheme. The efficacy of the proposed interpolation scheme is demonstrated for a single particle passing through a standing shock in Appendix~\ref{app:numerics}. It was found that interpolating the gas-phase stress tensor, $\nabla \cdot \left( \boldsymbol{\tau} - p \boldsymbol{\mathrm I}\right)$, without filtering is necessary to avoid excessive smearing of discontinuities in the presence of shocks (see figure~\ref{fig: filter_challenges}) and give reasonable estimate of the undisturbed gradients.

\subsection{Particle injection}{\label{subsec:vel_inj}}

\begin{table}
  \begin{center}
  \begin{tabular}{ccccccccc}
    \toprule
      \multirow{3}{*}{Case}&  $\dot{m}_p$ & {$U_p$} & {$\sigma_U$}  &   {$\sigma_V$} & $\overline{d}_p$ & {$\sigma_d$} & min({$d_p$}) & max({$d_p$}) \\
     \cmidrule(lr){2-2} \cmidrule(lr){3-5} \cmidrule(lr){6-9}
    &g/s &  \multicolumn{3}{c}{m/s} & \multicolumn{4}{c}{$\upmu$m}  \\
    \midrule
    A1 & 0.4 & 159 & 9.4 & 10 & 29.6 & 5.8 & 6 & 70 \\
    A2 & 0.96 & 154.8 & 8.1 & 12 & 29.6 & 5.8 & 6 & 70\\ 
    B1 & 1.02 & 151.3 & 11.2 & 11 & 42 & 5.0 & 10 & 70\\
    B2 & 2.2 & 145.9 & 13 & 11.1 & 42 & 5.0 & 10 & 70\\
    B3 & 3.7 & 114 & 12.1 & 7.8 & 96 & 11.8 & 50 & 150\\
    B4 & 4.2 &  113 & 12.6 & 7.3 & 96 & 11.8 & 50 & 150\\
    \bottomrule
  \end{tabular}
  \caption{Particle injection parameters. Velocities and diameters are sampled from normal and lognormal distributions, respectively.}
  \label{table: inj_cond}
  \end{center}
\end{table}
Particles are seeded in the computational domain at the nozzle exit after the gas phase has reached a statistically stationary state. The number of particles injected per timestep, $N(t)$, is determined from the experimentally-measured mass flow rate according to
\begin{equation}
    \dot m_p = \sum_{i=1}^{N(t)}\dfrac{\pi {d_p^{(i)}}^3 \rho_p}{6\Delta t}.
\end{equation}
Particles are injected uniformly random throughout the cross-section of the nozzle exit plane according to
\begin{equation}
x_p^{(i)} = 0,\quad  y_p^{(i)} = \dfrac{D_e}{2}\sqrt{\mathcal{R}} \cos(\mathcal{R}2\pi),\quad z_p^{(i)} = \dfrac{D_e}{2}\sqrt{\mathcal{R}} \sin(\mathcal{R}2\pi),
\end{equation}
where $\mathcal{R}\in[0,1]$ is a uniform random real number.

\begin{figure}
\centering
\includegraphics[width=0.6\linewidth]{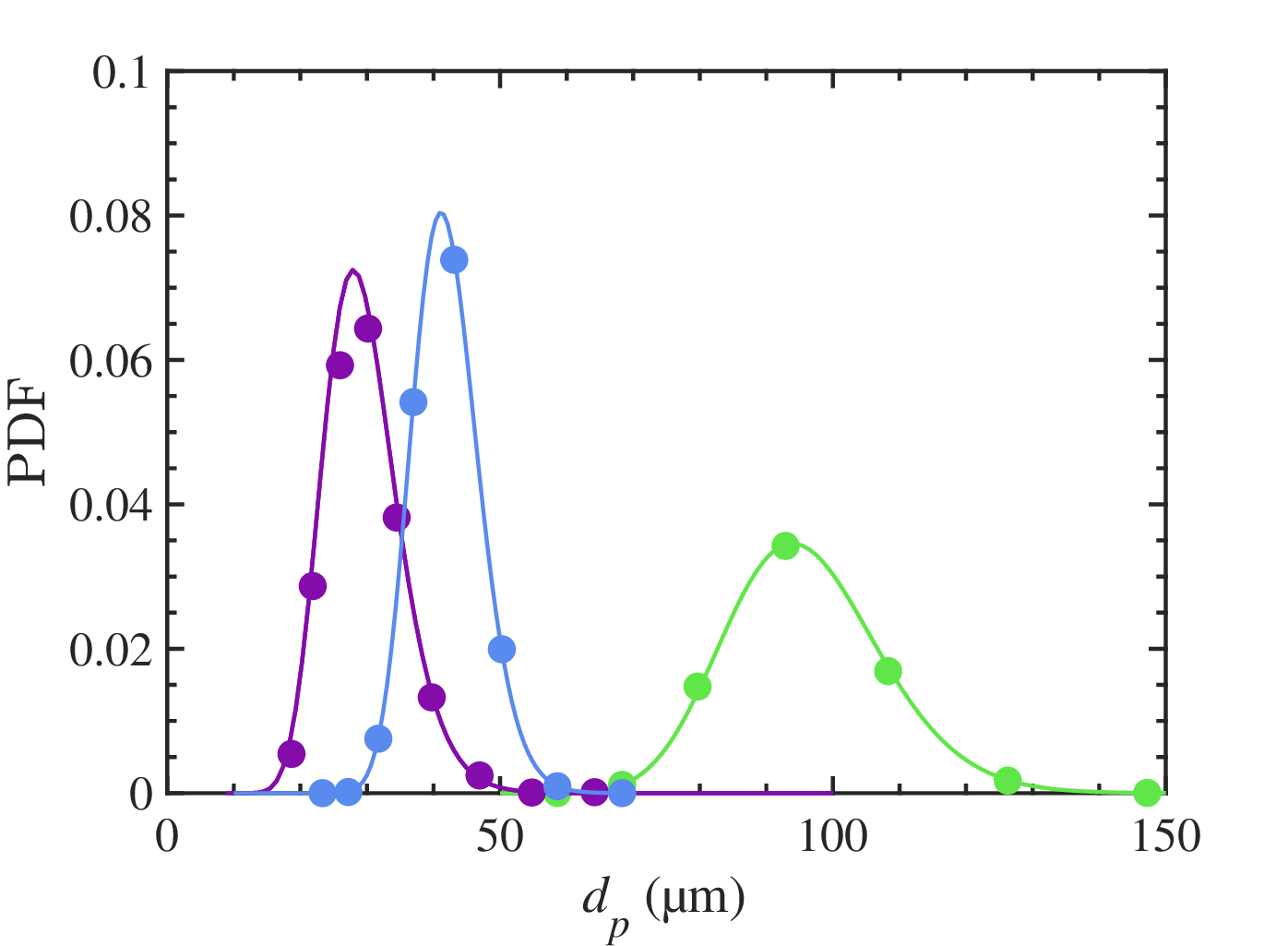}
\caption{Particle size distribution. Raw data: $\overline{d}_p=29.6~\upmu$m~(\protect \circlefillPurple),
$\overline{d}_p=42~\upmu$m~(\protect \circlefillBluetwo)
$\overline{d}_p=96~\upmu$m~(\protect \circlefillGreentwo). Lines are fitted lognormal distributions using the parameters listed in table~\ref{table: inj_cond}.}
\label{fig:psd}
\end{figure}
Particle diameters are sampled randomly from lognormal distributions, $d_p{\sim} \\ \text{Lognormal}(\overline{d}_p,\sigma_d)$, where the mean ($\overline{d}_p$) and standard deviation ($\sigma_d$) are determined from particle size distributions obtained experimentally (see figure~\ref{fig:psd}). The values for each case are reported in table~\ref{table: inj_cond}.

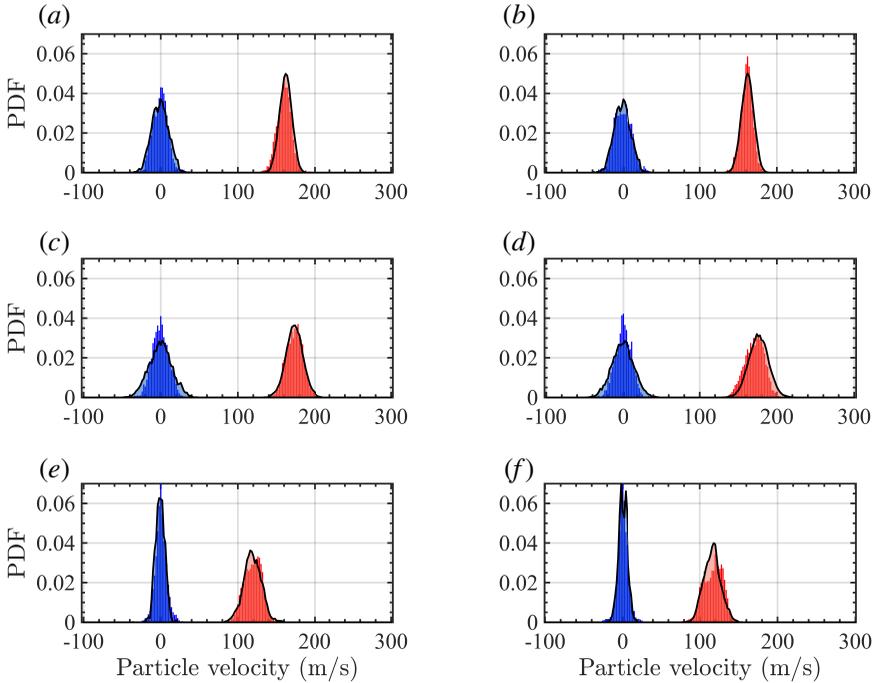
\begin{figure}
    \setlength{\lineskip}{0pt}
    \centering
    \begin{tikzpicture}
        \node[anchor=north west,inner sep=0pt] at (0,0){\includegraphics[width= 0.85\textwidth]{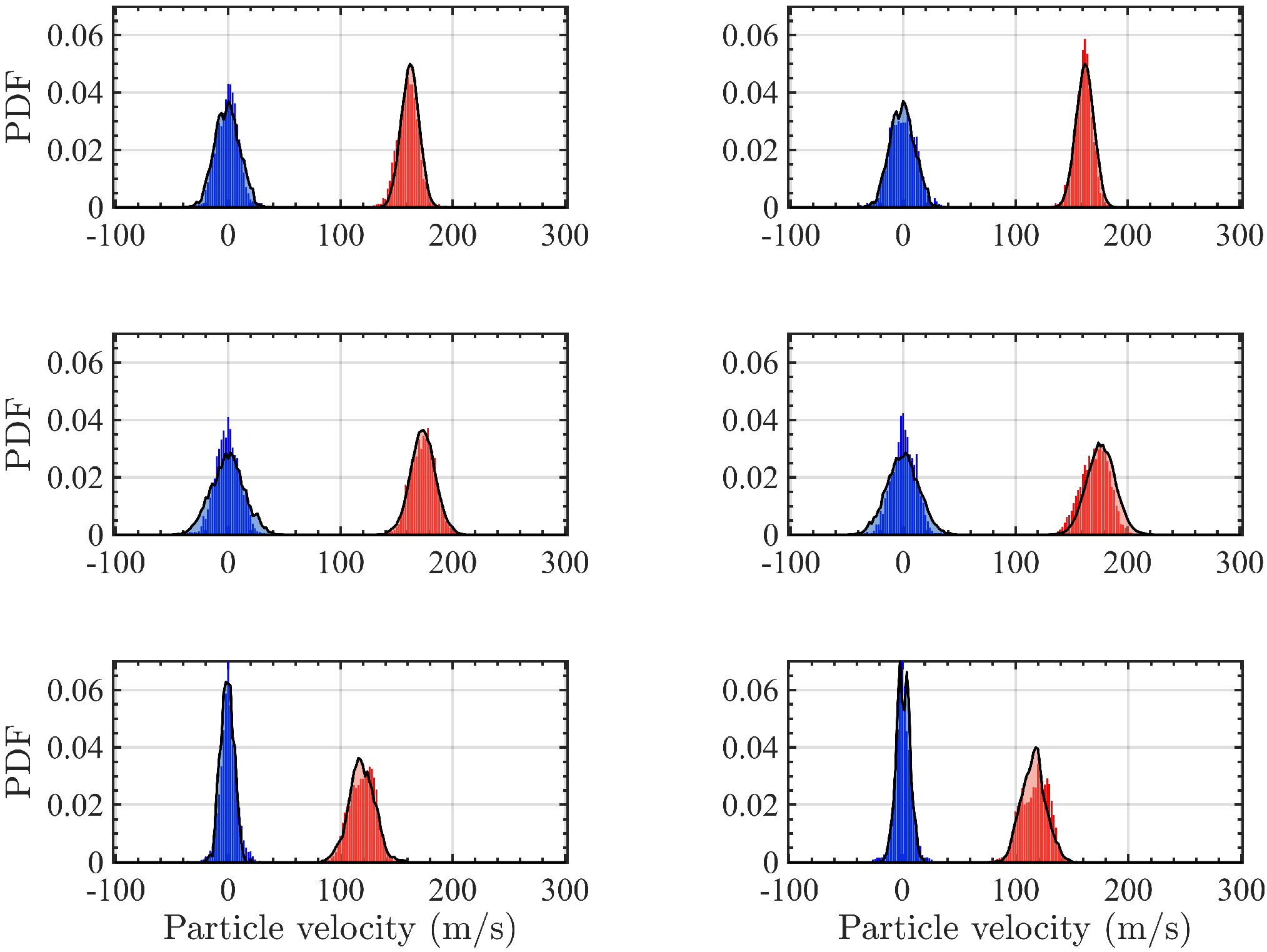}};
        \node[] at (4ex,1ex) {(\textit{a})};
        \node[] at (41ex,1ex) {(\textit{b})};
        \node[] at (4ex,-17ex) {(\textit{c})};
        \node[] at (41ex,-17ex) {(\textit{d})};
        \node[] at (4ex,-35ex) {(\textit{e})};
        \node[] at (41ex,-35ex) {(\textit{f})};
    \end{tikzpicture}
\caption{Particle velocity distributions near the nozzle exit ($0\le x/D_e\le 0.25$). (a) Case A1, (b) Case A2, (c) Case B1, (d) Case B2, (e) Case B3, (f) Case B4. Experiments: streamwise velocity (\protect \rectanglefillredvel) and spanwise velocity (\protect \rectanglefillbluevel). Simulations: streamwise velocity  (\protect \rectanglefillredtvel) and spanwise velocity  (\protect \rectanglefillbluetvel).}
\label{fig: injecting velocity pdfs}
\end{figure}

Similarly, newly seeded particles are assigned velocities that are randomly sampled from normal distributions $\mathcal{N}(\boldsymbol{U}_p,\sigma_{\boldsymbol{U}})$ fit to the experimental data with mean $\boldsymbol{U}_p=(U_p,V_p)$ and standard deviation $\sigma_{\boldsymbol{U}}=(\sigma_U,\sigma_V)$. Here, the streamwise particle velocities are sampled from $v_{p,x}{\sim}\mathcal{N}(U_p,\sigma_{U})$ and due to axial symmetry the spanwise velocities are sampled from $v_{p,y}{\sim}\mathcal{N}(V_p,\sigma_{V})$ and $v_{p,z}{\sim}\mathcal{N}(V_p,\sigma_{V})$. Probability density functions (PDFs) of the streamwise and spanwise particle velocities are collected experimentally near the nozzle exit within a window $0\le x/D_e\le 0.25$ (see figure~\ref{fig: injecting velocity pdfs}). The PDFs are found to be Gaussian, and parameters used in the simulations are chosen so that the velocity statistics match within this region (values are provided in table~\ref{table: inj_cond}).

\section{Results and discussion}\label{sec:results}

\subsection{Single-phase jet}
In this section, comparisons are made between the experiments and simulations of the single-phase jet. Shock structures are visualised in the experiments using an inline-type schlieren imaging system. Filters and converging lenses are employed to generate a spatially filtered collimated beam. At the focal point of the beam, a horizontal knife-edge cut-off is used to enhance the density gradient in the flow. The ensemble average is calculated using 256 instantaneous images in both the experiments and simulations, corresponding to 138 $\upmu$s of data. The numerical schlieren is produced by first integrating the vertical density gradient along the field-of-view ($z$-direction), $\psi(x,y)=\int \partial \rho/\partial y~{\rm d}z$, then applying the scaling proposed by \cite{QuirkSchlieren}:
\begin{equation}\label{eq:Schlieren}
 \xi = \exp(-k \psi_s)\quad\text{with}\quad \psi_s = \dfrac{\psi-k_0(\partial \rho/\partial y)_{\text{max}}}{k_1(\partial \rho/\partial y)_{\text{max}}- k_0(\partial \rho/\partial y)_{\text{max}}}, 
\end{equation}
where $k=5$, $k_0=-0.001$, and $k_1 = 0.05$.

\begin{figure}
  \centerline{\includegraphics[width = 0.9\textwidth]{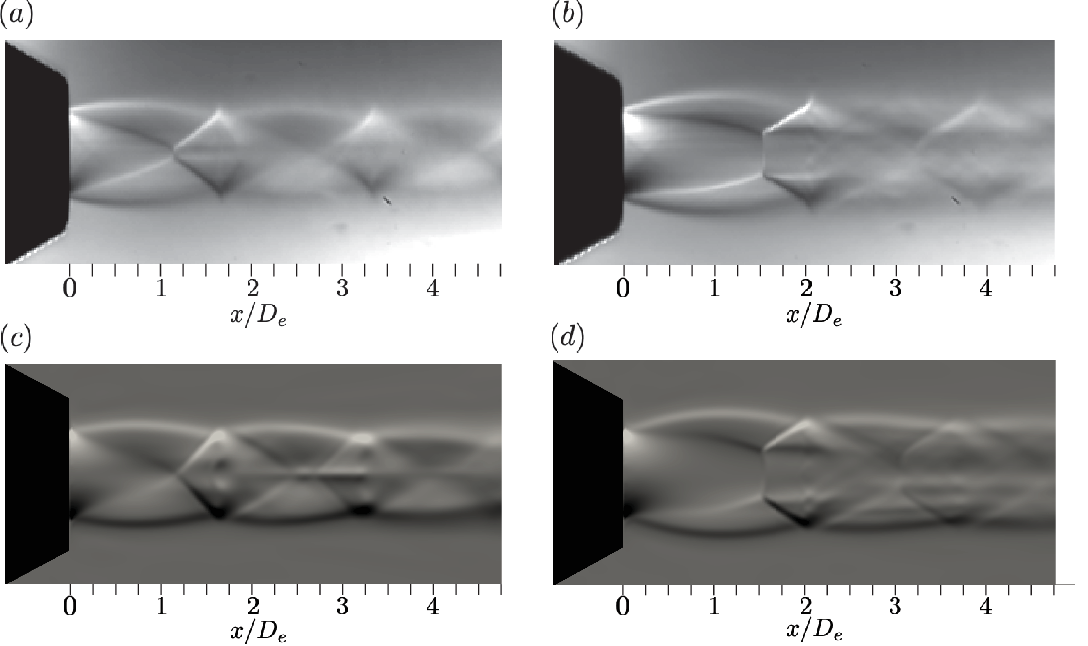}}
  \caption{Ensemble averaged Schlieren obtained from experiments (top) and simulations using the scaling given in \eqref{eq:Schlieren} (bottom) for $\eta_0 = 3.40$ (left) and $\eta_0 = 6.46$ (right).}
\label{fig:Schlieren images}
\end{figure}

Figure~\ref{fig:Schlieren images} shows schlieren images produced experimentally and numerically for nozzle pressure ratios $\eta_0 = 3.40$ and $6.46$. Regions of expansion fans, compression waves, oblique shocks and their spatial repetition are evident. 
At a pressure ratio $\eta_0 = 3.40$, the jet has a diamond shock structure with merged oblique shocks at the centreline near $x/D_e = 1.1$. Increasing the pressure ratio to $\eta_0 = 6.46$ causes the Mach disk to grow in size and move further downstream. The new shock structure resembles a barrel shock cell. The numerical schlieren shows overall good agreement with experiments. Some discrepancy can be observed at the lower pressure ratio of $\eta_0=3.4$ (figure~\ref{fig:Schlieren images}a and c) downstream of the first Mach disk.

\begin{figure}
\centering
\includegraphics[width = 0.6\textwidth]{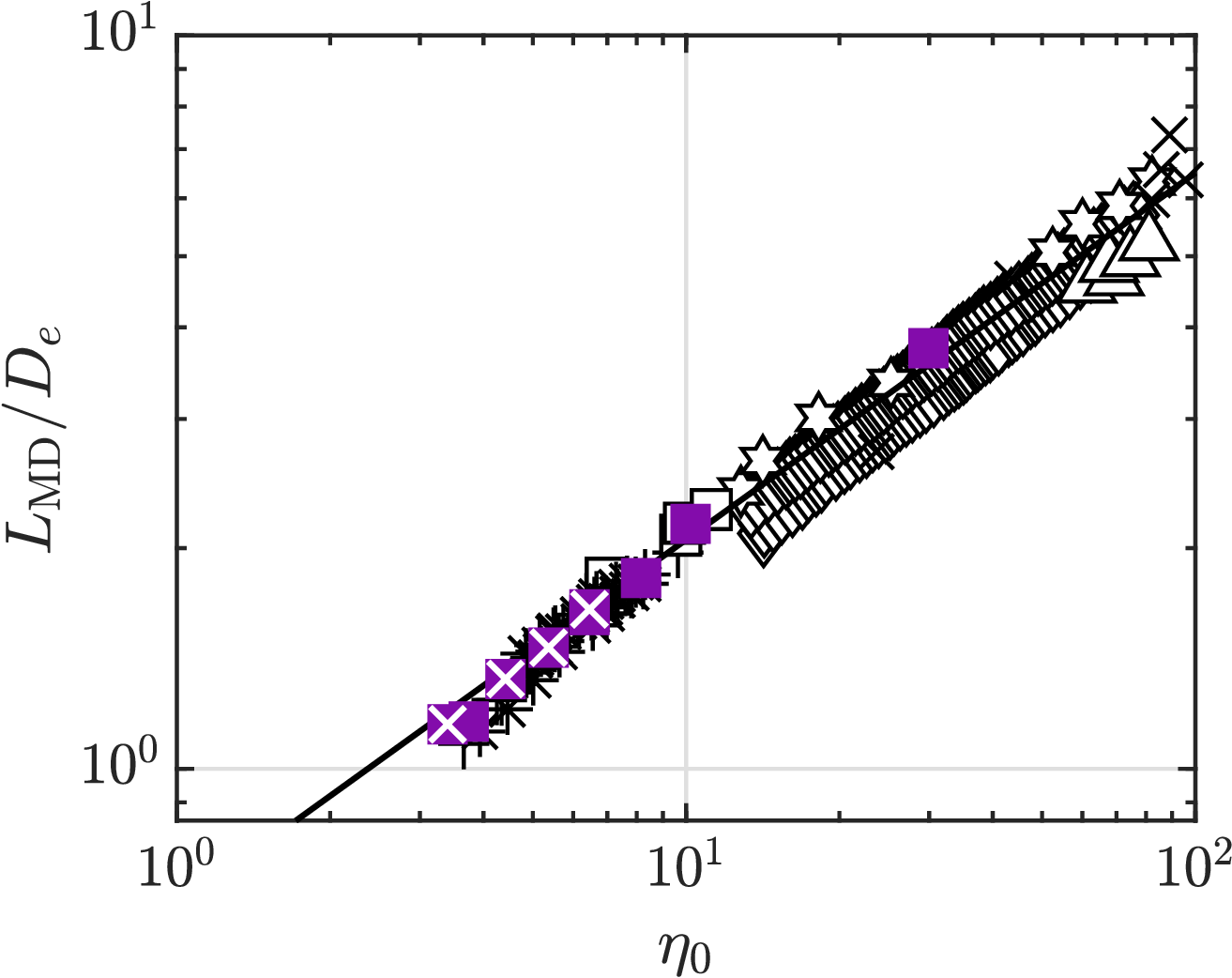}
\caption{Position of the Mach disk as a function of the total pressure ratio for single-phase flow. Current experiments (\protect \SquareCrossPurple[scale=0.2]) and simulations (\protect \rectanglefillPurplelarge), \cite{ashkenas1965structure} (\protect \Cross),~\cite{Crist1966study} (\protect \starhollowblack),~\cite{hatanaka2012influence} (\protect\tikz \protect \node [rectangle,draw,xscale=0.6,yscale=0.7,rotate=315, black,solid,line width = 0.5pt]{};),~\cite{tabei1992density} (\protect \rectanglehollow),~\cite{sommerfeld1994structure} (\protect \trianglehollow),~\cite{gibbings1972flow} (\protect \plussign),~\cite{addy1981effects} (\protect \asterisksign), correlation by \cite{Crist1966study} \eqref{eq:Crist} (\protect \blacklinesolidthickk).}
\label{fig: Mach disk movement}
\end{figure}

The normalized Mach disk location is extracted from averaged schlieren images and reported for a wide range of pressure ratios in figure~\ref{fig: Mach disk movement}. The validation is extended by comparing against previous works in the literature, including the empirical correlation proposed by \citet{Crist1966study}, given by
\begin{equation}\label{eq:Crist}
\dfrac{L_{\mathrm {MD}}}{D_{\text{e}}} = \sqrt{\dfrac{\eta_0}{2.4}}.
\end{equation}

Both experiments and simulations show good agreement with the correlation and previous experiments from the literature. Simulations were conducted at elevated pressure ratios ($\eta_0=[8, 10, 30]$) for further validation. Overall, excellent agreement is observed. It should be noted that significant variation has previously been reported at pressure ratios below $\eta_0<5$ \citep{Franquet2015}. Consequently, \eqref{eq:Crist} is less valid at such low pressure ratios, though good agreement is still observed with the present experiments and simulations. This might also explain the discrepancy in the shock structures observed in figure~\ref{fig:Schlieren images} at $\eta_0=3.4$.

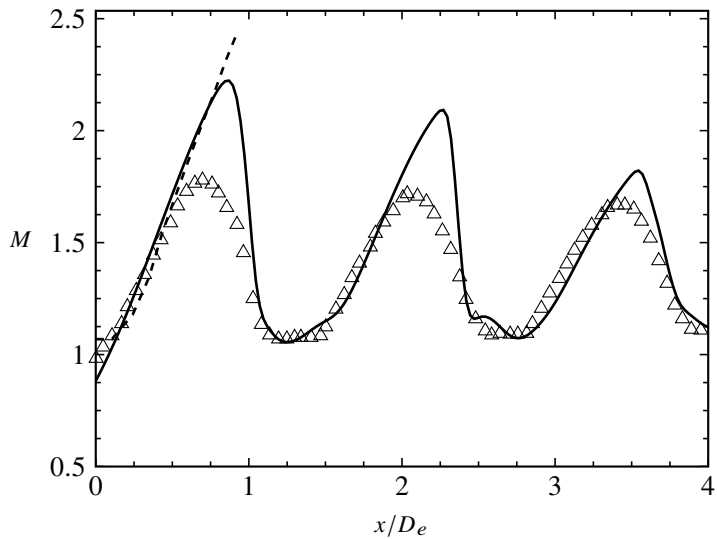
\begin{figure}
	\centering
	\setlength\figureheight{7cm}
	\setlength\mylinewidth{0.7pt}
	\setlength\mymarksize{2.5pt}
\setlength\figurewidth{0.75\textwidth}
	\setlength\figureheight{6cm}
	\setlength\mylinewidth{0.7pt}
	\setlength\mymarksize{2.5pt}
\setlength\figurewidth{0.6\textwidth}
	\input{tikz/Henderson_comp.tikz}
	\caption{Centerline Mach number obtained from experiments of a single-phase sonic jet from \citet{henderson2005experimental}  (\ref{leg:henderson}), method of characteristics solution~\citep{owen1948flow} (~\ref{leg:theory}) and numerical simulations performed in the present study (\ref{leg:sim_henderson}) with $\eta_0=3.40$.}
\label{fig: Mach vs x/D}
\end{figure}

Figure~\ref{fig: Mach vs x/D} shows the average centreline Mach number obtained from the simulation for $\eta_0=3.4$. Comparisons are made against experiments by \cite{henderson2005experimental} and a solution obtained from the method of characteristics (MOC) \citep{owen1948flow}. The MOC solution assumes steady supersonic inviscid flow with rotational symmetry and $\eta_0 = \infty $. Both simulations and experiments agree well with the MOC solution in the near-field region up to the first Mach disk. Beyond this point, the MOC solution is no longer valid due to the presence of discontinuities.

While the location of Mach diamonds predicted by the simulation are in good agreement with the experiment shown in figure~\ref{fig: Mach vs x/D}, the local Mach number is over-estimated by $\approx$ 20\%. The discrepancy with the experiment can be attributed to the non-negligible Stokes number associated with the seeded particles used with the experimental PIV system (0.6 $\upmu {\rm m}$ diameter particles were used). The authors estimate $0.04D_e-0.12D_e$ of particle lag downstream of the Mach disk, which is orders of magnitude larger than the shock thickness. Similar discrepancies in the centreline Mach number profile between simulations and experiments have previously been reported \citep[e.g.][]{saddington2004experimental,gojon2017numerical}.

\subsection{Two-phase jet: particle dynamics}{\label{subsec: particle_vel_pdfs}}

\begin{figure}
 \setlength{\lineskip}{0pt}
    \centering
    \begin{tikzpicture}
        \node[anchor=north west,inner sep=0pt] at (0,0){\includegraphics[width= 0.9\textwidth]{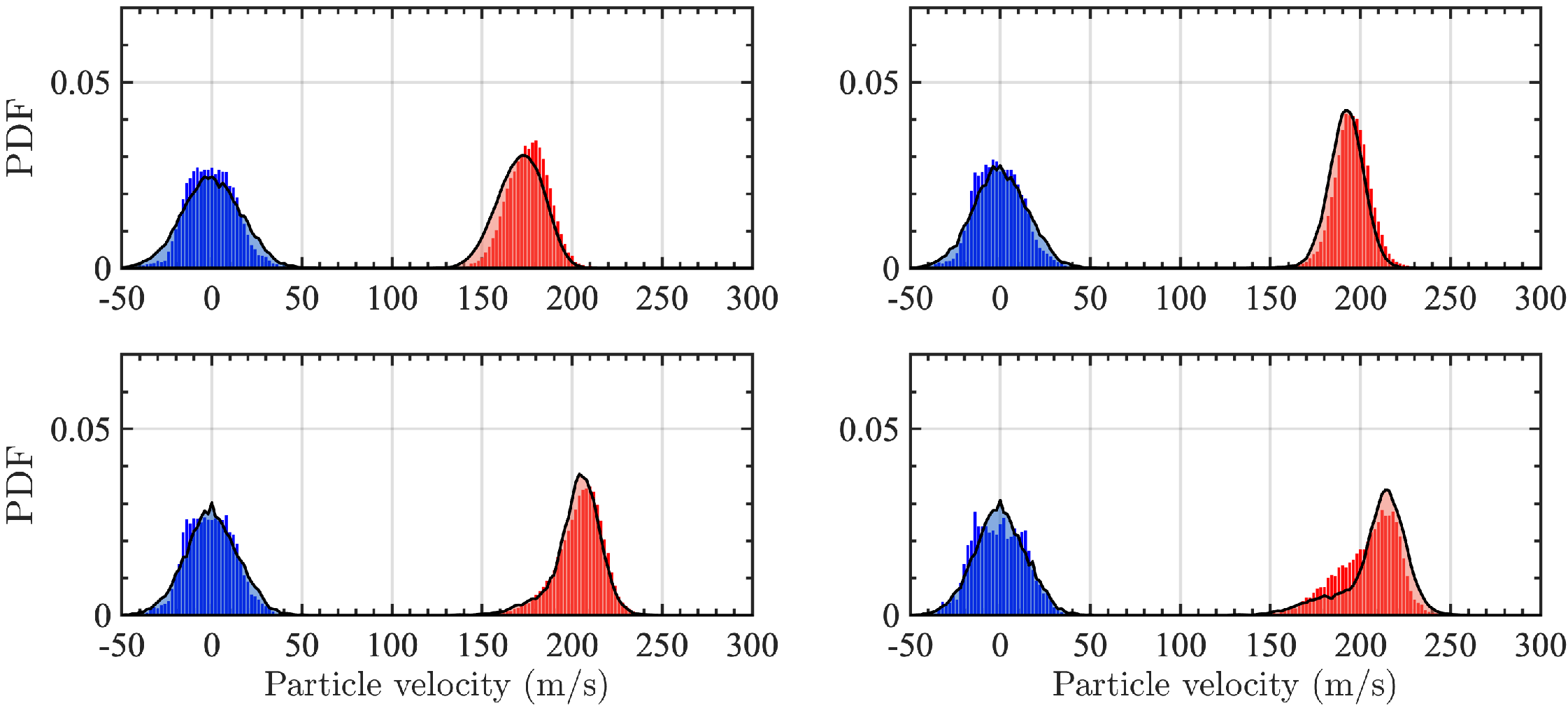}};
        \node[] at (3.5ex,-0.5ex) {(\textit{a})};
        \node[] at (39.5ex,-0.5ex) {(\textit{b})};
        \node[] at (3.5ex,-17ex) {(\textit{c})};
        \node[] at (39.5ex,-17ex) {(\textit{d})};
    \end{tikzpicture}
\caption{Velocity PDFs for Case A2 measured in the region (a) $0\leq x/D_e <1$, (b) $1 \leq x/D_e <2$, (c) $2 \leq x/D_e <3$ and (d) $3 \leq x/D_e <4$. Same legend as figure~\ref{fig: injecting velocity pdfs}.}
\label{fig: case_A2}
\end{figure}
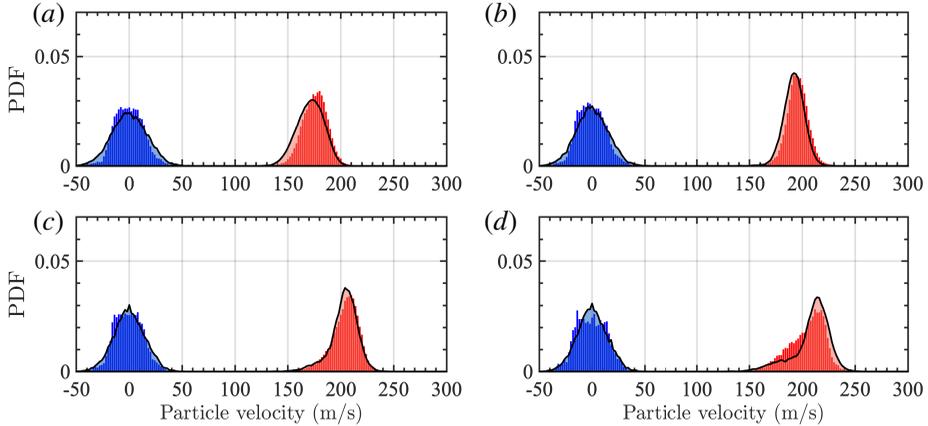

Particle velocity statistics are reported at different downstream locations from $0 \leq x/D_e \leq 4$ divided into windows with width of $D_e$. Simulation results are collected at the same frame rate of the experiments, corresponding to 256 snapshots taken in a 130 $\upmu$s time window. Figure~\ref{fig: case_A2} shows the probability density functions (PDFs) for the streamwise and spanwise particle velocities for Case A2 ($\mathit{\Phi_m}= 0.37$, $\overline{d}_p = 29 ~\upmu$m). Particles are injected into a rapidly accelerating gas phase, causing them to gain momentum through drag. This is evident in figure~\ref{fig: case_A2} as the particle streamwise velocity increases from the mean of $\approx$175 m/s to $\approx$220 m/s. Note that the shape of the streamwise velocity distribution (red) also changes with the streamwise distance. As particles travel downstream, the spanwise velocity distribution remains Gaussian (with zero mean) with marginal increase in variance owing to rotational symmetry of the flow.

On the other hand, the streamwise velocity distribution exhibits skewness which is captured by both simulations and experiments. However, the skewness is more pronounced for the simulations. These distributions are biased towards fast-moving particles, suggesting a higher number density for the faster-moving particles. The ability to capture this skewness is important for applications such as plume surface interactions during landing where these fastest-moving particles are the ones that could cause damage to the landing platform. The mean and variance of particle velocities with respect to streamwise distance are shown in Appendix~\ref{app:vel-pdf}.


\begin{figure}
\setlength{\lineskip}{0pt}
    \centering
    \begin{tikzpicture}
        \node[anchor=north west,inner sep=0pt] at (0,0){\includegraphics[width= 0.9\textwidth]{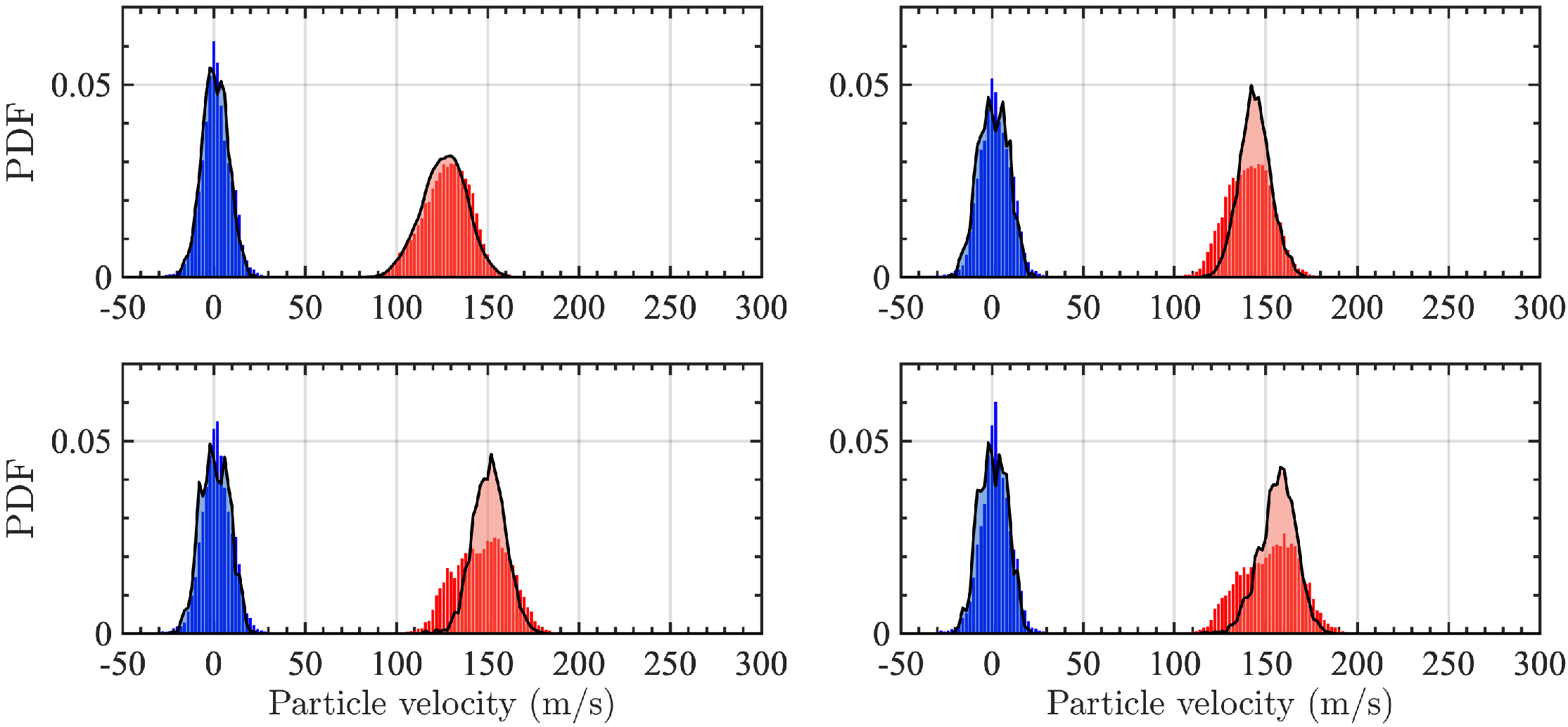}};
        \node[] at (3.5ex,-0.5ex) {(\textit{a})};
        \node[] at (39.5ex,-0.5ex) {(\textit{b})};
        \node[] at (3.5ex,-17ex) {(\textit{c})};
        \node[] at (39.5ex,-17ex) {(\textit{d})};
    \end{tikzpicture}
\caption{Velocity PDFs for Case B4 measured in the region (a) $0\leq x/D_e <1$, (b) $1 \leq x/D_e <2$, (c) $2 \leq x/D_e <3$ and (d) $3 \leq x/D_e <4$. Same legend as figure~\ref{fig: injecting velocity pdfs}.}
\label{fig: case_B4}
\end{figure}
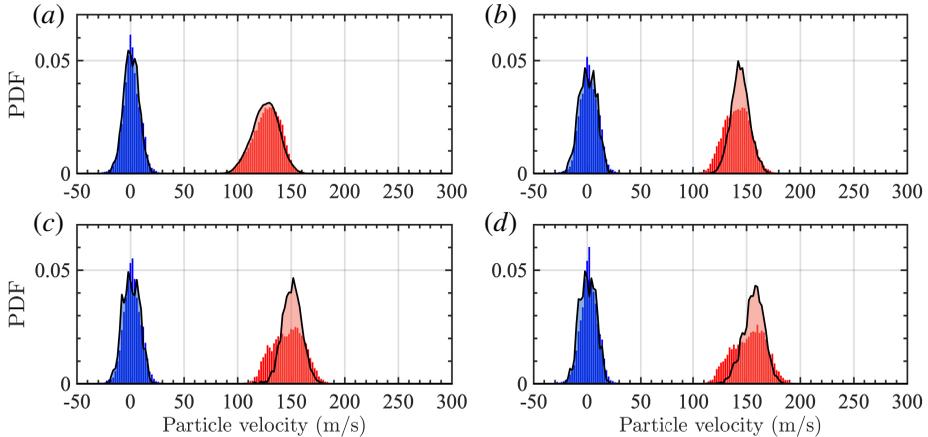

When the particle size and mass loading are increased (figure~\ref{fig: case_B4}; Case B4), particles do not accelerate as far downstream. This is attributed to the higher Stokes number compared to Case A2 resulting in increased lag. It can also be observed that the particles accelerate the most in the vicinity of the Mach disk ($0 \leq x/D_e \leq 2$) where the gas-phase acceleration is greatest. Similar to Case A2, there is an overall good agreement in the mean particle velocity observed between simulations and experiments. However, the streamwise velocity variance is slightly underpredicted at $x/D_e>2$. It is interesting to note that the skewness in the velocity distribution is less pronounced in Case B4 compared to Case A2. This is attributed to the ballistic nature of large-sized particles and is consistent with the velocity distributions observed in the other cases considered (see Appendix~\ref{app:vel-pdf}). These results also reveal that there is a significant momentum exchange between the phases, especially in the initial shock cells. 

\subsection{Changes in Mach disk characteristics}
The focus of this study lies in examining the movement of the Mach disk location due to the presence of particles. As mentioned in \S\ref{sec:intro}, the introduction of particles leads to an upstream movement of the Mach disk toward the nozzle exit. In this section, we demonstrate how the Mach disk and surrounding gas structures undergo modifications due to two-way coupling by particles.

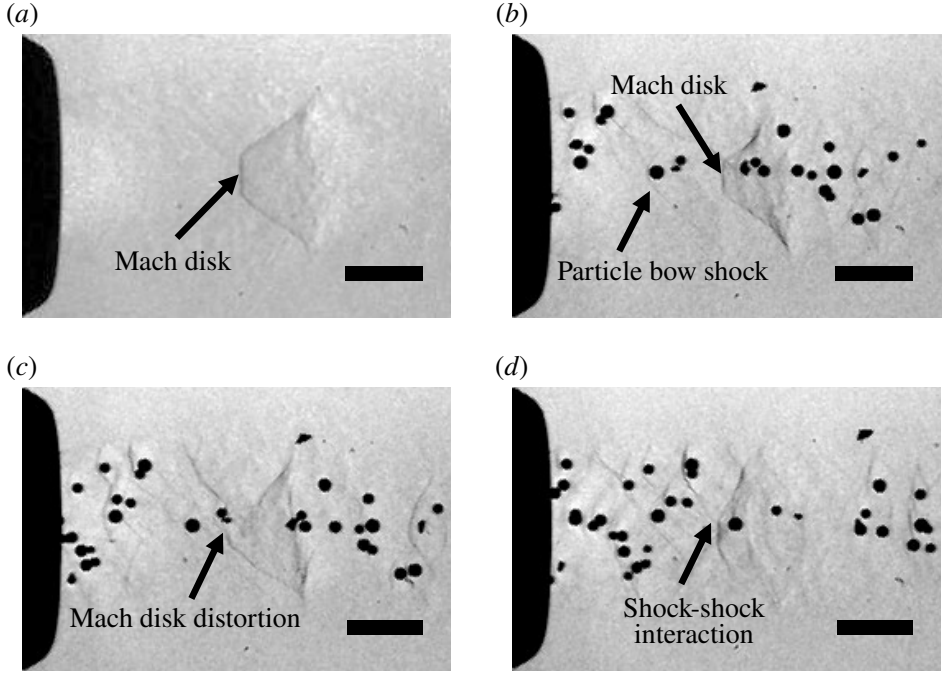
\begin{figure} 
\centering
\subfloat{%
  \begin{tikzpicture}
        \node[anchor=north west,inner sep=0pt] at (0,0){\includegraphics[width= 0.42\textwidth]{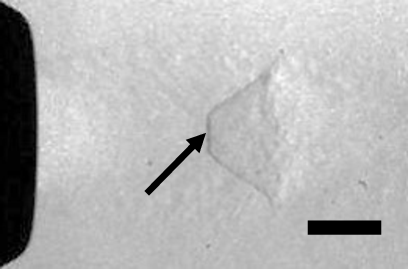}\label{subfig: exp_a}};
        \node[] at (0ex,1.5ex) {(\textit{a})};
        \node[] at (12ex,-18ex) {Mach disk};
    \end{tikzpicture}
}
\quad
\subfloat{%
  \begin{tikzpicture}
        \node[anchor=north west,inner sep=0pt] at (0,0){\includegraphics[width= 0.42\textwidth]{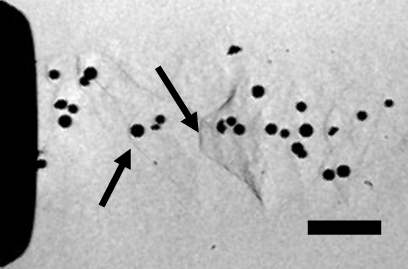}\label{subfig: exp_b}};
        \node[] at (0ex,1.5ex) {(\textit{b})};
        \node[] at (12.5ex,-4ex) {Mach disk};
        \node[] at (12ex,-18.8ex) {Particle bow shock};
    \end{tikzpicture}
}

\subfloat{%
  \begin{tikzpicture}
        \node[anchor=north west,inner sep=0pt] at (0,0){\includegraphics[width= 0.42\textwidth]{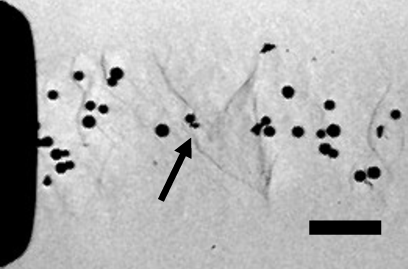}\label{subfig: exp_c}};
        \node[] at (0ex,1.5ex) {(\textit{c})};
        \node[] at (13ex,-18.3ex) {Mach disk distortion};
    \end{tikzpicture}
}
\quad
\subfloat{%
  \begin{tikzpicture}
        \node[anchor=north west,inner sep=0pt] at (0,0){\includegraphics[width= 0.42\textwidth]{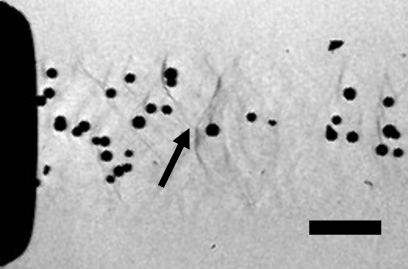}\label{subfig: exp_d}};
        \node[] at (0ex,1.5ex) {(\textit{d})};
        \node[] at (14.5ex,-17.5ex) {Shock-shock};
        \node[] at (14.5ex,-19.5ex) {interaction};
    \end{tikzpicture}
}
\caption{Instantaneous experimental snapshots showing observations of two-way coupling using particles with a mean diameter of $\overline{d}_p = 140 \ \upmu$m, $\eta_0 = 4.42$, and $\Phi_m = 0.37$. The particles and the gas flow from left to right. (a) $t = 79 \ \upmu$s, (b) $t = 84 \ \upmu$s, (c) $t = 89 \ \upmu$s, (d) $t = 94 \ \upmu$s. The scale bar is 1 mm in length for all images.}
\label{fig: shock-particle interactions}
\end{figure}

Figure~\ref{fig: shock-particle interactions} shows experimental snapshots of a flow with soda lime glass beads with a mean particle diameter of $\overline{d}_p = 140 \ \upmu$m and $\mathit{\Phi_m} = 0.37$, visualised by placing a vertical knife-edge at the focal plane to enhance density gradients in the horizontal direction. Figure~\ref{fig: shock-particle interactions}(a) shows the unladen jet at $\eta_0 = 4.42$, representing a highly underexpanded jet with a large Mach disk downstream of the nozzle. When particles are present, the relatively low mass loading significantly affects the structure of the jet, as evident in figure~\ref{fig: shock-particle interactions}(b), where the shock cell and Mach disk have moved closer towards the nozzle, and the oblique shocks appear more distorted. However, the shock cell remains intact. In figure~\ref{fig: shock-particle interactions}(c), a pair of particles interact with the Mach disk, distorting a portion of the Mach disk, yet the shock cell remains intact. However, in figure~\ref{fig: shock-particle interactions}, a particle passing through the jet centreline results in significant distortion of the shock cell and Mach disk. Despite the low mass loading, as evidenced by the few particles emanating from the nozzle, just two particles are enough to fully distort the Mach disk and surrounding shock structures.

In addition to the modification of the Mach disk, the presence of particles in the high-speed and accelerating flow results in the generation of bow shocks upstream of individual particles as shown in figure~\ref{fig: shock-particle interactions}(b). The emergence of bow shocks is due to the large slip Mach number $M_p > 1$. These structures manifest upstream of the Mach disk, within the supersonic flow regime, and dissipate downstream of the Mach disk where the flow is subsonic. Farther downstream of the shock cell, the flow accelerates to supersonic speeds, and bow shocks emerge once again around the particles. The presence of bow shocks could potentially alter the surrounding flow properties further, as these structures extend to the jet boundary, as seen for some of the particles. In figure~\ref{fig: shock-particle interactions}(c), a reflected shock is clearly visible above the pair of particles interacting with the Mach disk, which could be emanating from near the nozzle region where several particles are located. In addition, the time sequence shows the interaction of bow shocks between neighbouring particles near the nozzle exit. This was previously observed and analysed in experimental and numerical work on shock-shock interaction in the two-particle system by \cite{laurence2012surfing}. Figure~\ref{fig: shock-particle interactions}(d) also shows an instance of a bow shock interacting with the distorted Mach disk. 

\begin{figure}
\subfloat{%
  \begin{tikzpicture}
        \node[anchor=north west,inner sep=0pt] at (0,0){\includegraphics[width= 0.42\textwidth]{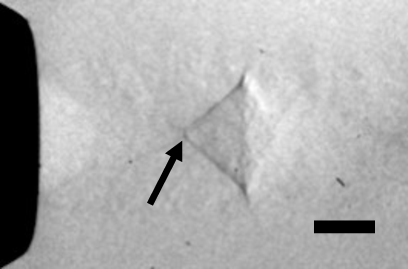}\label{subfig: exp_aa}};
        \node[] at (0ex,1.5ex) {(\textit{a})};
        \node[] at (12.5ex,-18.5ex) {Mach disk};
    \end{tikzpicture}
}
\quad
\subfloat{%
  \begin{tikzpicture}
        \node[anchor=north west,inner sep=0pt] at (0,0){\includegraphics[width= 0.42\textwidth]{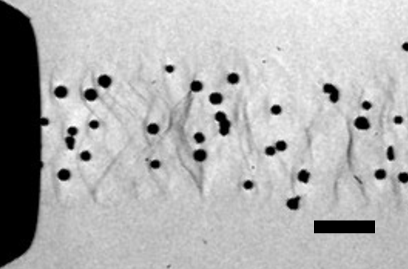}\label{subfig: exp_bb}};
        \node[] at (0ex,1.5ex) {(\textit{b})};
    \end{tikzpicture}
}

\subfloat{%
  \begin{tikzpicture}
        \node[anchor=north west,inner sep=0pt] at (0,0){\includegraphics[width= 0.42\textwidth]{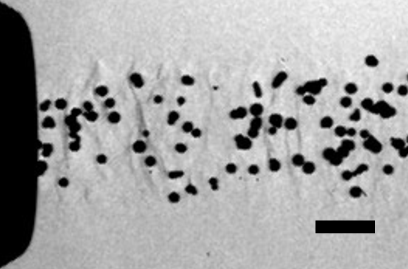}\label{subfig: exp_cc}};
        \node[] at (0ex,1.5ex) {(\textit{c})};
    \end{tikzpicture}
}
\quad
\subfloat{%
  \begin{tikzpicture}
        \node[anchor=north west,inner sep=0pt] at (0,0){\includegraphics[width= 0.42\textwidth]{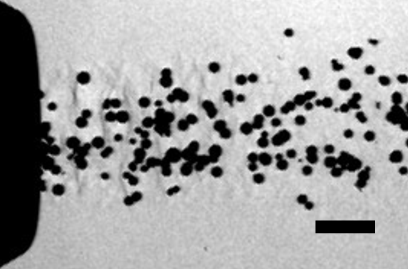}\label{subfig: exp_dd}};
        \node[] at (0ex,1.5ex) {(\textit{d})};
    \end{tikzpicture}
}
\caption{Experimental images showing comparisons between the (a) single-phase jet and particle-laden case with $\overline{d}_p = 140 \ \upmu$m particles and $\eta_0 = 3.40$ with (b) $\mathit{\Phi_m} = 0.80$, (c) $\mathit{\Phi_m} = 1.73$, and (d) $\mathit{\Phi_m} = 2.26$. The scale bar is 1 mm for all is 1 mm in length.}
\label{fig: effect of mass loading}
\end{figure}

The distortion of the Mach disk and shock cell at low mass loadings clearly show the strength of the two-way coupling. When $\mathit{\Phi_m}$ becomes larger, the momentum of the flow is further reduced due to drag by the particles. Figure~\ref{fig: effect of mass loading} shows instantaneous experimental images with $\eta_0 = 3.40$ and particles of $\overline{d}_p = 140 \ \upmu$m. Figure~\ref{fig: effect of mass loading}(a) shows the single-phase jet, with a small Mach disk diameter and visible oblique and reflected shocks. When particles are present at $\mathit{\Phi_m} = 0.80$, a shock cell is no longer visible, and the Mach disk appears to be broken down, a similar behavior as in figure~\ref{fig: shock-particle interactions}. As the mass loading increases beyond $\mathit{\Phi_m} > 1$, both shock cell and Mach disk completely disappear as seen in figures~\ref{fig: effect of mass loading}~(c,d). The only remaining shock structures in the flow are the bow shocks around particles, which wrap around the particles at $\mathit{\Phi_m} < 1$ (see figure~\ref{fig: effect of mass loading}(b)) and become a curtain of shock structures as the mass loading is increased (see figure~\ref{fig: effect of mass loading}(c,d)). 

\begin{figure}
 \setlength{\lineskip}{0pt}
    \centering
    \begin{tikzpicture}
        \node[anchor=north west,inner sep=0pt] at (0,0){\includegraphics[width= 0.66\textwidth]{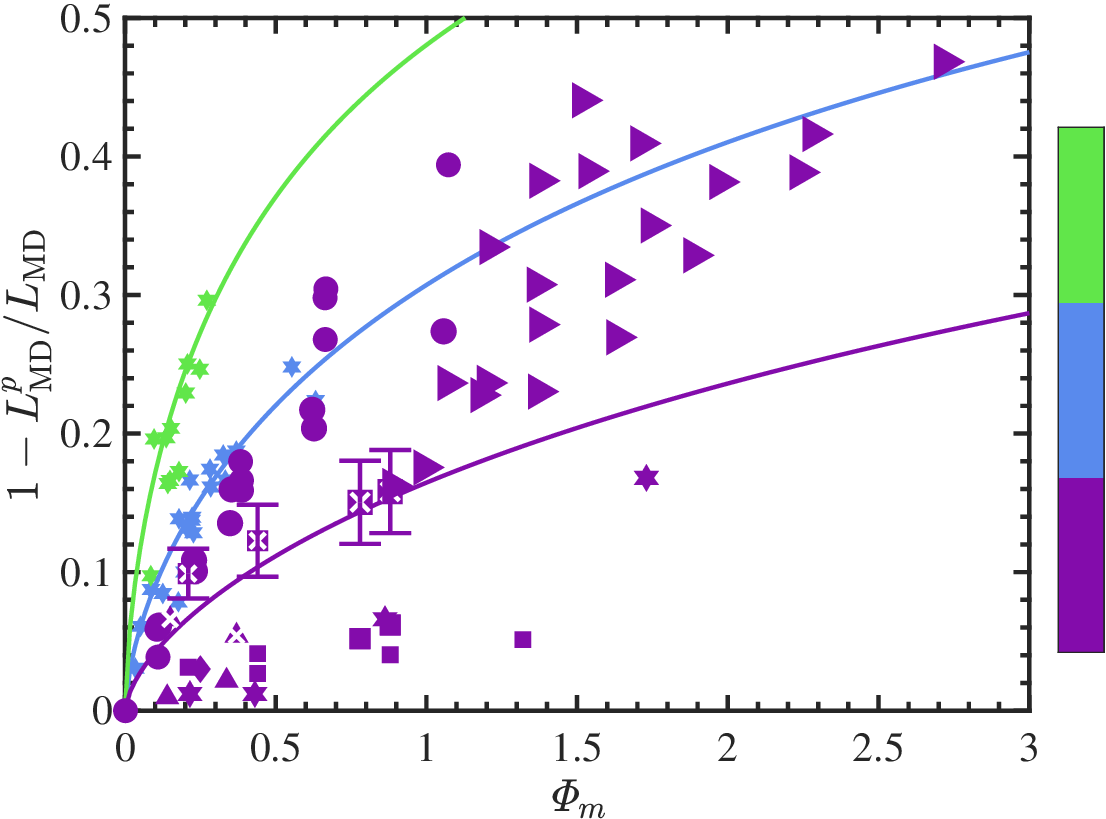}};
        \node[] at (53ex,-5ex) {$M_e$};
        \node[] at (53.5ex,-11ex) {-};
        \node[] at (57ex,-11ex) {2.905};
        \node[] at (53.5ex,-19.25ex) {-};
        \node[] at (57ex,-19.25ex) {1.75~~};
        \node[] at (53.5ex,-27.5ex) {-};
        \node[] at (57ex,-27.5ex) {1~~~~~~~~};
    \end{tikzpicture}\\
 \vspace{.05in}
  \caption{Relative shift in the Mach disk location as a function of mass loading.  Experiments in the present study: $\text{NPR} \ {=} \ 3.4$, $\overline{d}_p \ {=} \ 29~\upmu$m  (\protect \triangleCrossPurple[scale=0.2]),
    $\text{NPR} \ {=} \ 6.46$, $\overline{d}_p \ {=} \ 42~\upmu$m and $96~\upmu$m (\protect \SquareCrossPurple[scale=0.2]). Simulations: $\text{NPR} \ {=} \ 3.4$, $\overline{d}_p \ {=} \ 29~\upmu$m (\protect \trianglefillPurple),  $\text{NPR} \ {=} \ 6.46$, $\overline{d}_p \ {=} \ 42~\upmu$m and $96~\upmu$m (\protect \rectanglefillPurplelarge),
    $\text{NPR} \ {=} \ 30$, $\overline{d}_p \ {=} \ 49~\upmu$m (\protect\tikz \protect \node [rectangle,draw,xscale=0.6,yscale=0.7,rotate=315, fill=pat_purple, pat_purple,solid,line width = 1pt]{};).
   Experiments by \cite{sommerfeld1994structure} with $\text{NPR}{\approx}30$ (\protect \circlefillPurple). \cite{jain2024experimental} with $\text{NPR} \ {=} \ 3.46-6.23$, $\overline{d}_p \ {=} \ 116~\upmu$m ($\,$\protect \righttrianglefillPurple), simulations from \cite{carcano2013semi} (\protect \starfillpurplethree), experiments by \cite{lewis1964normal}: ${M}_e=2.905$~(\protect \starfillgreenthree), ${M}_e=1.75$~(\protect \starfillbluethree). Solid lines correspond to the empirical correlation \eqref{eq: emp_corr}. Larger symbols indicate bigger particles.}  
  \label{fig: percent change of Mach disk}
\end{figure}

To quantify these shock structure changes, the location of the Mach disk is extracted from both experiments and simulations. This is performed by averaging every image together within each experimental and simulation run (a total of 130 $\upmu$s with 256 frames). After obtaining the ensemble average image in the experiments, the background (an image in which the disk was filtered out) is subtracted, and a low-pass FFT filter is applied to enhance the signal-to-noise ratio of the shock structure \citep{ref:Shapiro2001a}. Similar to \cite{ref:Shekhtman2021c}, the Gaussian peak finding algorithm from \cite{ref:OHaver1997a} is then used to determine the locations of the Mach disk and oblique shock waves that shoulder it. The endpoints of the Mach disk were determined by intersections of linear fits for the Mach disk and oblique shocks. The error analysis for the Mach disk (the upper and lower intersection points of the disk with an oblique shock) was performed by using the confidence intervals for lines fitted on signal peaks. This error is propagated along with the standard deviation obtained from multiple runs and is converted to real units using the corresponding spatial resolution for each case studied. Such treatment is not required in the simulations since the particles can be easily decoupled from the gas phase during post-processing. The Mach disk location is extracted in this manner for all of the runs for each case. After obtaining the Mach disk location from experiments and simulations, the relative shift in the Mach disk location as a function of mass loading is computed and shown in figure~\ref{fig: percent change of Mach disk}.

Experiments by \citet{lewis1964normal} were performed at supersonic exit Mach numbers for a single particle size and particle density under the conditions listed in table~\ref{table:one}. They proposed an empirical correlation for the Mach disk location in the presence of particles, given by
\begin{equation}{\label{eq: emp_corr}}
    {L^p_\mathrm{MD}}/{L_\mathrm{MD}}=
      \left({1+0.197{ M}_e^{1.45} \mathrm{\mathit{\Phi_m}}^{0.65}}\right)^{-1}.
\end{equation}
It should be noted that the correlation is independent of nozzle pressure ratio. It returns the single-phase value in the limit $M_e$ and $\Phi_m\rightarrow 0$, and predicts a maximum shift in the Mach disk at the nozzle exit when $M_e$ and $\Phi_m\rightarrow\infty$.

The experimental results of \cite{sommerfeld1994structure} and \cite{jain2024experimental} are also included in figure~\ref {fig: percent change of Mach disk}.  It can be seen that the empirical correlation fails to capture the trends from these two studies. \cite{sommerfeld1994structure} observed that smaller size particles have an increased effect on Mach disk shift compared to large particles. This was attributed to the larger spreading angle associated with the smaller particle, consequently affecting the larger portion of the jet core until the first Mach disk. \citet{jain2024experimental} observed a similar trend. They also found the Mach disk shift to scale with $\eta_0$, counter to the observations made by \cite{lewis1964normal}. The experiments in the present study show reasonable agreement with the correlation. The corresponding error bars show the 95\% confidence interval.

The numerical simulations severely under-predict the shift in the Mach disk location. This is primarily attributed to an underprediction in the two-way coupling source terms. As shown in Appendix~\ref{app:2way}, the distortion of the Mach disk is highly sensitive to the details of the projection scheme in \eqref{eq: vf2way}--\eqref{eq: heat2way}. The wake and shock structures around an individual particles are smeared out when When the filter size used in the Lagrangian projection is too large. This leads to an underestimated pressure drop in the wake region which directly correlates to disturbances in the standing shock. Experimental uncertainty in the particle injection system may also contribute to the differences observed. 


\floatsetup[figure]{style=plain,subcapbesideposition=top}

\section{A semi-analytical model for the Mach disk location}\label{sec:model}

The addition of particles in the flow results in the loss of stagnation pressure, stagnation temperature, and rise in entropy since particles introduce irreversibilities through drag and work exchange in an otherwise isentropic flow. In this section, a one-dimensional model of the two-way coupled flow is formulated to understand the mechanisms contributing to the shift in Mach disk when particles are added to the flow. We consider a one-dimensional constant-area flow with friction (Fanno flow). The canonical friction term resulting from wall shear stress is replaced with the contribution of drag from the particles. The derivation for Fanno flow (without particles) can be found elsewhere \cite[e.g.][]{Hill1992}. 

The two-phase flow is assumed to be steady and one-dimensional with a non-constant fluid velocity $u(x)$. If the particle material density, $\rho_p$, is constant, the continuity equation for the disperse phase can be expressed as ${\rm d}(\alpha_p u_p)=0$, or
\begin{equation}{\label{eq: 1d_disp_mass}}
    \frac{{\rm d}\alpha_p}{\alpha_p}+\frac{{\rm d}u_p}{u_p}=0.
\end{equation}
Thus, $\alpha_p u_p=\text{const.}$ and since $\alpha_p\ll1$, the gas-phase volume fraction $\alpha\approx 1$, and is assumed to be constant and equal to $1-\mathit{\Phi}_v$. Consequently, $u_p$ is taken to be constant.

Under these assumptions and invoking the ideal gas law, the continuity, momentum, and energy equations for the gas phase \eqref{eq:main_gas} reduce to
\begin{equation}{\label{eq: 1d_mass}}
    \frac{{\rm d}\rho}{\rho}+\frac{{\rm d}u}{u}=0,
\end{equation}

\begin{equation}{\label{eq: 1d_momentum}}
    \frac{{\rm d}p}{p} = -\frac{3}{4}\gamma M^2\frac{\mathit{\Phi}_v }{(1-\mathit{\Phi}_v) d_p} C_D \phi {\rm d}x  - \frac{u {\rm d}u}{RT}
\end{equation}
and
\begin{equation}{\label{eq: 1d_energy}}
    \frac{{\rm d}T}{T} = -\dfrac{\gamma-1}{\gamma} \dfrac{u {\rm d}u}{RT}
    -\frac{3}{4}(\gamma-1) \left(M- M_p\right) M\frac{\mathit{\Phi}_v }{(1-\mathit{\Phi}_v) d_p} C_D \phi {\rm d}x - \dfrac{6 (\gamma-1)}{\gamma}\dfrac{\mathit{\Phi}_v \kappa {\rm Nu}(T-T_p)}{d_p^2  p u}{\rm d}x,
\end{equation}
where $\phi = |u-u_p|(u-u_p)/u^2$, $M = |u|/\sqrt{\gamma RT}$ and $M_p = |u-u_p|/\sqrt{\gamma RT}$. $\phi>0$ corresponds to the case when particles lag the fluid (i.e. $u_p<u$) and $\phi<0$ corresponds to flows where particles are traveling faster than the fluid. The third term on the right-hand side of \eqref{eq: 1d_energy} represents heat exchange between two phases. For simplicity, added mass and lift are neglected. Numerical experiments showed these contributions have a negligible effect on velocity statistics and the shift in Mach disk.

The relative contribution of heat exchange to the work done by drag (the last two terms in \eqref{eq: 1d_energy}) is
\begin{equation}
    \dfrac{8(1-\mathit{\Phi}_v)\kappa\mathit{Nu}\left(T-T_0\right)}{\gamma d_p p u (M-M_p)M C_D \phi}\ll 1.
\end{equation}
Substituting in representative values of $\kappa, \mathit{Nu}, M,M_p, C_D, p, \mathit{\Phi}_v,$ and $d_p$, the ratio is found to be $\mathcal{O}\left(10^{-3}\right)$. Therefore, we neglect heat transfer between the phases for the subsequent analysis.

Substituting \eqref{eq: 1d_energy} into \eqref{eq: 1d_momentum} yields
\begin{equation}{\label{eq: 1d_momentum2}}
    \frac{{\rm d}p}{p} = -\frac{3}{4}\gamma M^2\frac{\mathit{\Phi}_v }{(1-\mathit{\Phi}_v) d_p} C_D \phi {\rm d}x  + \frac{\gamma}{\gamma-1} \frac{{\rm d}T}{T} + \frac{3}{4}\gamma \left(M- M_p\right)  M\frac{\mathit{\Phi}_v }{(1-\mathit{\Phi}_v) d_p} C_D \phi {\rm d}x,
\end{equation}
which simplifies to


\begin{equation}{\label{eq: 1d_momentum3}}
    \frac{{\rm d}p}{p} = {\frac{\gamma}{\gamma-1} \frac{{\rm d}T}{T}} - {\frac{3}{4}\gamma \frac{\mathit{\Phi}_v }{(1-\mathit{\Phi}_v) d_p} C_D \phi   MM_p{\rm d}x}. 
\end{equation}

Using the isentropic relations for pressure and temperature (see Appendix~\ref{app:nozzle}), yields
\begin{equation}{\label{eq: stagnation1}}
    \frac{{\rm d}p_0}{p_0} = \frac{{\rm d}p}{p} - \frac{\gamma}{\gamma-1}\frac{{\rm d}m}{m},
\end{equation}
and
\begin{equation}{\label{eq: stagnation2}}
    \frac{{\rm d}T_0}{T_0} = \frac{{\rm d}T}{T} - \frac{{\rm d}m}{m},
\end{equation}
where $m = 1 + (\gamma-1)M^2$.

Substituting \eqref{eq: stagnation1} and \eqref{eq: stagnation2}  into \eqref{eq: 1d_momentum3} yields

\begin{equation}{\label{eq: 1d_momentum4}}
    \frac{{\rm d}p_0}{p_0} = {\frac{\gamma}{\gamma-1} \frac{{\rm d}T_0}{T_0}}  - {\frac{3}{4}\gamma \frac{\mathit{\Phi}_v }{(1-\mathit{\Phi}_v) d_p} C_D \phi   MM_p{\rm d}x}. 
\end{equation}

Finally, integrating \eqref{eq: 1d_momentum4} to the location of the Mach disk, $L^p_{{\rm MD}}$, results in the expression


\begin{equation}{\label{eq: pressure_ratio2}}
    \dfrac{p^{p}_0}{p_0} = \left(\dfrac{T^{p}_0}{T_0}\right)^{\gamma/(\gamma-1)} {\rm exp}\left(-b \int_0^{L^p_{{\rm MD}}} G(x)~{\rm d}x\right),
\end{equation}
where $b = {3} {\gamma}/({4(1-\mathit{\Phi}_v) d_p})$ and $G(x) ={ {\mathit{\Phi}_v } C_D \phi   MM_p} $. As previously stated, the above equation assumes no heat exchange between the two phases, but it does take into account the work due to drag.

Here, we assume the Mach disk location in a particle-laden flow, $L^p_{\rm {MD}}$, is equivalent to its single phase counterpart but with a modified pressure ratio $\eta_0^p$. Using the correlation \eqref{eq:Crist}, the location of the Mach disk in the presence of particles can be expressed as
\begin{equation}{\label{eq: crist_modified}}
    \frac{L^p_{\rm {MD}}}{D_{\text{e}}} = \frac{L_{\rm {MD}}}{D_{\text{e}}} \sqrt{\dfrac{p^p_{0}}{p_0}} = \sqrt{\dfrac{\eta_0}{2.4}\bigg(\dfrac{p^p_{0}}{p_0}\bigg)}.
\end{equation}
Substituting \eqref{eq: pressure_ratio2} into \eqref{eq: crist_modified} and assuming $T_0^p\approx T_0$ since interphase heat transfer was found to be negligible, yields
\begin{equation}{\label{eq: pressure_ratio3}}
    \dfrac{L^p_{\rm {MD}}}{D_e} =  \sqrt{\dfrac{\eta_0}{2.4}}\exp\left(-\frac{b}{2} \int_0^{L^p_{{\rm MD}}} G(x)~{\rm d}x\right).
\end{equation}
Note that because the upper limit of integration $L^p_{{\rm MD}}$ is not known a-priori, \eqref{eq: pressure_ratio3} must be solved iteratively. 

It can be seen that ${L^p_{\rm {MD}}}/{D_{\text{e}}}$ varies with the particle-phase volume fraction, gas phase Mach number, slip Mach number, and varies inversely with the gas-phase volume fraction and particle diameter. Additionally, it can be seen that the maximum extent of the Mach disk shift due to the particles (i.e. when the Mach disk moves up to the nozzle exit; $L_{\rm MD}^p=0$) occurs in the limit of infinite mass loading and infinite Mach number. 

\begin{figure}
\centering
\includegraphics[width = 0.5\textwidth]{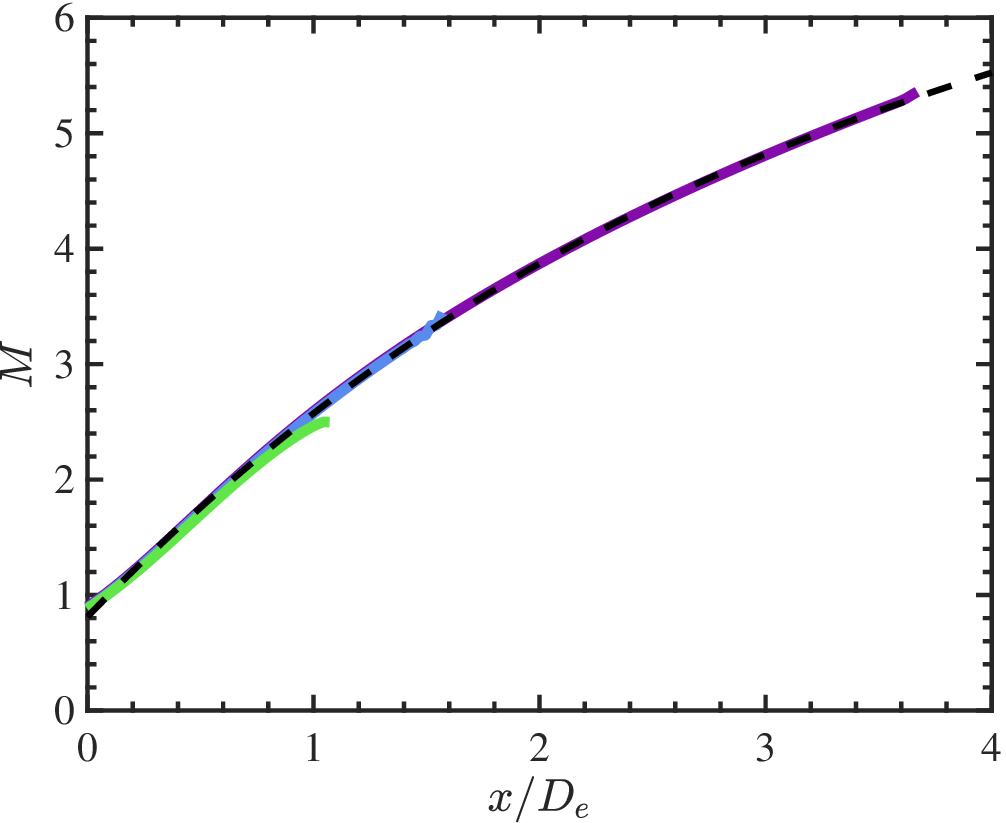}
\caption{$\eta_0 = 3.4$~(\protect \greenlinesolidthicker),  $\eta_0 = 6.46$~(\protect \bluelinesolidthicker), $\eta_0 = 30$~(\protect \purplelinesolidthicker), polynomial fit~\eqref{eq: Ma_dist_sim}~(\protect \blacklinedashedthickk).} 
\label{fig:Mach_fit}
\end{figure}

Obtaining the new Mach disk location from \eqref{eq: pressure_ratio3} requires knowledge of how $C_D$, $M$ and $M_p$ vary in $x$. The gas-phase Mach number is estimated from a polynomial fit informed by the numerical simulations (see figure~\ref{fig:Mach_fit}), given by
\begin{equation}{\label{eq: Ma_dist_sim}}
  M\left({{x}/D_e}\right) =  0.01839\left(\frac{x}{D_e}\right)^3 -0.2863\left(\frac{x}{D_e}\right)^2 + 2.029\left(\frac{x}{D_e}\right) + 0.8123 .
\end{equation}

Table~\ref{table:three} shows a comparison of the percent shift in the Mach disk between the one-dimensional model \eqref{eq: pressure_ratio3} and experiments using the parameters reported in Table~\ref{table: params}. Here, the drag law of \cite{osnes2023comprehensive} is used for $C_D$.  The shift in the Mach disk location is approximately $10-20$\% lower than what is seen in the experiments. This is likely due to the assumption of one-dimensional flow, since the shock structures and particle dynamics are inherently three dimensional. Despite its simplicity, the model predicts the correct trends and gives overall good agreement against the experimental data.

\begin{table}
  \begin{center}
  \def~{\hphantom{0}}
  \begin{tabular}{lccccc}
   \hline 
       Case & $\eta_0$ &~$\mathit{\Phi}_v$  & $d_p$  & \% shift  & \% shift   \\
        & &~$(\times ~10^{-3})$  & $(\upmu {\rm m})$  &  Exp. &  Model 
 \\ \hline
       A1 &3.4&0.29& 29   & 6.61 & 1.75  \\ 
       A2 &3.4&0.77& 29    & 5.44 & 4.28  \\ 
       B1 &6.46&0.84& 42    &9.9 & 6.61  \\ 
       B2 &6.46&1.7  &42    & 12.26 & 11.37  \\ 
       B3 &6.46&3.1  & 96    &15.04 & 11.46  \\ 
       B4 &6.46&3.5  &96   & 15.81& 12.51 \\ 
       \hline
  \end{tabular}
  \caption{Comparison between experiments and the model given in \eqref{eq: pressure_ratio3}.}
  \label{table:three}
  \end{center}
\end{table}

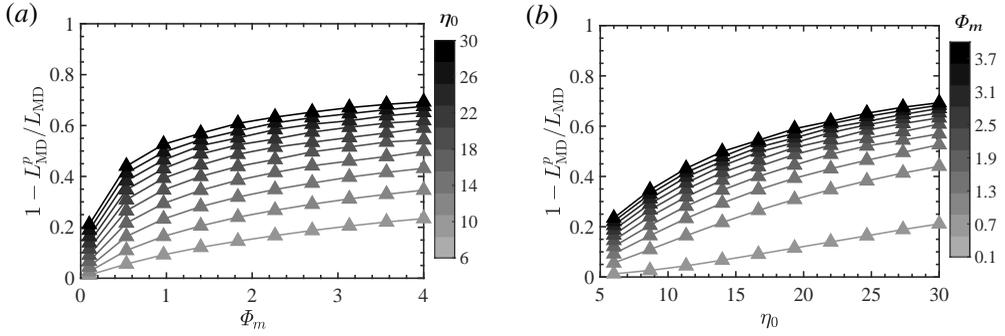
\begin{figure}\centering
\subfloat{%
  \begin{tikzpicture}
        \node[anchor=north west,inner sep=0pt] at (0,0){\includegraphics[width= 0.45\textwidth]{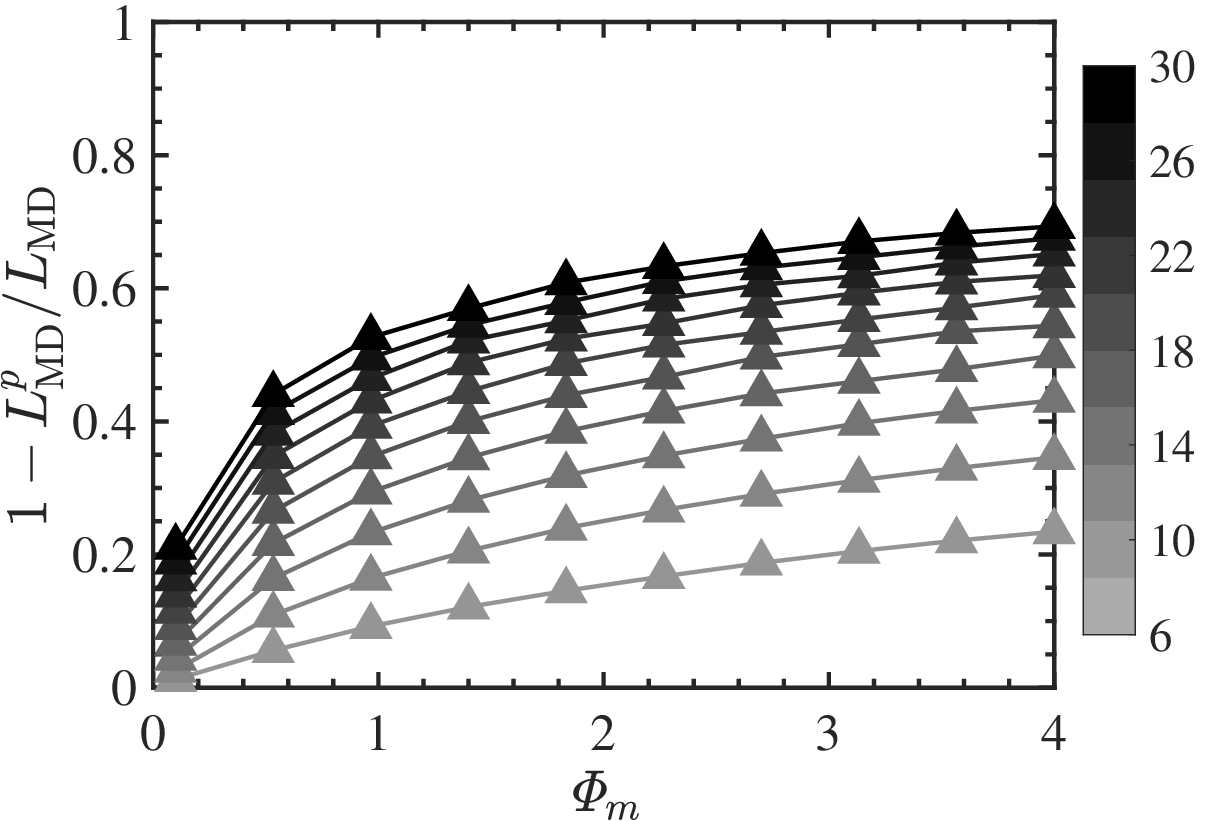}\label{subfig: phi_NPR}};
        \node[] at (0ex,0ex) {(\textit{a})};
        \node[] at (34ex,-0.5ex) {\scriptsize{$\eta_0$}};
    \end{tikzpicture}
}
\quad
\subfloat{%
  \begin{tikzpicture}
        \node[anchor=north west,inner sep=0pt] at (0,0){\includegraphics[width= 0.45\textwidth]{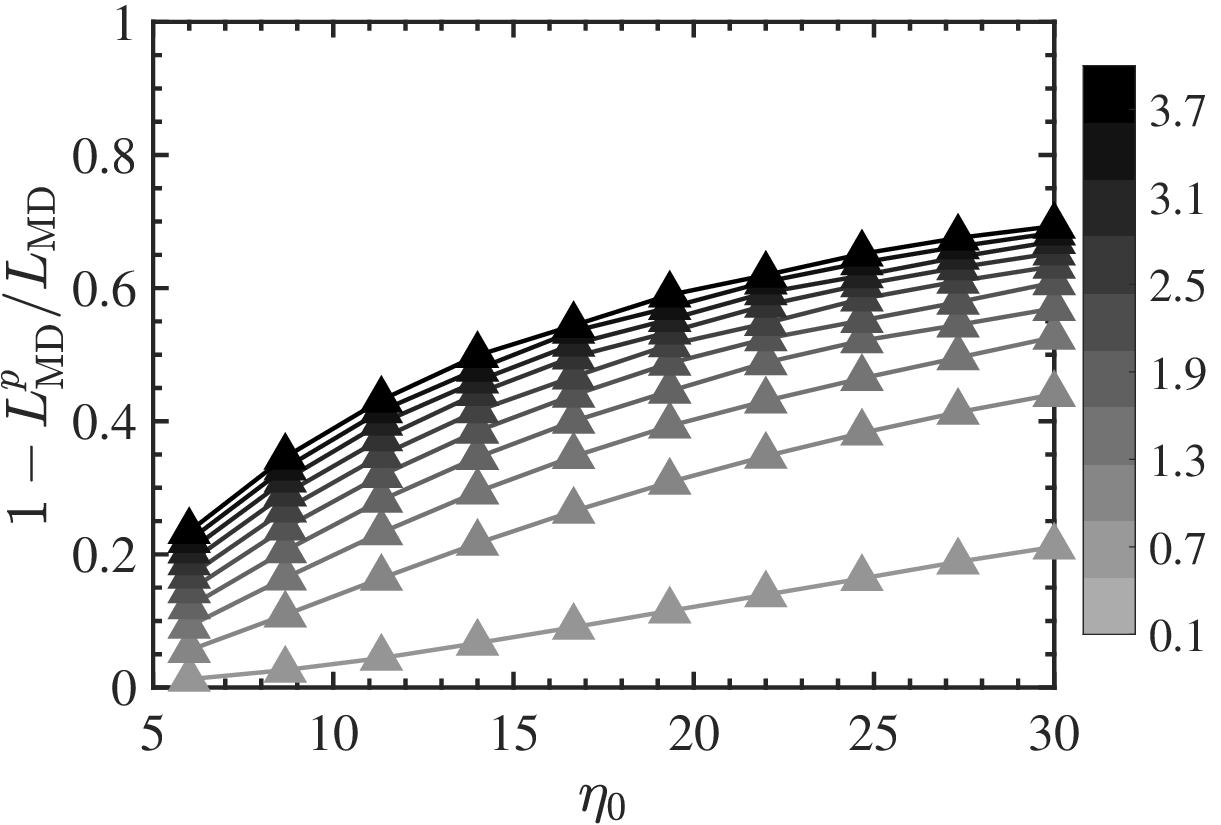}\label{subfig: NPR_phi}};
        \node[] at (0ex,0ex) {(\textit{b})};
        \node[] at (34ex,-0.5ex) {\scriptsize{$\mathit{\Phi}_m$}};
    \end{tikzpicture}
}
\caption{Effect of mass loading and nozzle pressure ratio on Mach disk shift.}
\label{fig:model_NPR_phi}
\end{figure}

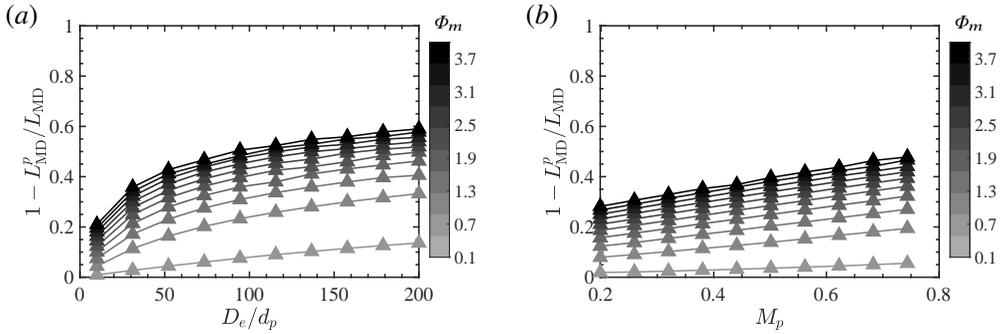
\begin{figure}\centering
\subfloat{%
  \begin{tikzpicture}
        \node[anchor=north west,inner sep=0pt] at (0,0){\includegraphics[width= 0.45\textwidth]{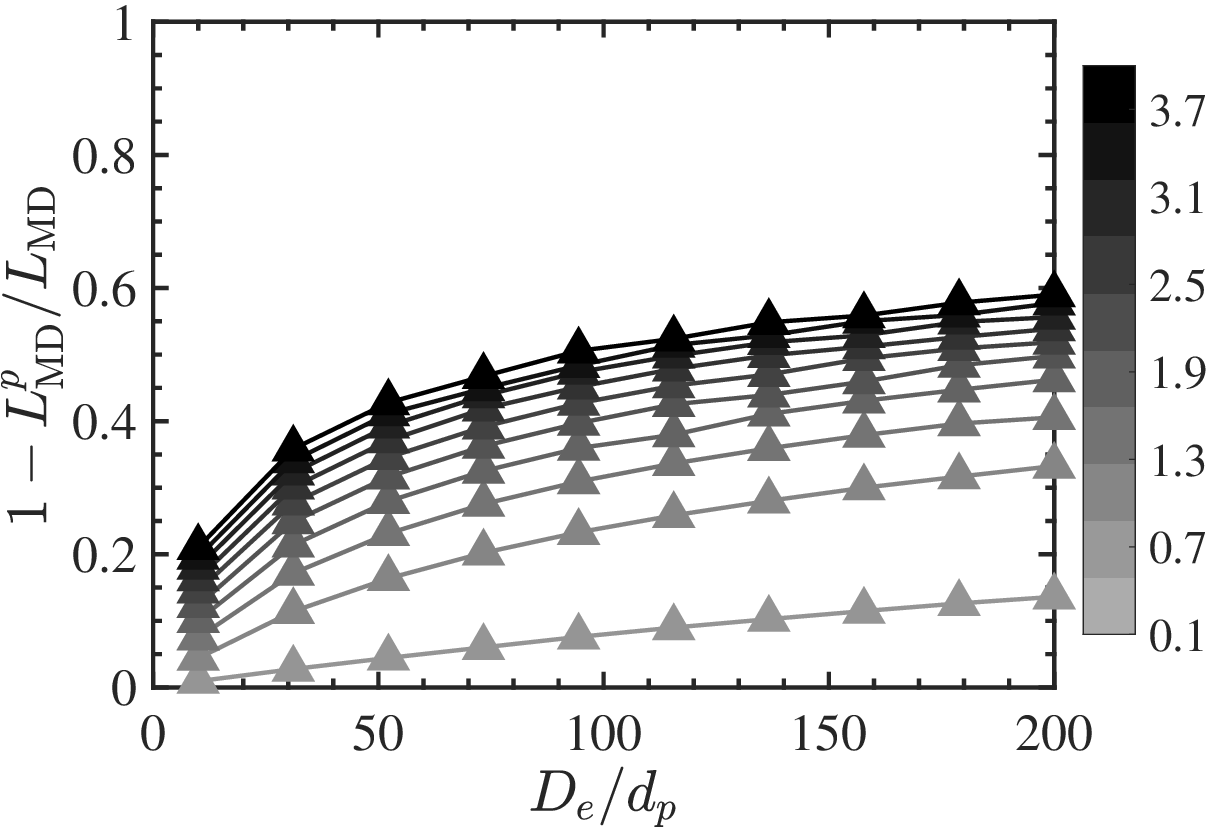}\label{subfig: diameter_ratio}};
        \node[] at (0ex,0ex) {(\textit{a})};
        \node[] at (34ex,-0.5ex) {\scriptsize{$\mathit{\Phi}_m$}};
    \end{tikzpicture}
}
\quad
\subfloat{%
  \begin{tikzpicture}
        \node[anchor=north west,inner sep=0pt] at (0,0){\includegraphics[width= 0.45\textwidth]{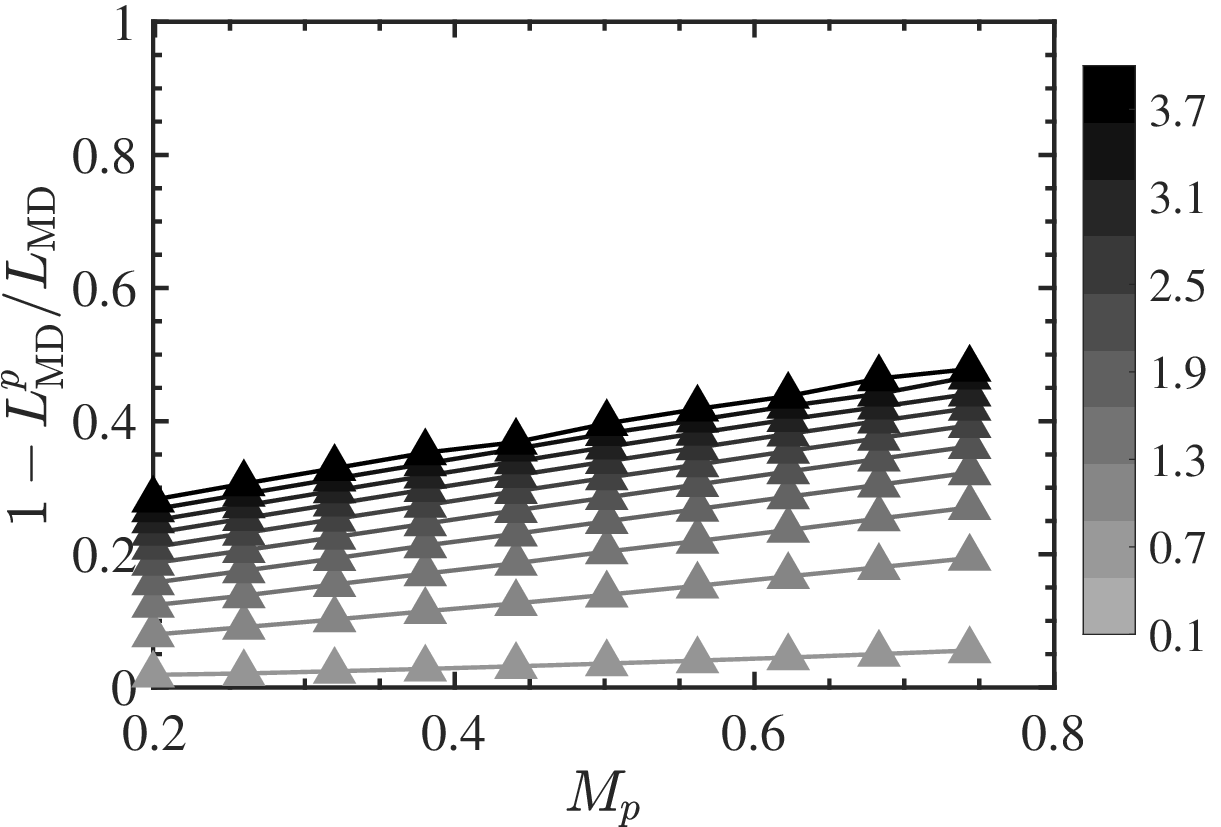}\label{subfig: slip_Mach}};
        \node[] at (0ex,0ex) {(\textit{b})};
        \node[] at (34ex,-0.5ex) {\scriptsize{$\mathit{\Phi}_m$}};
    \end{tikzpicture}
}
\caption{Effect of particle size and inlet slip Mach number on the Mach disk shift.}
\label{fig:model_Dedp_Mp_phi}
\end{figure}

To understand the effect of the parameters appearing in \eqref{eq: pressure_ratio3},  figures~\ref{fig:model_NPR_phi} and~\ref{fig:model_Dedp_Mp_phi} report the Mach disk shift obtained from the model by varying each parameter independently. Figure~\ref{fig:model_NPR_phi}(a) shows how the shift scales with mass loading for a diameter ratio $D_e/d_p=40$ and $M_p=0.42$ at the nozzle exit, corresponding to a particle velocity $u_p= 160$ m/s. At low $\eta_0$, the shift scales with the mass loading linearly and with an increase in $\eta_0$ the shift scales according to a fractional power of $\mathit{\Phi}_m$. Figure~\ref{fig:model_NPR_phi}(b) reveals a strong correlation with $\eta_0$, consistent with the experimental observations by \citet{jain2024experimental}.

From figure~\ref{fig:model_Dedp_Mp_phi}(a), it is evident that smaller particles result in a larger shift. This is consistent with observations from experiments in this work and in the literature. It is interesting to note that \citet{sommerfeld1994structure} attributed this to a Stokes number effect, whereby smaller particles influence a larger portion of the jet cross section due to increase in lateral spreading. Meanwhile, the one-dimensional model does not account for lateral spreading but predicts a similar trend nonetheless. 

Figure~\ref{fig:model_Dedp_Mp_phi}(b) shows the effect of the velocity lag between the phases at the nozzle exit characterized by $M_p$. Unlike $\mathit{\Phi}_m$, $\eta_0$, and $D_e/d_p$, the effect of $M_p$ is relatively less prominent and exhibits a linear dependence on the Mach disk shift. The results shown in figure~\ref{fig:model_Dedp_Mp_phi}(a) are with constant $\eta_0=10$ and $M_p=0.48$. The results shown in figure~\ref{fig:model_Dedp_Mp_phi}(b) were obtained with $\eta_0=10$ and $D_e/d_p=40$. 
The range of $M_p$ spans the values considered in the present work and experiments reported in the literature. Here, $M_p=0.2$ corresponds to particles injected from the nozzle with velocity $u_p=250$ m/s and $M_p=0.8$ corresponds to $u_p=62$ m/s.

\begin{figure}
\centering
\includegraphics[width = 0.6\textwidth]{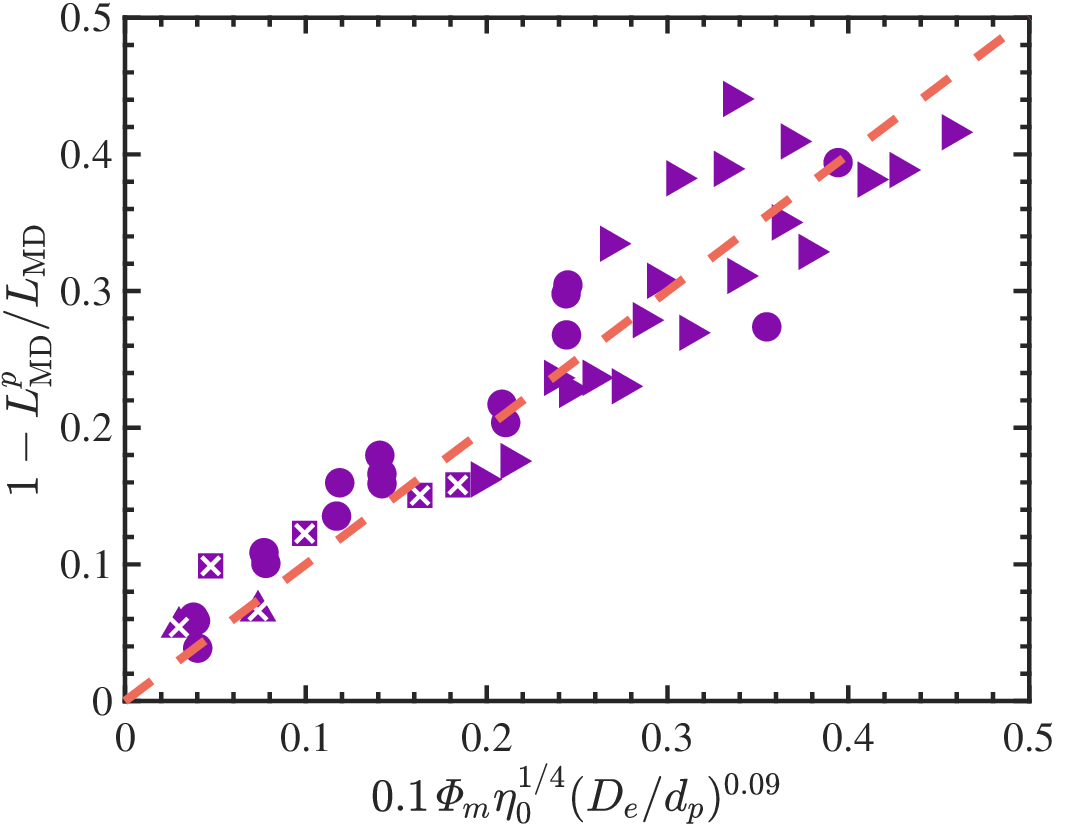}
\caption{Mach disk shift as a function of normalized mass loading from experiments of sonic jets ($M_e=1$). Symbols same as figure~\ref{fig: percent change of Mach disk}. Slope of 1 (\protect \redlinedashedthickk) with $R^2 = 0.89$.}
\label{fig: Mach disk movement collapse}
\end{figure}

As described above, the one-dimensional model suggests the shift in Mach disk caused by particles depends on $\mathit{\Phi}_m$, $\Phi_v$, $D_e/d_p$, $M$, and $M_p$. In order to find an adequate scaling based on the experimental datasets from the literature \citep[e.g.][]{sommerfeld1994structure,jain2024experimental} and the present work, we only consider $\mathit{\Phi}_m$, $\eta_0$ and $D_e/d_p$ since these are readily attainable. It should be noted that in most cases the particle size and slip Mach number are mutually dependent. Smaller particles typically exhibit lower slip Mach numbers due to reduced inertial lag. A least-squares fit reveals the relative shift in Mach disk scales according to
\begin{equation}
    1-\frac{L^p_{\rm MD}}{L_{\rm MD}}\propto\mathit{\Phi}_m\eta_0^{1/4} \left(D_e/d_p\right)^{0.09}.
\end{equation}
As shown in figure~\ref{fig: Mach disk movement collapse}, this results in a modest collapse of the data and a significant improvement over the correlation developed for supersonic jets \eqref{eq: crist_modified} (see figure~\ref{fig: percent change of Mach disk}).

\section{Conclusions}\label{sec:conclusions}
The study of particle-laden sonic and supersonic jets has a long history, stemming from research on solid rockets in the 1960s. This flow configuration showcases a distinct jet structure consisting of shocks, compression, and expansion waves. The introduction of inertial particles significantly modifies the structure of the jet due to strong two-way coupling between the phases. A key feature is the upstream shift in the Mach disk location towards the nozzle. This shift is amplified as the particle mass flow rate increases.

In this work, we performed high-resolution experiments and simulations of sonic jets under two pressure ratios and a wide range of particle sizes and mass loadings. Optical measurements of both phases were obtained using an ultra-high-speed camera system operating at 2 million frames per second. The gas phase was visualised using a schlieren imaging system and particle phase statistics were obtained via Lagrangian particle tracking. Concurrent three-dimensional Eulerian-Lagrangian simulations were performed using a high-order, low-dissipative discretization of the gas phase. An efficient and simple two-way coupling strategy was proposed to handle the interphase exchange term in the presence of discontinuities. Particle size and velocity distributions at the nozzle exit were measured experimentally and used to inform boundary conditions in the simulations.


The focus of this work was on the near-field region within the first four diameters downstream of the exit of the nozzle. Numerical simulations were found to accurately predict the gas-phase shock structures in the unladen cases and particle velocity distributions in the laden cases. The streamwise and spanwise velocity distributions of the particles exiting the nozzle were close to Gaussian with increasing variance as they spread downstream. The particles exit the nozzle with velocity that lags the gas phase, resulting in a net drag force that removes energy from the carrier flow. Although the simulations were capable of reproducing the particle velocity statistics, they  were found to underpredict the shift in Mach disk location by as much as 50\%. This points to a need for improved models to handle two-way coupling in flows with significant acceleration and discontinuities. 


Despite the relatively low volume fractions of $\mathcal{O}(10^{-4}-10^{-3})$, strong two-way coupling between particles and the gas phase was observed. This coupling manifests into short-range disturbances, such as localized bow shocks around individual particles, and long-range interactions that modulate the shock structure far from the particle. This is distinct from incompressible jets that exhibit only mild changes to the carrier flow under such dilute conditions.

To better understand the mechanism associated with the shift in the Mach disk, a one-dimensional model based on Fanno flow was derived that accounts for the effect of volume displacement and momentum and energy exchange between the phases. It was found that the shift in the Mach disk location scales with the mass loading, volume fraction, nozzle pressure ratio, and slip velocity, and inversely with the particle size. Predictions from the model show overall good agreement with the experiments. Based on the results from this study and data collected from the literature, a simple scaling was proposed to collapse the shift in Mach disk location observed in particle-laden sonic jets.

	



\backsection[Acknowledgements]{The authors would like to acknowledge Dr. Yuan Yao and Dr. Taehoon Kim for their early contributions to this work.}

\backsection[Funding]{This work was supported by the National Aeronautics and Space Administration (NASA) grant no. 80NSSC20K0295.}

\backsection[Declaration of interests]{The authors report no conflict of interest.}

\backsection[Data availability statement]{The data that support the findings of this study are openly available in GitHub at https://github.com/jessecaps/jCODE \citep{jCODE}}

\backsection[Author ORCIDs]

\noindent \orcidlink{0000-0002-1952-1960} Meet Patel\hspace{2mm}
\href{https://orcid.org/0000-0002-1952-1960}{https://orcid.org/0000-0002-1952-1960}\\
\noindent \orcidlink{0000-0003-0835-5386} Juan Sebastian Rubio\hspace{2mm}\href{https://orcid.org/0000-0003-0835-5386}{https://orcid.org/0000-0003-0835-5386}\\
\noindent \orcidlink{0000-0003-0648-5435} David Shekhtman\hspace{2mm}\href{https://orcid.org/0000-0003-0648-5435}{https://orcid.org/0000-0003-0648-5435}\\
\noindent \orcidlink{0000-0001-9880-1727} Nick Parziale\hspace{2mm}\href{https://orcid.org/0000-0001-9880-1727}{https://orcid.org/0000-0001-9880-1727}\\
\noindent \orcidlink{0000-0002-1914-7964} Jason Rabinovitch\hspace{2mm}\href{https://orcid.org/0000-0002-1914-7964}{https://orcid.org/0000-0002-1914-7964}\\
\noindent \orcidlink{0000-0003-0835-5386} Rui Ni\hspace{2mm}\href{https://orcid.org/0000-0003-0835-5386}{https://orcid.org/0000-0003-0835-5386}\\
\noindent \orcidlink{0000-0003-0835-5386} Jesse Capecelatro\hspace{2mm}\href{https://orcid.org/0000-0002-1684-4967}{https://orcid.org/0000-0003-0835-5386}

\appendix

\section{Further details on the numerics}\label{app:numerics}
This section provides further details on the numerics than what was presented in the main text.
\subsection{Nozzle parameterization}\label{app:nozzle}
\begin{figure}
\centering
\includegraphics[width = 0.45\textwidth]{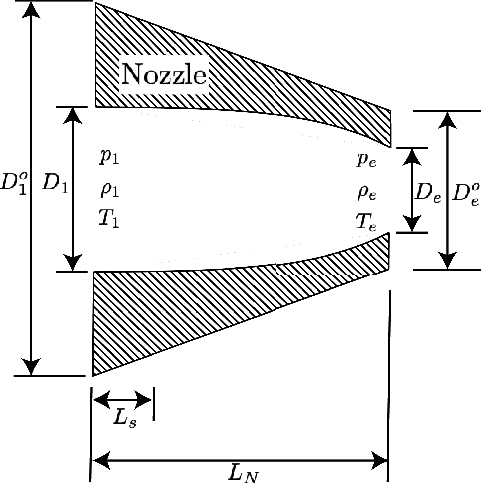}
\caption{Nozzle parameters considered in the numerical simulations.}
\label{fig: nozzle_details}
\end{figure}

A converging nozzle with an analytic interior contour based on a hyperbolic tangent profile is chosen. The ratio of inlet to exit diameters is kept same as the nozzle used in experiments. The inlet geometrical and fluid parameters are denoted by $(\cdot)_1$ and corresponding exit parameters by $(\cdot)_e$. Relationship between inlet and exit conditions are based on the isentropic relations, more details on the derivation can be found in \citet{anderson1990modern}. The relationship between inlet quantities ($D_1, p_1, M_1, \rho_1$) shown in figure~\ref{fig: nozzle_details} is given by 

\begin{align}
    p_1 &= p_0\left(1 + (\gamma-1) M^2_1/2\right)^{-\gamma/(\gamma-1)}, \\
    \rho_1 &= \rho_0\left(1 + (\gamma-1) M^2_1/2\right)^{-1/(\gamma-1)}
\end{align}
and
\begin{align}
    D^2_1 &= D^2_e/M_1\left({\gamma+1}/{2}\right)^{(\gamma+1)/(2\gamma-2)}\left(1 + (\gamma-1) M^2_1/2\right)^{(\gamma+1)/(2\gamma-2)},
\end{align}
where $L_N=4.35D_e$ is the nozzle length, $L_s=0.15D_e$ is the straight section length. The inner contour of the nozzle follows the analytical profile of
\begin{align}{\label{eq: nozzle_contour_eqs}}
    D(x) = & \text{max}\Big[1.1D_e + (1.1D_e-D_1)\text{tanh}\Big(20\sigma_1(x-L_N+L_s)\Big), \\
    &D_e + (D_e-D_1)\text{tanh}\Big(25\sigma_2(x-L_N+L_s)\Big)\Big],
\end{align}
with
\begin{align}
    \sigma_1 = 0.015\quad\text{and}\quad\sigma_2 =\sigma_1\dfrac{D_1 -1.1D_e}{0.1D_e}.
\end{align}
In figure~\ref{fig: nozzle_details}, $D_1^o$ and $D_e^o$ are outer diameters at the inlet and nozzle exit, respectively, that are set to $D_1^o=6.5D_e$ and $D_e^o=2D_e$.

\subsection{Immersed boundary method}\label{app:ibm}
\begin{figure}
\centering
\includegraphics[width=0.55\textwidth]{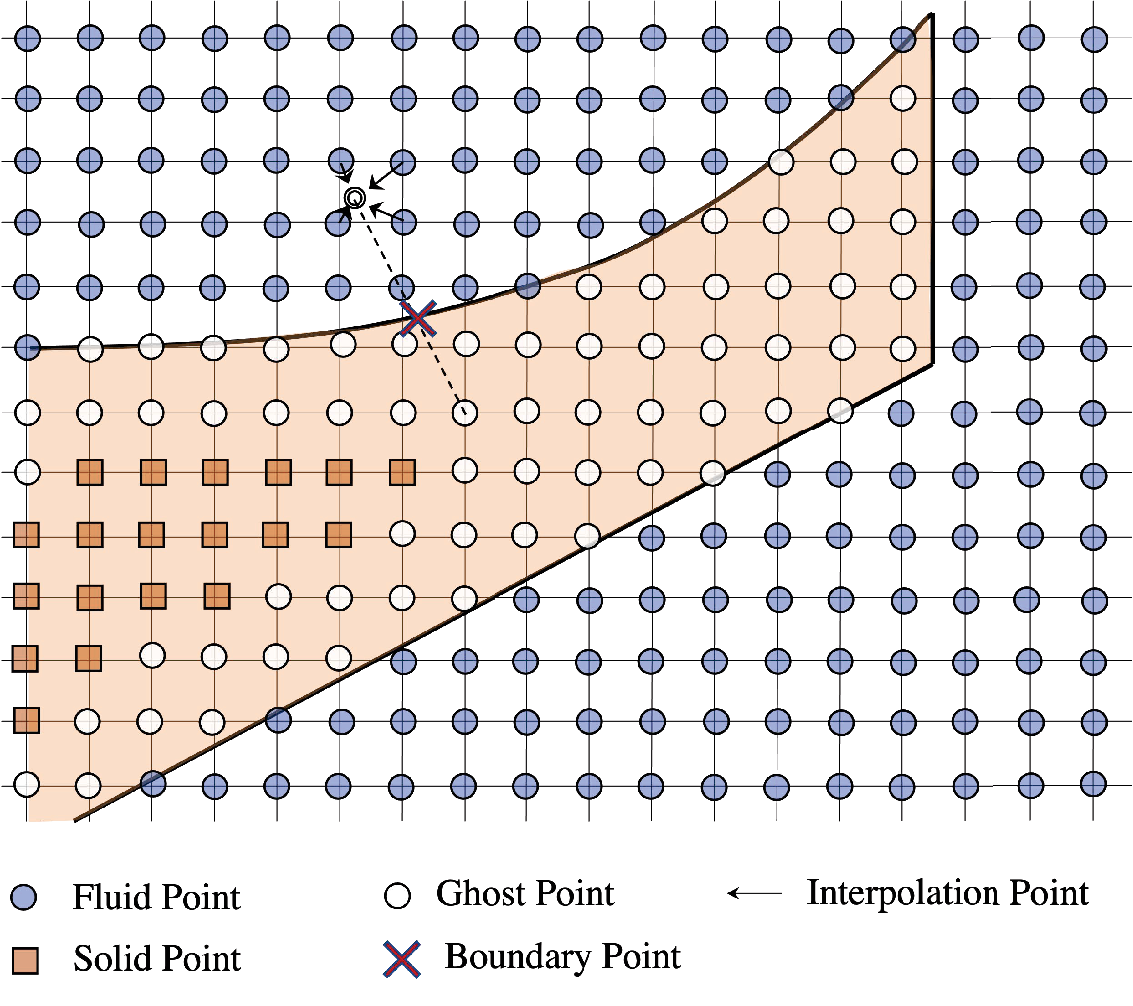}
\caption{Schematic of the ghost point immersed boundary method with two layers of ghost points.}
\label{fig:IBM_approch}
\end{figure}

Boundary conditions are enforced at the surface of the nozzle using a ghost-point immersed boundary method. As shown in figure~\ref{fig:IBM_approch}, the values of the conserved variables at ghost points residing within the solid are assigned based on the quantities at the image point. The image point is identified through a normal vector outward from the surface, $\bm{n}=\nabla G$, where $G$ is a signed distance levelset function. This is performed after each Runge-Kutta sub-iteration. Because image points do not align with grid points, fluid quantities are interpolated to image points via an inverse distance weighting scheme proposed by \citet{chaudhuri2011use}. The number of layers of ghost points grows with increasing order of accuracy of the scheme. For the sixth-order interior finite difference stencil used herein, requires three layers of grid points for the first derivatives. However, for the second derivatives, the number of layers are further extended. 

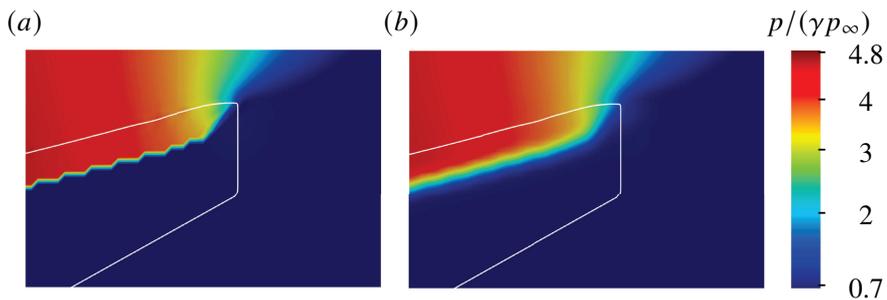
\begin{figure}
 \setlength{\lineskip}{0pt}
    \centering
    \begin{tikzpicture}
        \node[anchor=north west,inner sep=0pt] at (0,0){\includegraphics[width= 0.8\textwidth]{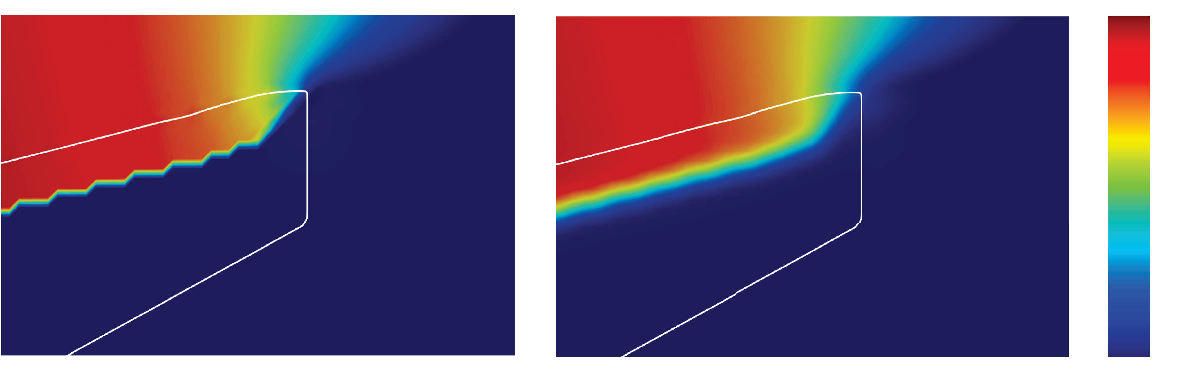}};
        \node[] at (0ex,1.25ex) {$(a)$};
        \node[] at (30ex,1.25ex) {$(b)$};
        \node[] at (63.25ex,1.25ex) {$p/(\gamma p_\infty$)};
        \node[] at (63.5ex,-1.3ex) {-};
        \node[] at (67ex,-1.2ex) {4.8};
        \node[] at (63.5ex,-5ex) {-};
        \node[] at (67ex,-5ex) {4};
        \node[] at (63.5ex,-9ex) {-};
        \node[] at (67ex,-9ex) {3};
        \node[] at (63.5ex,-14ex) {-};
        \node[] at (67ex,-14ex) {2};
        \node[] at (63.5ex,-19.75ex) {-};
        \node[] at (67ex,-19.75ex) {0.7};
    \end{tikzpicture}\\
 \vspace{.05in}
  \caption{Pressure field in the vicinity of the nozzle lip (a) before filtering and (b) after filtering. The white line shows a contour of the zero levelset (nozzle surface).}  
  \label{fig: filter_pressure}
\end{figure}

It should be noted that the nozzle lip introduces additional challenges because of the zeroth order continuity at the sharp corner. This results in a singularity that can result in stability issues and ill-defined normal vectors. \citet{boukharfane2018combined} proposed solving an additional set of equations using the stencils from all sides of the corner (three sides in three dimensions) to prescribe values at ghost-points. Alternatively, \citet{chaudhuri2011use} proposed storing multiple arrays of ghost-points to be used in each flux direction, however this is an expensive approach to evaluate derivatives and is potentially memory intensive. In the present work, we apply a truncated 9-point Gaussian filter \citep{cook2004high} to the flow field within the solid ($G<0$) to smooth the discontinuities within the nozzle. The resulting pressure field inside the nozzle before and after filtering is shown in figure~\ref{fig: filter_pressure}. This was found to reduce spurious oscillations and give better results against the experiments (i.e. the Mach disk location and Mach disk diameter).

\subsection{Shock capturing}\label{app:shock}
The bulk viscosity and thermal conductivity appearing in \eqref{eq: visc_stress} and \eqref{eq: thermal_cond} are augmented according to $\mu_b=\mu_f+\mu^*$ and $\kappa=\kappa_f+\kappa^*$, where the subscript $f$ and asterisks denote fluid and artificial transport coefficients, respectively. The artificial dissipation terms take the form
\begin{equation}\label{eq:LAD}
    \mu^*=C_\beta\overline{\rho f_{sw}|\nabla^4\theta|}\Delta^6,\quad \kappa^*=C_\kappa\overline{\frac{\rho c}{T}|\nabla^4 e|}\Delta^5,
\end{equation}
where $\theta=\nabla\cdot\bm{u}$, $e=(\gamma-1)^{-1}p/\rho$, $C_\beta=1$, and $C_\kappa=0.01$. The overbar denotes a truncated 9-point Gaussian filter \citep{cook2004high}. Fourth derivatives are approximated via a sixth-order
compact (Pad\`e) finite-difference operator \citep{lele1992compact}. 
To limit the artificial bulk viscosity to regions of high compression (shocks), we employ a similar sensor originally proposed by \citet{ducros1999large} and later improved by \citet{hendrickson2018improved}, given by $f_{sw}=\min\left(\frac{4}{3}H(-\theta)\times\frac{\theta^2}{\theta^2+\Omega^2+\epsilon},1\right)$, where $H$ is the Heaviside function, $\epsilon-10^{-32}$ is a small positive constant to prevent division by zero, and $\Omega=\max\left(|\nabla\times\bm{u}|,0.05 c/\Delta \right)$ is a frequency scale that ensures the sensor tends to zero where vorticity is negligible.

It is important to note that in the presence of strong discontinuities, the artificial diffusivity terms used for shock capturing \eqref{eq:LAD} may induce a severe time-step restriction. To avoid introducing unphysical discontinuities near the immersed interface, $\beta^*$ and $\kappa^*$ are defined at every grid point within the domain (interior and exterior), but values inside the solid ($G<0$) are not used when computing ${\rm CFL}_v$.

\subsection{Particle velocity statistics}\label{app:vel-pdf}
This section contains PDFs of particle velocities for all of the cases considered in the current work (Case A1, and Case B1--B3) not reported in the main text. Similar to Case A2 discussed in \S\ref{subsec: particle_vel_pdfs}, Case A1 shows comparable trends in velocities. Additionally, similar to Case B4, Case B3 shows similar trends. For Cases B1 and B2, the overall velocities and the acceleration (comparing velocities between successive windows) appear to be higher than in Cases B3--B4 and lower than observed in Cases A1--A2. This is because Cases B3--B4 have larger size particles with mean diameter of $42~\upmu \mathrm{m}$, compared to $29~\upmu \mathrm{m}$ considered in Cases A1--A2 and $98~\upmu \mathrm{m}$ in B3--B4. Overall, simulation results show good agreement with the experiments, though fail to predict the bi-modal distribution seen in the streamwise velocity in B2.

\begin{figure}
 \setlength{\lineskip}{0pt}
    \centering
    \begin{tikzpicture}
        \node[anchor=north west,inner sep=0pt] at (0,0){\includegraphics[width= 0.9\textwidth]{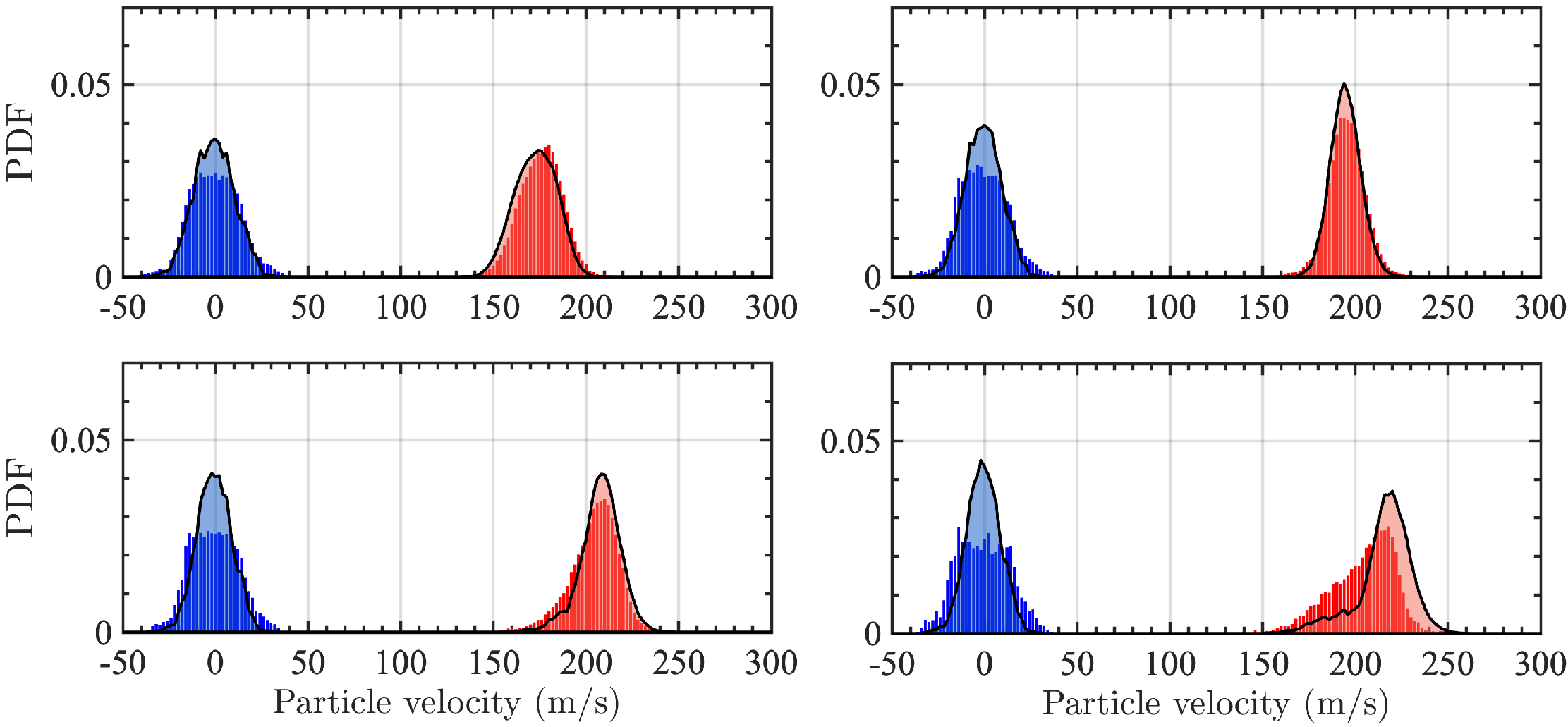}};
        \node[] at (3.5ex,-0.5ex) {(\textit{a})};
        \node[] at (39.5ex,-0.5ex) {(\textit{b})};
        \node[] at (3.5ex,-17ex) {(\textit{c})};
        \node[] at (39.5ex,-17ex) {(\textit{d})};
    \end{tikzpicture}
\caption{Velocity PDFs for Case A1 measured in the region (a) $0\leq x/D_e <1$, (b) $1 \leq x/D_e <2$, (c) $2 \leq x/D_e <3$ and (d) $3 \leq x/D_e <4$. Same legend as figure~\ref{fig: injecting velocity pdfs}.}
\label{fig: case_A1}
\end{figure}
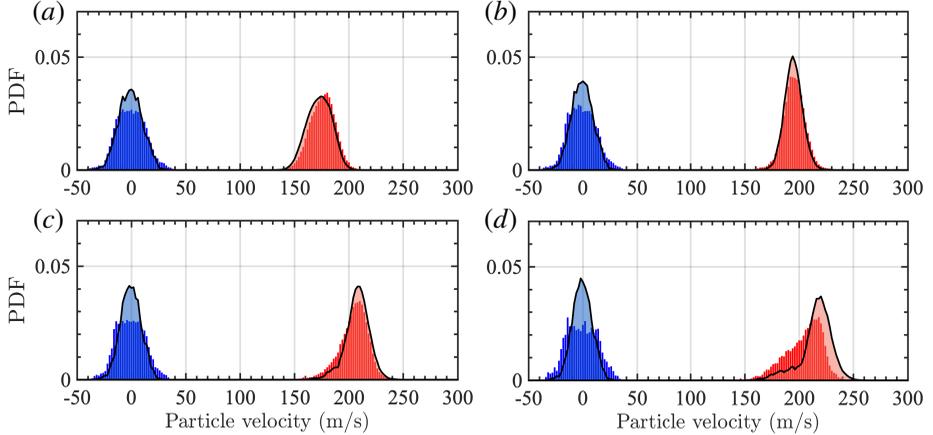

\begin{figure}
 \setlength{\lineskip}{0pt}
    \centering
    \begin{tikzpicture}
        \node[anchor=north west,inner sep=0pt] at (0,0){\includegraphics[width= 0.9\textwidth]{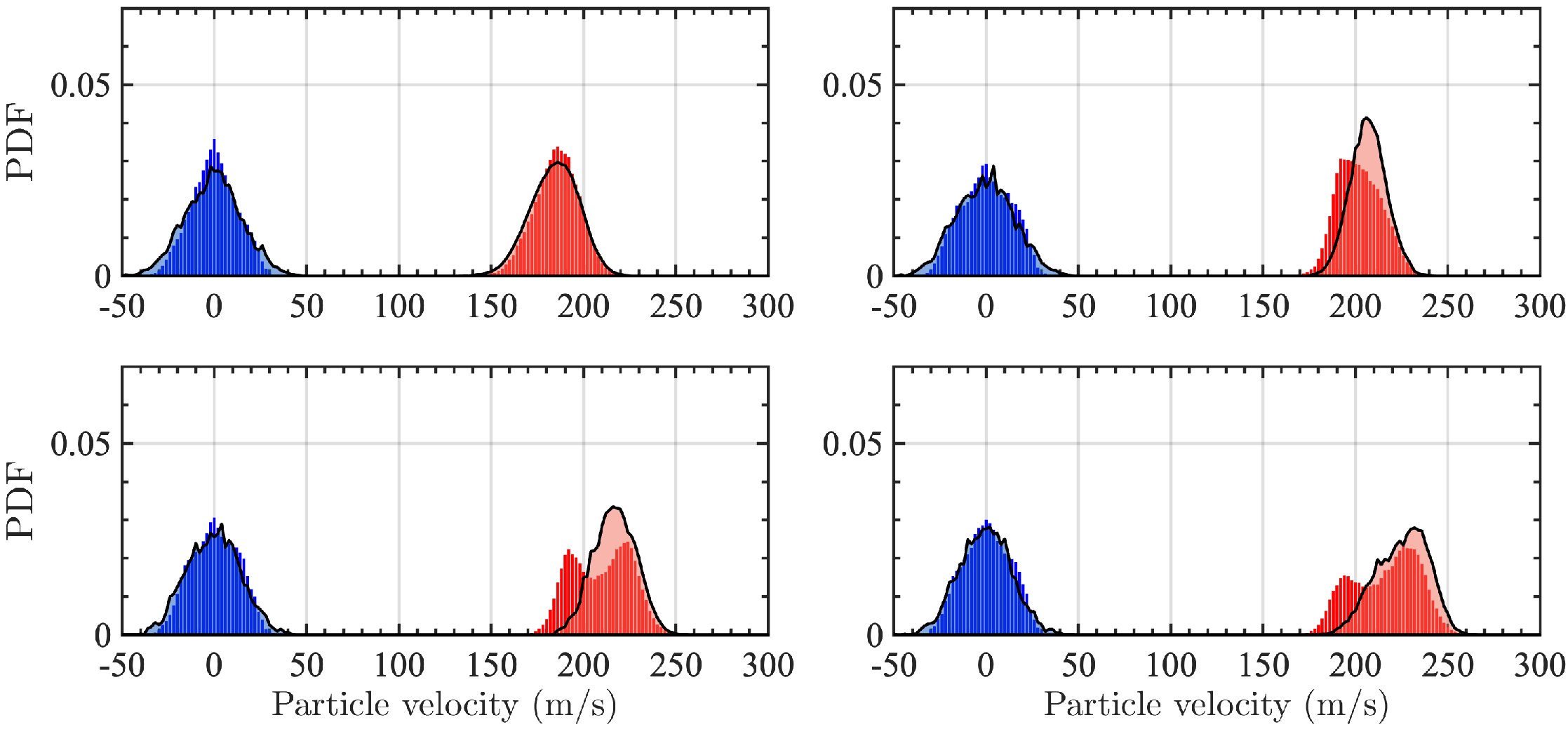}};
        \node[] at (3.5ex,-0.5ex) {(\textit{a})};
        \node[] at (39.5ex,-0.5ex) {(\textit{b})};
        \node[] at (3.5ex,-17ex) {(\textit{c})};
        \node[] at (39.5ex,-17ex) {(\textit{d})};
    \end{tikzpicture}
\caption{Velocity PDFs for Case B1 measured in the region (a) $0\leq x/D_e <1$, (b) $1 \leq x/D_e <2$, (c) $2 \leq x/D_e <3$ and (d) $3 \leq x/D_e <4$. Same legend as figure~\ref{fig: injecting velocity pdfs}.}
\label{fig: case_B1}
\end{figure}

\begin{figure}
 \setlength{\lineskip}{0pt}
    \centering
    \begin{tikzpicture}
        \node[anchor=north west,inner sep=0pt] at (0,0){\includegraphics[width= 0.9\textwidth]{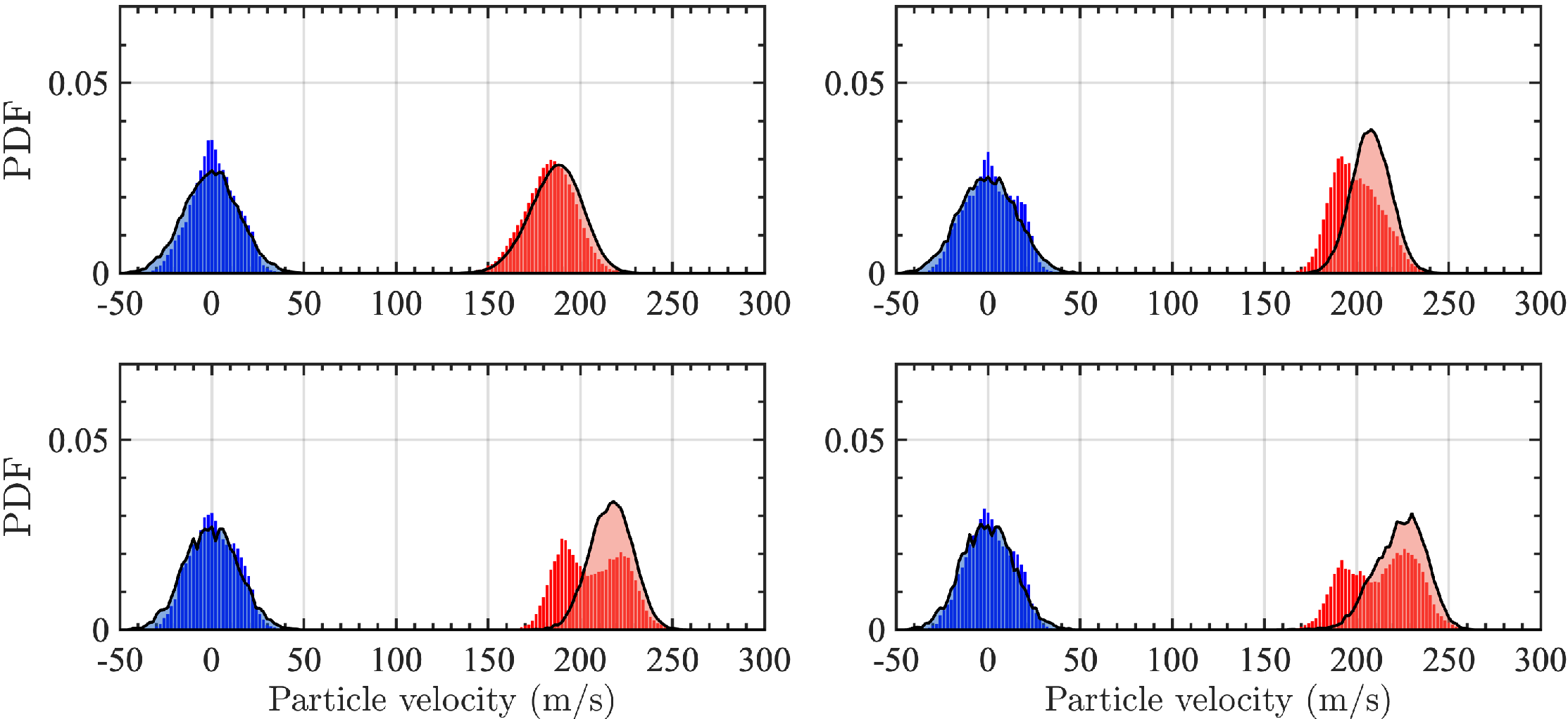}};
        \node[] at (3.5ex,-0.5ex) {(\textit{a})};
        \node[] at (39.5ex,-0.5ex) {(\textit{b})};
        \node[] at (3.5ex,-17ex) {(\textit{c})};
        \node[] at (39.5ex,-17ex) {(\textit{d})};
    \end{tikzpicture}
\caption{Velocity PDFs for Case B2 measured in the region (a) $0\leq x/D_e <1$, (b) $1 \leq x/D_e <2$, (c) $2 \leq x/D_e <3$ and (d) $3 \leq x/D_e <4$. Same legend as figure~\ref{fig: injecting velocity pdfs}.}
\label{fig: case_B2}
\end{figure}

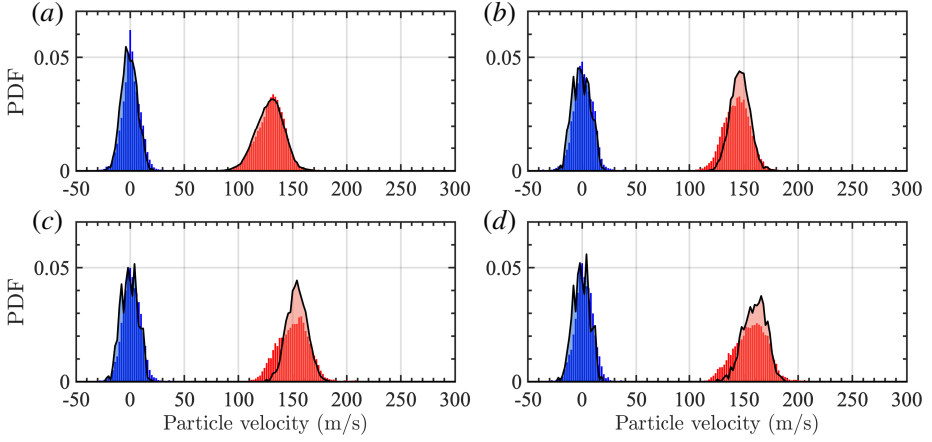
\begin{figure}
 \setlength{\lineskip}{0pt}
    \centering
    \begin{tikzpicture}
        \node[anchor=north west,inner sep=0pt] at (0,0){\includegraphics[width= 0.9\textwidth]{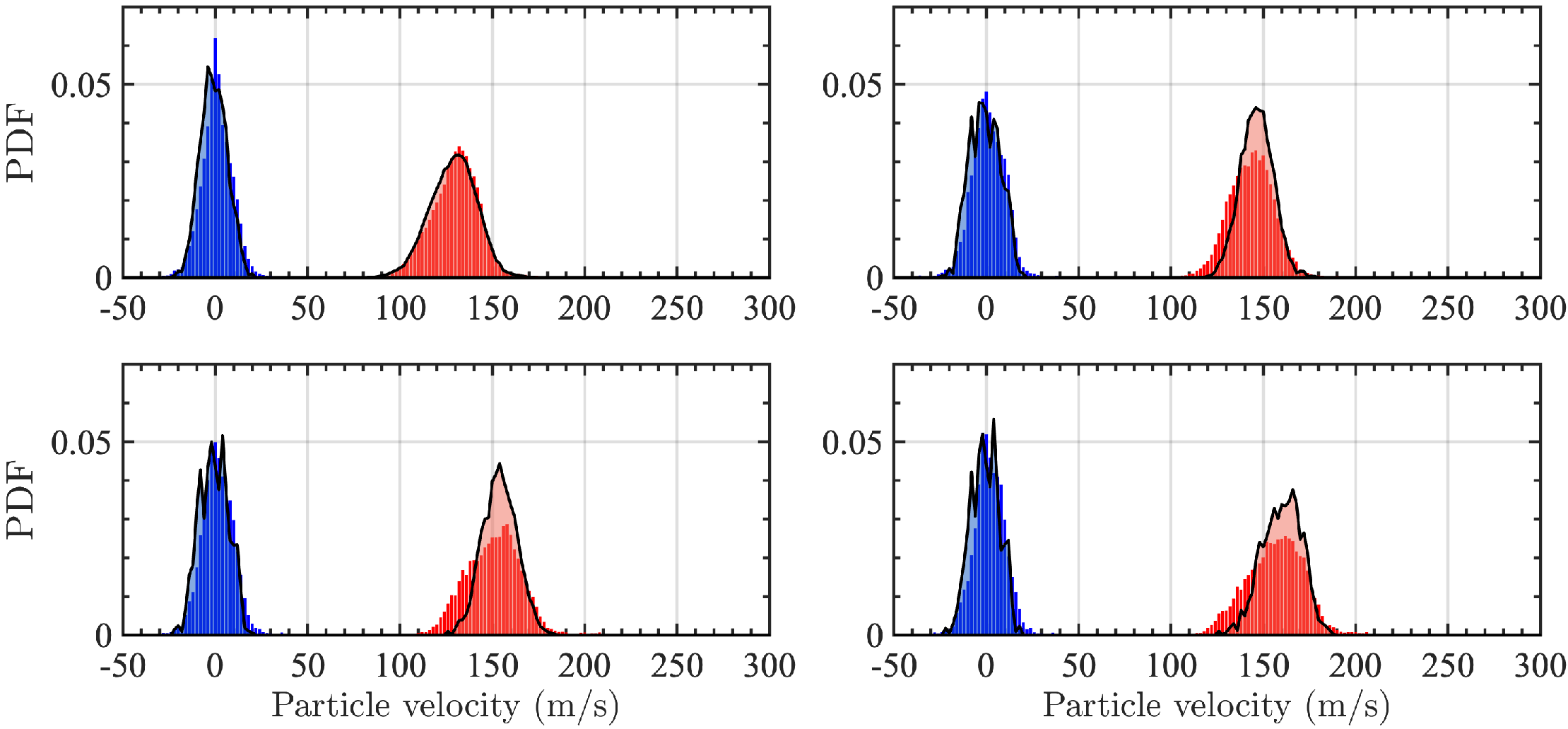}};
        \node[] at (3.5ex,-0.5ex) {(\textit{a})};
        \node[] at (39.5ex,-0.5ex) {(\textit{b})};
        \node[] at (3.5ex,-17ex) {(\textit{c})};
        \node[] at (39.5ex,-17ex) {(\textit{d})};
    \end{tikzpicture}
\caption{Velocity PDFs for Case B3 measured in the region (a) $0\leq x/D_e <1$, (b) $1 \leq x/D_e <2$, (c) $2 \leq x/D_e <3$ and (d) $3 \leq x/D_e <4$. Same legend as figure~\ref{fig: injecting velocity pdfs}.}
\label{fig: case_B3}
\end{figure}

\subsection{Effect of filtering during two-way coupling}\label{app:2way}
As discussed in \S\ref{sec:2way}, the particle models are based on correlations that depend on \textit{undisturbed} flow quantities. In two-way coupled simulations, the particle modifies the local flow and thus the undisturbed quantities must be reconstructed. Although strategies exist for incompressible flows \citep[e.g.][]{horwitz2016accurate,horwitz2018correction,balachandar2019self,liu2019self,pakseresht2020correction,pakseresht2021disturbance}, no formal approach yet exists for compressible flows.

In this work, we propose to remove the local disturbance caused by the particle by applying a low-pass filter to the Eulerian field prior to interpolation per \eqref{eq:interp}. Such an approach has shown success in previous studies~\citep{evrard2020euler,superspnic_EL_filter}. Numerical experiments revealed that two-way coupling has minimal effect on local gradients in gas-phase pressure and velocity. In the presence of shock waves, filtering will severely dampen the undisturbed gradients. Thus, for the particle force balance in \eqref{eq:dvdt}, we propose to filter the gas-phase velocity prior to interpolation but not the pressure gradient or viscous stress. 

\begin{figure}\centering
\subfloat{%
  \begin{tikzpicture}
        \node[anchor=north west,inner sep=0pt] at (0,0){\includegraphics[width= 0.45\textwidth]{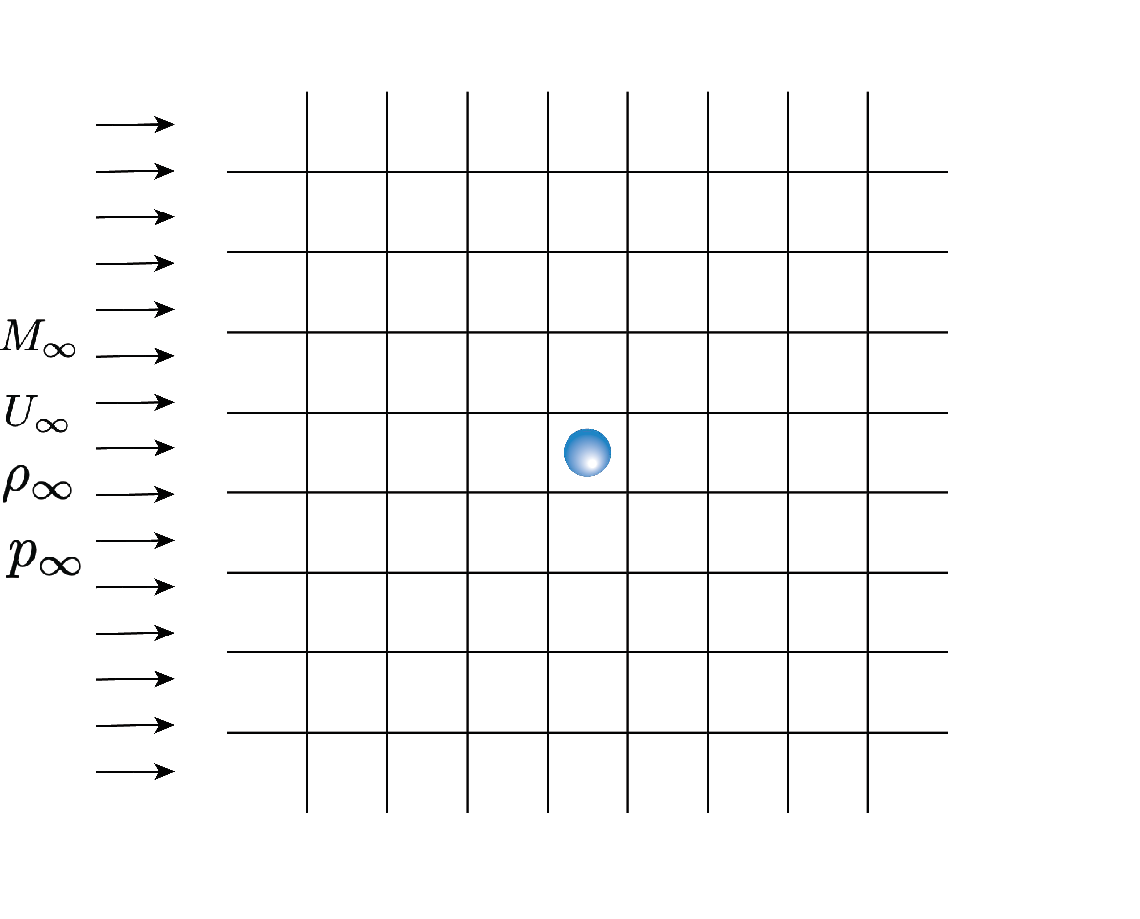}\label{subfig: uniform_filter_test}};
        \node[] at (0ex,0ex) {(\textit{a})};
    \end{tikzpicture}
}
\quad
\subfloat{%
  \begin{tikzpicture}
        \node[anchor=north west,inner sep=0pt] at (0,0){\includegraphics[width= 0.45\textwidth]{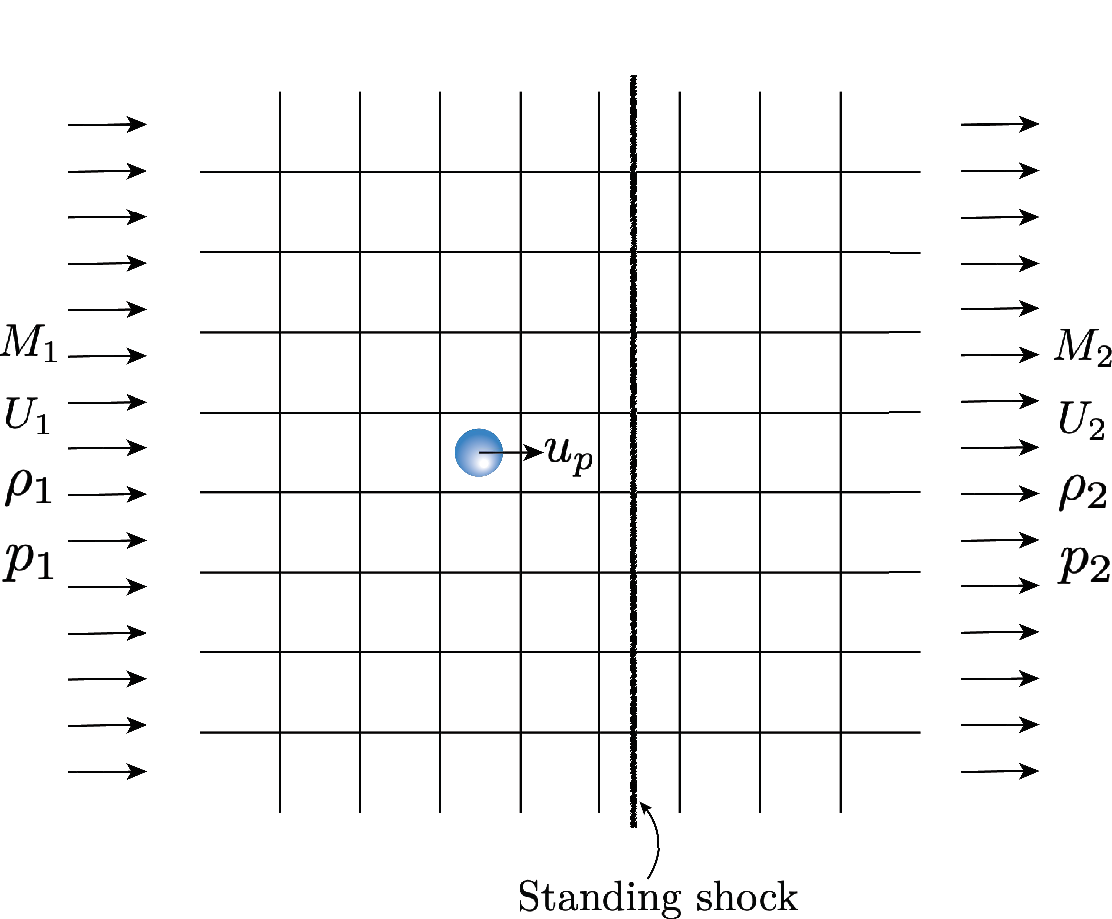}\label{subfig: standing_shock_filter_test}};
        \node[] at (0ex,0ex) {(\textit{b})};
    \end{tikzpicture}
}
\caption{Two configurations considered to assess the effect of filtering during of two-way coupling. (a) Uniform flow past a frozen particle and (b) particle passing through a standing shock.}
\label{fig: filter_testing_schematics}
\end{figure}

To test the efficacy of this approach, we consider two canonical three-dimensional cases (see figure~\ref{fig: filter_testing_schematics}): Case C1 corresponds to a uniform supersonic flow past a stationary particle and Case C2 corresponds to a particle moving with constant velocity (with $M_p=2$) passing through a standing shock. The parameters listed in table~\ref{table:five} are chosen to be representative of the conditions particles experience in the underexpanded jet.

The accuracy of the scheme is measured by comparing the interpolated values to the corresponding values in a one-way coupled flow. We evaluate the streamwise interpolated velocity, $u_x$, and streamwise component of the resolved stress, $P_x  = \nabla_x \cdot \left( \boldsymbol{\tau} - p \boldsymbol{\mathrm I}\right)$ at the location of the particle. The corresponding errors, $\mathcal{E}_u$ and $\mathcal{E}_{P}$, are evaluated according to
\begin{equation}
    \mathcal{E}_u = \dfrac{u_x-{u}^{\text{un}}_x}{{u}^{\text{un}}_x}
\end{equation}
and 
\begin{equation}
    \mathcal{E}_{P} = \dfrac{P_x-P^{\text{un}}_x}{P^\text{un}_x},
\end{equation}
 where $(\cdot)^{\text{un}}$ denotes an interpolated quantity obtained from the corresponding undisturbed (one-way coupled) simulations. We report the  maximum error, which occurs at steady state in C1 and when the particle is at the location of the shock in C2.

\begin{table}
  \begin{center}
  \def~{\hphantom{0}}
  \begin{tabular}{cccc|cccccc}
   \hline 
       \multicolumn{4}{c}{C1}  &  \multicolumn{5}{c}{C2}    \\
       \hline
        $M_{\infty}=M_p$&$Re_p$&$\rho_{\infty}~(\mathrm{kg/m^3})$ & $p_{\infty}~(\mathrm{kPa})$& &$M_1$ & $M_p$ & $Re_p$ &$p_2/p_1$ &$\rho_p/\rho_1$  \\
        \hline
       2 & 2000&0.96&35.4 & & 2.8 & 2 & 2000 &14.12 &2018   \\ 
       \hline
  \end{tabular}
  \caption{Simulation parameters associated with the test cases shown in figure~\ref{fig: filter_testing_schematics}.}
  \label{table:five}
  \end{center}
\end{table}

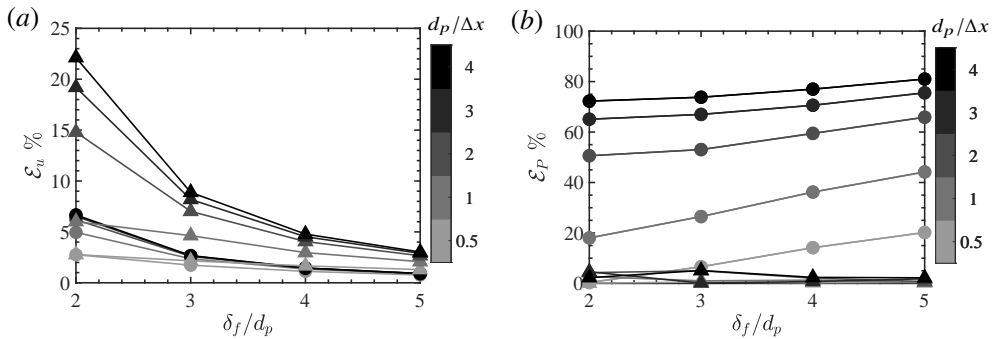
\begin{figure}\centering
\subfloat{%
  \begin{tikzpicture}
        \node[anchor=north west,inner sep=0pt] at (0,0){\includegraphics[width= 0.45\textwidth]{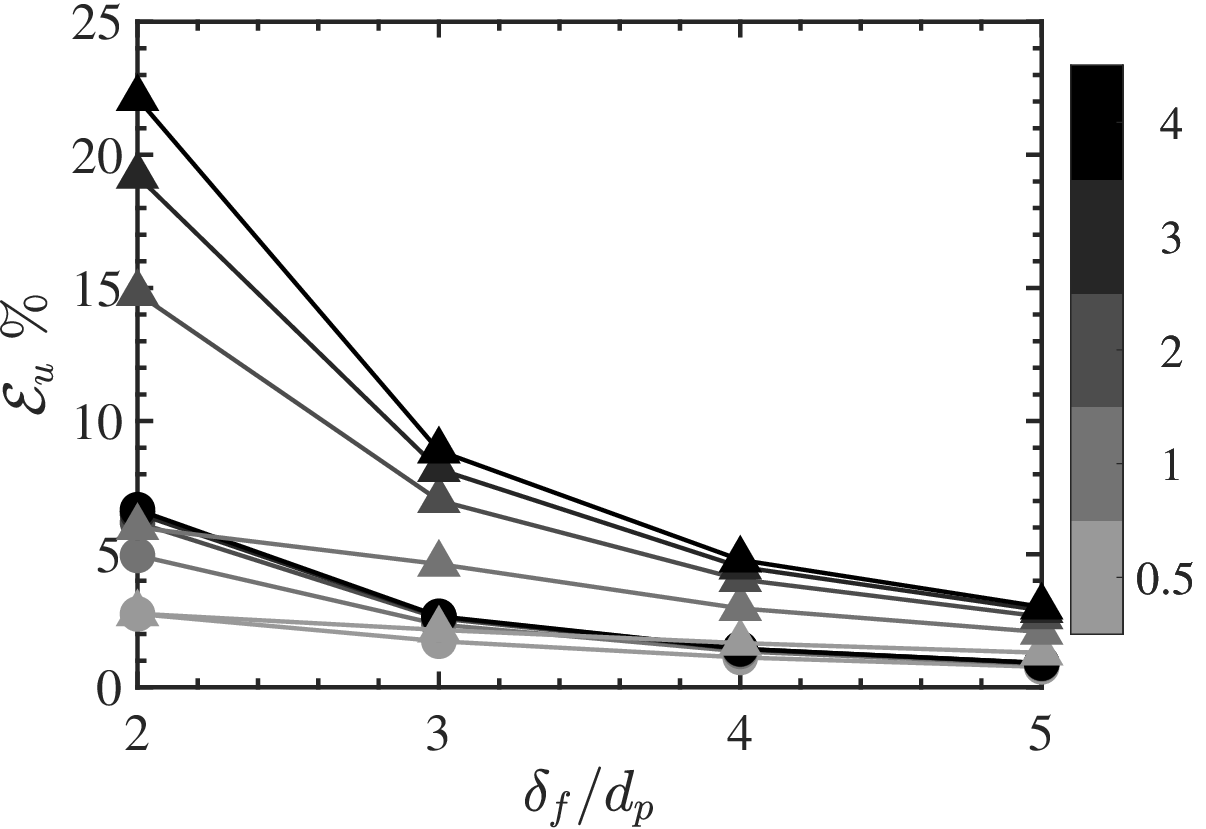}\label{subfig: uniform_deltaf_dia}};
        \node[] at (0ex,0ex) {(\textit{a})};
        \node[] at (34.5ex,-0.5ex) {\scriptsize{$d_p/\Delta x$}};
    \end{tikzpicture}
}
\subfloat{%
  \begin{tikzpicture}
        \node[anchor=north west,inner sep=0pt] at (0,0){\includegraphics[width= 0.45\textwidth]{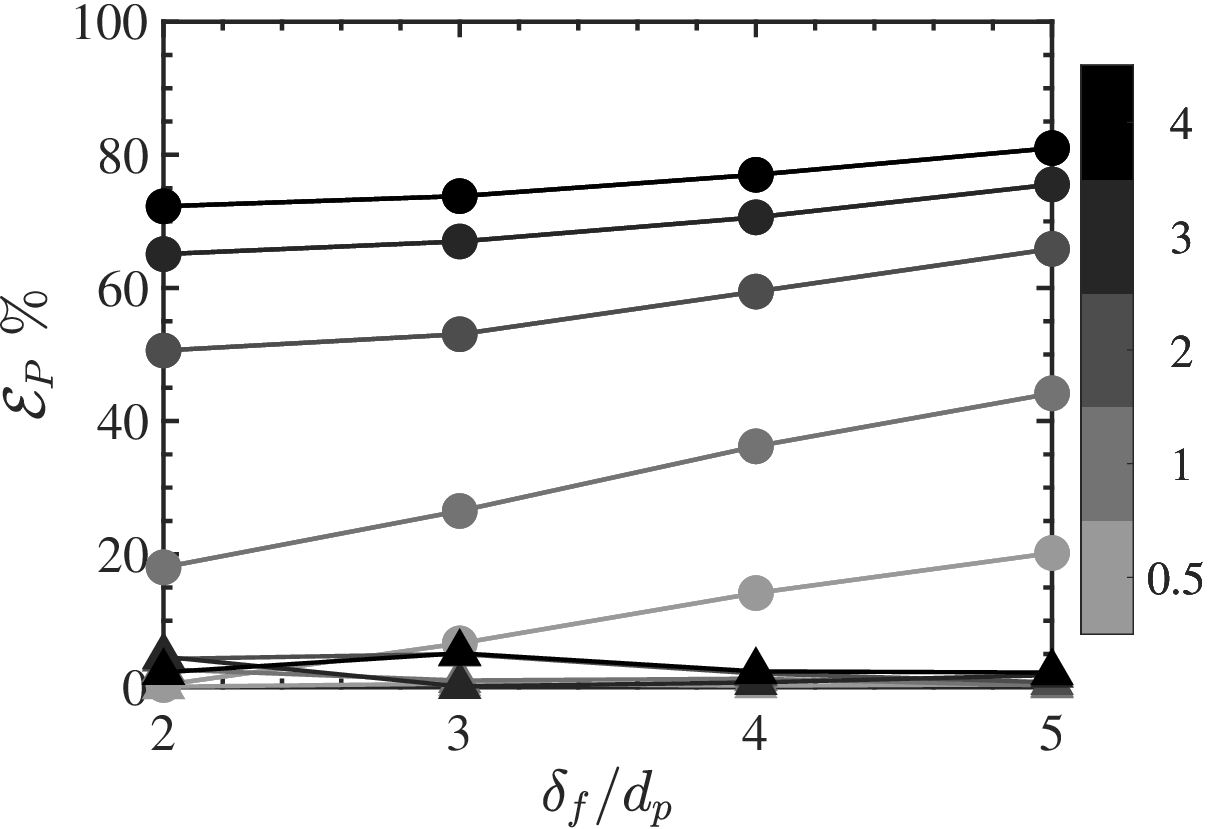}\label{subfig: standing_deltaf_dia}};
        \node[] at (0ex,0ex) {(\textit{b})};
        \node[] at (34.5ex,-0.5ex) {\scriptsize{$d_p/\Delta x$}};
    \end{tikzpicture}
}
\caption{Effect of filtering when evaluating gas-phase quantities at the particle location. (a) Error in the interpolated velocity for Case C1 and (b) error in the interpolated stress for Case C2. Circles indicate runs when filtering is applied prior to interpolation and triangles indicates runs without filtering.}
\label{fig: filter_testing_results}
\end{figure}

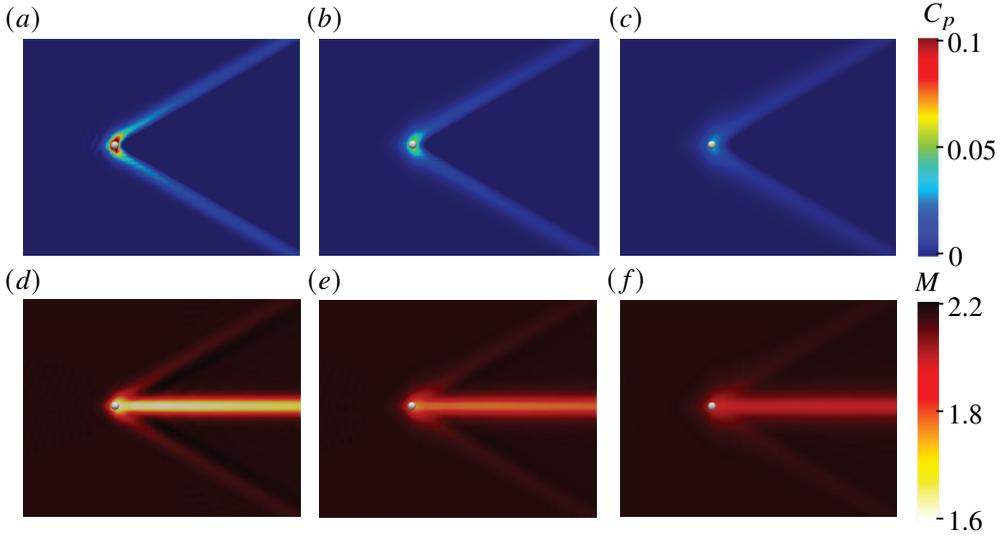
\begin{figure}\centering
\subfloat{%
  \begin{tikzpicture}
        \node[anchor=north west,inner sep=0pt] at (0,0){\includegraphics[width= 0.9\textwidth]{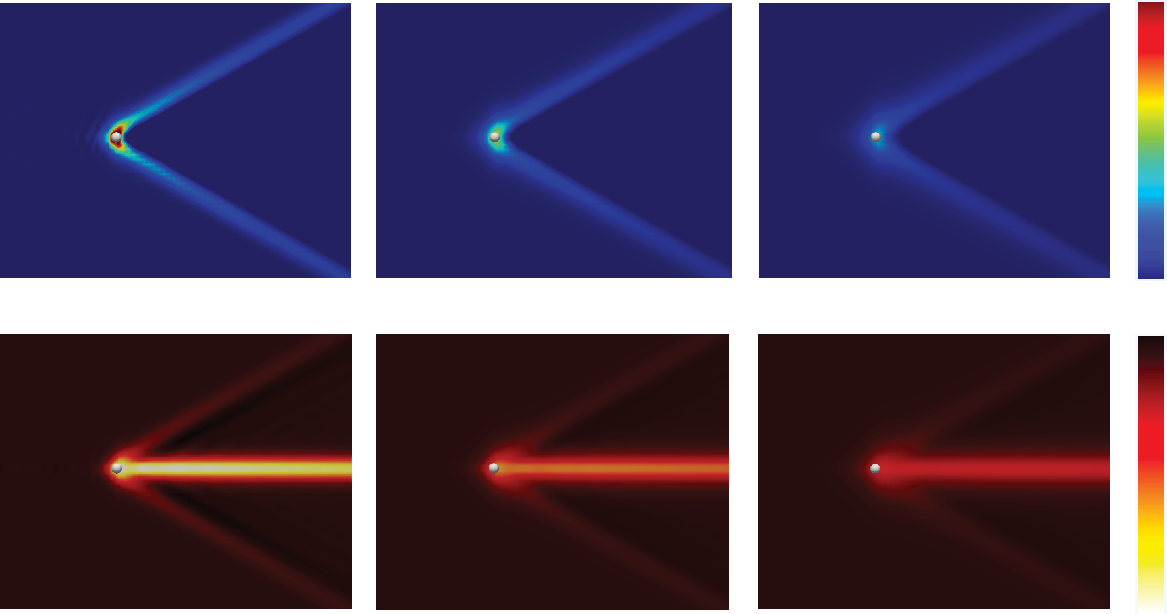}\label{subfig: uniform_deltaf_2_4dp}};
        \node[] at (0ex,1.25ex) {$(a)$};
        \node[] at (24ex,1.25ex) {$(b)$};
        \node[] at (48ex,1.25ex) {$(c)$};
        \node[] at (0ex,-19.5ex) {$(d)$};
        \node[] at (24ex,-19.5ex) {$(e)$};
        \node[] at (48ex,-19.5ex) {$(f)$};
        \node[] at (73ex,1.5ex) {$C_p$};
        \node[] at (72ex,-19.5ex) {$M$};
        \node[] at (73ex,-0.6ex) {-};
        \node[] at (75ex,-0.6ex) {0.1};
        \node[] at (73ex,-9.05ex) {-};
        \node[] at (75ex,-9.05ex) {~~0.05};
        \node[] at (73ex,-17.5ex) {-};
        \node[] at (74.25ex,-17.5ex) {0};
        \node[] at (73ex,-21.5ex) {-};
        \node[] at (75ex,-21.5ex) {2.2};
        \node[] at (73ex,-30ex) {-};
        \node[] at (75ex,-30ex) {1.8};
        \node[] at (73ex,-38.5ex) {-};
        \node[] at (75ex,-38.5ex) {1.6};
    \end{tikzpicture}
} 
\caption{Simulation results of a uniform flow past a stationary particle (Case C1) showing local coefficient of pressure (top) and gas-phase Mach number (bottom). The filter size employed in the two-way coupling increases from left-to-right with (a,d) $\delta_f  = 2d_p$, (b,e) $\delta_f  = 3d_p$ and (c,f) $\delta_f  = 4d_p$.}
\label{fig: filter_width_testing_uniform}
\end{figure}

Figure~\ref{fig: filter_testing_results} shows the errors in the gas-phase velocity and stress evaluated at the particle location as a function of filter size, $\delta_f$, and particle diameter. In the case of a uniform supersonic flow past a particle (Case C1), the velocity evaluated at the location of the particle should be equal to the free-stream velocity. As the filter size increases, the interphase source terms sent to the fluid are progressively smeared out and the error decreases. However, as shown in figure~\ref{fig: filter_width_testing_uniform}, this also reduces the wake and bow shock that should be present. Filtering the gas-phase velocity prior to interpolation (denoted by circles) significantly reduces the error, even when the filter size is relatively low.

Because the flow is uniform, the undisturbed pressure gradient and viscous stress are null. Thus, we evaluate the accuracy of evaluating the fluid stress at the particle location in Case C2. It can be seen that filtering the stress prior to interpolation (circles) results in large error regardless of the filter size. The error is significantly lower (<5\% $\forall \delta_f/d_p$ and $\forall \Delta x/d_p$ considered) if the stress remains unfiltered prior to interpolation (triangles), which is what was employed in the simulations reported in the main text.
 
Figure~\ref{fig: filter_width_testing_uniform} shows the local coefficient of pressure, $C_p = 2(p-p_{\infty})/(\rho_{\infty} U^2_{\infty})$, and Mach number for the uniform flow past a sphere (Case C1) with $d_p=2\Delta x$. A bow shock and wake are clearly visible when the coupling terms are localized to the particle (i.e. small filter sizes). As one might expect, increasing the filter mollifies the momentum and pressure deficit resulting in weaker flow structures. 

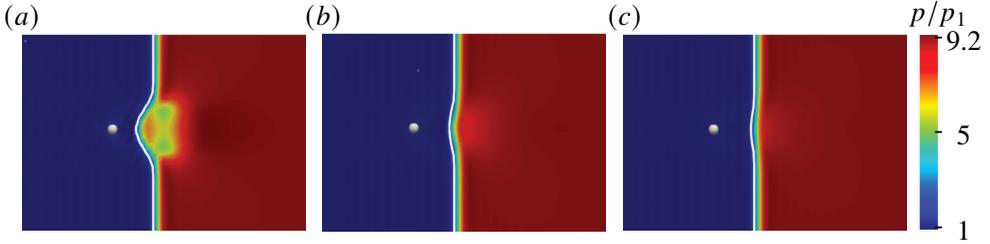
\begin{figure}\centering
\subfloat{%
  \begin{tikzpicture}
        \node[anchor=north west,inner sep=0pt] at (0,0){\includegraphics[width= 0.9\textwidth]{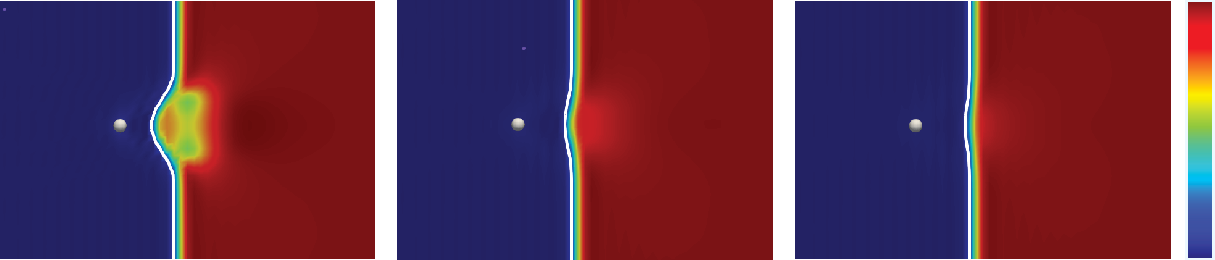}\label{subfig: standing_deltaf_2_4dp}};
        \node[] at (0ex,1.25ex) {$(a)$};
        \node[] at (24ex,1.25ex) {$(b)$};
        \node[] at (48ex,1.25ex) {$(c)$};
        \node[] at (73ex,1.5ex) {$p/p_1$};
        \node[] at (73ex,-0.6ex) {-};
        \node[] at (75ex,-0.6ex) {9.2};
        \node[] at (73ex,-8.1ex) {-};
        \node[] at (75ex,-8.1ex) {5};
        \node[] at (73ex,-15.6ex) {-};
        \node[] at (75ex,-15.6ex) {1};
    \end{tikzpicture}
} 
\caption{Instantaneous snapshots of Case C2 showing local gas-phase pressure (color) and particle position (white). The images correspond to the instant the particle is $4.5d_p$ upstream of the shock. The filter size employed in the two-way coupling increases from left-to-right with (a) $\delta_f  = 2d_p$, (b) $\delta_f  = 3d_p$ and (c) $\delta_f  = 4d_p$. Solid white line shows the extracted shock location.}
\label{fig: filter_width_testing_standing_shock}
\end{figure}

Figure~\ref{fig: filter_width_testing_standing_shock} shows normalized local pressure contours in Case C2 when the particle is $4.5d_p$ upstream of the standing shock. As with Case C1, the effect of two-way coupling is less pronounced as the filter size increases. Significant distortion of the shock can be observed when $\delta_f=2 d_p$. Small increases in the filter size have a pronounced effect, with almost no distortion when $\delta_f=4 d_p$.

It is important to note that a requirement of the volume-filtered Eulerian-Lagrangian framework is that $\delta_f> d_p$ when $d_p>\Delta x$ so that the volume fraction remains bounded $\alpha\in[0,1]$ and to avoid numerical instabilities associated with the source terms. Because the jet simulations considered a polydisperse distribution of particles, $\delta_f\gg d_p$ for the majority of particles, which likely contributes to the underprediction in the shift in Mach disk observed. This points to a need for improved two-way coupling strategies for Eulerian-Lagrangian simulations involving shock-particle interactions.

\clearpage
\bibliography{jfm.bib}
\bibliographystyle{jfm}

\end{document}

%% file: tikz/Henderson_comp.tikz
\begin{tikzpicture}[scale=1]
\pgfmathsetlengthmacro\MajorTickLength{
      \pgfkeysvalueof{/pgfplots/major tick length} * 0.75
    }
\pgfplotsset{every axis y label/.append style={yshift=-0.3cm}}
\begin{axis}[%
	width=1.0\figurewidth,
	height=\figureheight,
	scale only axis,
        axis line style = thick,
	xmin=0, xmax=4,
	xtick={0,1,...,4},
	xminorticks=true,
	xlabel={$x/D_e$},
	ymin=0.5, ymax=2.35,
	enlarge y limits=upper,
	yminorticks=true,
        major tick length=\MajorTickLength,
        minor tick length=0.6*\MajorTickLength,
        every tick/.style={
                black,
                semithick,
              },
	ylabel={$M$},
        y label style={at={(-0.125,0.5)},rotate=-90},
	minor tick num=3
	]
	\addplot [
		color=black,
		solid,
		only marks,
		mark=triangle,
		mark color=white,
		mark size=\mymarksize,
		mark options={scale=1.2,fill=white},
		forget plot
		]
	table[x expr=\thisrowno{0}, y expr=\thisrowno{1}] {./tikz/henderson.dat};
	\label{leg:henderson}
	\addplot [
	    smooth,
	    tension={0.02},
	    ultra thick,
		color=black,
		solid,
		line width=1,
		forget plot
		]
	table[x expr=\thisrowno{0}, y expr=\thisrowno{1}] {./tikz/sim_Ma.dat};
	\label{leg:sim_henderson}
	\addplot [
	    smooth,
	    tension={0.02},
	    ultra thick,
		color=black,
		dashed,
		line width=1,
		forget plot
		]
	table[x expr=\thisrowno{0}, y expr=\thisrowno{1}] {./tikz/theory_edit.dat};
	\label{leg:theory}
\end{axis}

\end{tikzpicture}